\documentclass[reprint,pdftex,nofootinbib]{revtex4-1}
\usepackage{amsmath,amssymb,amsthm,undertilde}
\usepackage{graphicx,color}
\usepackage[pdftex,colorlinks]{hyperref}

\newcommand{\bra}[1]{\left\langle #1 \right|}
\newcommand{\ket}[1]{\left| #1 \right\rangle}

\newcommand{\scpi}[0]{\mbox{\normalsize$\pi$}}
\numberwithin{equation}{section}
\begin{document}

\title{Non-Markovian Dynamics of Open Quantum Systems:\\Stochastic Equations and their Perturbative Solutions}
\author{C.~H.~Fleming}
\affiliation{Joint Quantum Institute and Maryland Center for Fundamental Physics, University of Maryland, College Park, Maryland 20742}
\author{B.~L.~Hu}
\affiliation{Joint Quantum Institute and Department of Physics, University of Maryland, College Park, Maryland 20742}
\date{\today}

\begin{abstract}
We treat several key stochastic equations for non-Markovian open quantum system dynamics and present a formalism for finding solutions to them via canonical perturbation theory, without making the Born-Markov or rotating wave approximations (RWA).
This includes master equations of the (asymptotically) stationary, periodic, and time-nonlocal type.
We provide proofs on the validity and meaningfulness of the late-time perturbative master equation and on the preservation of complete positivity despite a general lack of Lindblad form.
More specifically, we show how the algebraic generators satisfy the theorem of Lindblad and Gorini, Kossakowski and Sudarshan, even though the dynamical generators do not.
These proofs ensure the mathematical viability and physical soundness of solutions to non-Markovian processes.
Within the same formalism we also expand upon known results for non-Markovian corrections to the quantum regression theorem.
Several directions where these results can be usefully applied to are also described, including the analysis of near-resonant systems where the RWA is inapplicable and the calculation of the reduced equilibrium state of open systems.
\end{abstract}

\maketitle

\tableofcontents

\section{Introduction}

An open quantum system (OQS) is a quantum system that interacts with some environment whose degrees of freedom have been coarse-grained over.%
\footnote{Alternatively called `integrated over' or `traced out', but note that their effects are far from being eliminated: the environment backreacts on the system imparting dissipative dynamics and quantum decoherence effects in the, thus reduced, `open system'.
The main concern for OQS studies is to describe these effects on the otherwise closed quantum system}
A general dynamical non-Markovian environment generates noise, and brings forth dissipation and decoherence on the system in a way far more complex than the commonly encountered Markovian (white noise) counterpart, whose treatment can be argued from purely phenomenological grounds.
Specifically for the reduced density matrix $\boldsymbol{\rho}$ and corresponding single-time correlations of the open system, the non-Markovian dynamics are described by a master equation.
Exact master equations for the stochastic dynamics of open quantum systems are, in general, out of reach.
However, master equations in the form of an arbitrary perturbative order (in the system-environment interaction) can be derived in a variety of ways \cite{Kampen97,Breuer03,Strunz04}
and find application in many branches of physics and chemistry \cite{Pollard97,Carmichael99,Breuer02,Kampen07}.
Multi-time correlations add an extra layer of complexity in the non-Markovian regime (e.g. finite temperature, cutoff, etc.) as we shall discuss.

In this work we develop a formalism for quantum open systems which closely mirrors the more well-known canonical perturbation theory applied to Schr\"{o}dinger's equation. For completeness we include some essential background material, with the purpose of clarifying some previous confusions in concepts,  some arising from non-strict usage of terminology. We identify new results we obtained in the subsection below which require a thorough buildup of foundational material to derive and explain. We also point out connections with recent work derived from these results. We describe the organization of this paper and a brief summary of each section below. 
\subsection{Summary}

In Sec.~\ref{sec:Prelim} we start with some preliminary information on open systems pertaining to the mathematical structure of non-unitary (dissipative) evolution.
In particular, we wish to distinguish among the various usages of the term \emph{Markovian} with regard to their distinct properties.
We place the most emphasis on distinguishing the Markovian \emph{representation}, which can be rather superficial, and the Markovian \emph{process}, which is much more important.
Finally we carefully note the distinction between the instantaneous dynamical semi-group and all-time algebraic semi-group when employing the Lindblad-GKS theorem \cite{Lindblad76,Gorini76}.

In Sec.~\ref{sec:TLME} we present a new and much simplified derivation of the perturbative time-local master equation, which is fairly well-known, and focus primarily upon the second-order master equation.
It is often suspected that these perturbative master equations cannot be employed for significant lengths of time,
but we find this to be more specifically determined by the noise distribution.
We also explicitly demonstrate how one can test for the complete positivity of a non-Markovian master equation which is not of Lindblad form. The microscopically derived master equation indeed passes this test at second order.
Perturbative solutions are detailed for the asymptotically stationary and cyclo-stationary (periodic) master equation.
Later in Sec.~\ref{sec:QOSn} we derive the dual time-nonlocal master equation and determine it to be equivalent to the appropriate perturbative order, at least asymptotically.

In Sec.~\ref{sec:QRT} we rederive the perturbative non-Markovian Quantum Regression Theorem (QRT) corrections in a simple manner and examine their structure more carefully.
It is shown that, even at late times when the master equation has settled down into its stationary limit, the dynamics of the system remain non-Markovian as they are expected to.

In Sec.~\ref{sec:QLE} we briefly discuss the quantum Langevin equation, which we generalize to handle nonlinear systems and couplings.
The Langevin equation provides a better perspective for determining the effective forces and renormalization terms induced by the environment.
In the master-equation perspective, the effective forces and self interactions are not so easily delineated from other unitary generators induced by the environment.

In Sec.~\ref{sec:QNC} we give some categorization of the fundamental object of all second-order open-system dynamics, regardless of formalism: the environmental correlation function.
Highlighted are the decomposition into fluctuations and dissipation, and their relation.
Some passing mention is given to the correlation function's important role in categorizing the \emph{decoherence strength} and a newly discovered \emph{fluctuation-dissipation inequality}, described more thoroughly in Ref.~\cite{Decoherence,FDR} respectively.
Finally, in Sec.~\ref{sec:Equilibrium} we discuss the correlations of thermal reservoirs.
The correspondences between the fluctuation-dissipation, Kubo-Martin-Schwinger (KMS) relations, Boltzmann distribution, and detailed balance are drawn.

\subsection{New Results and Derivatives}
Here we wish to itemize our most important contributions in this work and cite derivative works which have spawned from these,
though this work contains numerous additional results.
\begin{itemize}
\item Our first result is the \emph{proof of perturbative complete positivity} for master equations which generate non-Markovian dynamics and therefore do not take the Lindblad-GKS form.
This result was expanded upon and applied in Ref.~\cite{Decoherence} to demonstrate that the environmental correlation function can be used as a comparative measure of the amount of irreversible dynamics induced upon the open system,
which we refer to as the \emph{decoherence strength} of the environment.
\item A key result of this work is the discovery of \emph{perturbative solutions} for general systems and environments by the application of canonical perturbation theory to non-Hermitian super operators.
This result allowed a careful analysis of the RWA in Ref.~\cite{RWA}, which precisely demonstrated its shortcomings as compared to the full second-order solutions.
Later this result was applied in Ref.~\cite{Accuracy} to focus upon the unfortunate discovery that \emph{all perturbative master equations generate solutions of lesser accuracy}.
Then, in Ref.~\cite{Correlations}, this result was applied to the analysis of the evolution of systems and environments with proper initial correlations, whereas the standard treatment is to consider an initially uncorrelated system and environment.
Most recently this result has been used to calculate the correct asymptotic state of the open system for non-vanishing interaction with a heat bath \cite{Dipole,Equilibrium},
which is, in general, not the equilibrium state of the closed system.
\item We have discovered a generalization of the fluctuation-dissipation relation (FDR), which is an inequality (FDI) that applies to arbitrary, non-equilibrium environments.
In Ref.~\cite{FDR} this result was expanded upon in great detail and a strict correspondence was drawn between our FDI and the Heisenberg uncertainty principle (HUP).
\end{itemize}

\section{Essential background in newer perspectives}

\subsection{Markovian versus Non-Markovian Dynamics}\label{sec:Prelim}

The environment's effect on the system can, under specified condition, be represented as noises or stochastic forces exerting different influences on the system.
These noises can have correlations (memory) which reflect differing timescales of the environment (e.g. the temperature of a heat bath).
Such noises are said to be \emph{colored} or \emph{correlated} and the processes containing memories they engender or are associated with are called \emph{non-Markovian}.%
\footnote{We urge to use `colored' or `correlated' for the nature of noise and save the word \emph{non-Markovian} for processes with memories.}
In the \emph{Markovian limit} the timescales of the environment are taken to be much shorter than the timescales of the system.
Such noise is said to \emph{white} and the process \emph{Markovian}; its timescales cannot be resolved by the system.
\emph{Quantum noise} correlations are complex, containing both real noise and dissipation (see Sec.~\ref{sec:NoiseDecomp}).
Thus we say that for quantum noise, the Markovian limit corresponds to all such memories being of insignificant duration.
This is the more physical notion of the Markov property, as given non-Markovian dynamics in this sense
(1) the quantum regression theorem does not apply \cite{Swain81},
(2) one cannot add Hamiltonian terms to the master equation post derivation (e.g. see \cite{QBM}: Sec.~7.1),
(3) one cannot add dissipative terms to the master equation post derivation (e.g. see \cite{Scala07}).

An open-system master equation is a linear map for the reduced density matrix generated via
\begin{align}
\dot{\boldsymbol{\rho}} &= \boldsymbol{\mathcal{L}} \{ \boldsymbol{\rho} \} \, , \label{eq:ME0} \\
\boldsymbol{\mathcal{L}}_0 \{ \boldsymbol{\rho} \} &= \left[ -\imath \mathbf{H} , \boldsymbol{\rho} \right] \, ,
\end{align}
where, in our open system formalism, $\mathbf{H}$ denotes the free system Hamiltonian and $\boldsymbol{\mathcal{L}}_0$ the free system Liouville operator in the Schr\"{o}dinger picture.%
\footnote{It is standard to consider Hilbert-space operators (e.g. $\boldsymbol{\rho}$) as \emph{super-vectors}
and linear operators (e.g. $\boldsymbol{\mathcal{L}}$) on super-vectors as \emph{super-operators}.
In this representation, the master equation \eqref{eq:ME0} is simply a linear differential equation akin to the Schr\"{o}dinger equation when written in terms of operators.}
The master equation is said to have \emph{time-local}, \emph{time-convolutionless}, or \emph{Markovian representation}
if $\boldsymbol{\mathcal{L}}(t)$ does not reference the past history of states $\boldsymbol{\rho}(\tau)$ for $\tau<t$.
A Markovian representation does not necessarily imply Markovian dynamics.
Non-Markovian processes can readily admit Markovian representations for their coarsegrained behavior, e.g. the exact \emph{Quantum Brownian Motion} master equation of HPZ~\cite{HPZ92,QBM}.
Moreover, for any non-Markovian master equation of the form
\begin{align}
\dot{\boldsymbol{\rho}}(t) &= \int_0^t \!\! d\tau \, \boldsymbol{\mathcal{K}}(t\!-\!\tau) \, \boldsymbol{\rho}(\tau) \, , \label{eq:MElocal}
\end{align}
with invertible propagator $\boldsymbol{\mathcal{G}}(t): \boldsymbol{\rho}(0) \to \boldsymbol{\rho}(t)$, there is a corresponding time-local master equation
\begin{align}
\dot{\boldsymbol{\rho}}(t) &= \boldsymbol{\mathcal{L}}(t) \, \boldsymbol{\rho}(t) \, , \label{eq:MEnonlocal}
\end{align}
where $\boldsymbol{\mathcal{L}}(t) = \dot{\boldsymbol{\mathcal{G}}}(t) \, \boldsymbol{\mathcal{G}}^{-1}(t)$ trivially and in this case $\boldsymbol{\mathcal{G}}(t)$ and thus the master equation can be formally determined via Laplace transformation.
Eq.~\eqref{eq:MElocal} and \eqref{eq:MEnonlocal} are fully equivalent and can both generate non-Markovian dynamics despite any Markovian representation;
all memory effects are encoded in the full time dependence of the master equation coefficients themselves.

Given a time-local master equation, an $\boldsymbol{\mathcal{L}}$ constant in time is said to be \emph{homogeneous} while an $\boldsymbol{\mathcal{L}}(t)$ variable in time is said to be \emph{inhomogeneous}.
Some authors in the physics community refer to this distinction as the Markovian /non-Markovian distinction as such processes can lead to this kind of evolution.
However, time-dependence in $\boldsymbol{\mathcal{L}}(t)$ is also not specific to non-Markovian dynamics,
as one can couple a time-dependent Hamiltonian system to a Markovian environment.
For time-dependent master equations which asymptote into a homogeneous form, this is known as the \emph{stationary limit} of the master equation.
A well-defined stationary limit is also not specific to Markovian processes, e.g. the exact QBM master equation for a regulated Ohmic coupling.
Nor does a stationary limit imply Markovian dynamics therein, as can be seen by a damped quantum oscillator's response to forces in Ref.~\cite{Xu03,QBM}.

In the context of quantum open systems, Markovian processes generally result in \emph{Lindblad equations}.
Lindblad equations generate completely-positive evolution for all states at all times.
However not all master equations which generate completely-positive evolution are of Lindblad form,
and not all Lindblad equations arise from Markovian processes.
The relationship between Markov, Lindblad, and complete positivity will be discussed more thoroughly in Sec.~\ref{sec:CP}.
A conflation of the stochastic process and representation often results in properties of Markovian dynamics being incorrectly applied to master equations of Markovian representation.

\subsubsection{Quantum Regression Theorem}
For a closed system, or open system driven by Markovian processes, all multi-time correlations of the system can be generated by the Liouville operator for the master equation $\boldsymbol{\mathcal{L}}(t)$, or more specifically its super-adjoint $\boldsymbol{\mathcal{L}}^\dagger(t)$, Sec.~\ref{sec:Ldagger}.
This is what is known as the \emph{Quantum Regression Theorem} (QRT), e.g. for two-time correlations we have
\begin{equation}
\left\langle \dot{\mathbf{X}}_1(t_1) \, \mathbf{X}_2(t_2) \right\rangle = \Bigl\langle \left\{ \boldsymbol{\mathcal{L}}^\dagger \, \mathbf{X}_1 \right\}\!(t_1) \, \{ \mathbf{X}_2 \}(t_2) \Bigr\rangle  \, .
\end{equation}
But for a non-Markovian process, the open-system master equation cannot generate all multi-time correlations despite whatever \emph{representation} the master equation may have.
This is, in fact, a defining characteristic of the Markovian process and has nothing to do with quantum mechanics specifically. In classical terms, the Markov property can be defined in terms of probabilities of events or expectation values of observables.
Extending this definition to the expectation values of quantum observables, the Markov property is rather synonymous with the QRT.
Other definitions of the (quantum) Markov property may not admit a classical correspondence with what is well understood to be Markovian and non-Markovian in classical probability theory.

The non-Markovian corrections to the QRT have been known for almost three decades~\cite{Swain81}.
This was first reported via the projection operator method and has recently been duplicated and expanded upon with stochastic Schr\"{o}dinger-equation techniques~\cite{Vega06}.
These corrections do not vanish in the weak coupling regime, but strictly in the white-noise limit.
In Sec.~\ref{sec:QRT} we rederive the non-Markovian corrections to the QRT in a simple perturbative fashion and
express them in a form which demonstrates their inherent non-Markovian character.

\subsubsection{The Validity of Combining Master Equations} 

As a Markovian process is without memory, its response is somewhat system-agnostic.
This allows one to combine open-system master equations by the simple addition of their individual individual super operators,
under the assumption that the different underlying processes are uncorrelated.
Let us consider the reduced density matrices of systems A and B coupled to identical dissipative environments.
We then have the open-system master equations
\begin{align}
\dot{\boldsymbol{\rho}}_\mathrm{A} &= -\imath[ \mathbf{H}_\mathrm{A} , \boldsymbol{\rho}_\mathrm{A} ] + \boldsymbol{\delta\! \mathcal{L}}_\mathrm{A} \{ \boldsymbol{\rho}_\mathrm{A} \} \, , \\
\dot{\boldsymbol{\rho}}_\mathrm{B} &= -\imath[ \mathbf{H}_\mathrm{B} , \boldsymbol{\rho}_\mathrm{B} ] + \boldsymbol{\delta\! \mathcal{L}}_\mathrm{B} \{ \boldsymbol{\rho}_\mathrm{B} \} \, ,
\end{align}
where the Hamiltonians are those of the free systems and the corrections to the Liouvillian are introduced via interaction with the dissipative environment.
In the Hamiltonian formalism one can simply add two Hamiltonians and arrive at another Hamiltonian, though one might be motivated to fix the energy spectrum through renormalization.
One cannot do this with non-Markovian Liouvillians, e.g. given some subsystem coupling $\mathbf{H}_\mathrm{AB}$ one cannot simply add dissipative terms.
\begin{align}
\dot{\boldsymbol{\rho}}_\mathrm{A+B} & \neq  -\imath[ \mathbf{H}_\mathrm{A} + \mathbf{H}_\mathrm{B} + \mathbf{H}_\mathrm{AB} , \, \boldsymbol{\rho}_\mathrm{A+B} ] \nonumber \\
& \phantom{\neq} + \boldsymbol{\delta\! \mathcal{L}}_\mathrm{A} \{ \boldsymbol{\rho}_\mathrm{A+B} \} + \boldsymbol{\delta\! \mathcal{L}}_\mathrm{B} \{ \boldsymbol{\rho}_\mathrm{A+B} \} \, , \label{eq:combo_master_eq}
\end{align}
The above (incorrect) master equation is in general completely different from the correct open-system master equation derived from first principles.
For non-Markovian processes, the environmental contributions have a nontrivial dependence (due to memory effects) upon the systems' dynamics through their couplings.
If one changes the system Hamiltonian, then one must also change the environmental contributions to be compatible with the history these new terms will create.
This is how memory exhibits itself in a time-local representation.
Moreover, one must also take into account whether or not the dissipative environments are separate or shared.
If the dissipative environment is shared then the two subsystems can interact via environmental back-reaction, even in the Markovian regime.
This effect is also missed when simply combining the Liouville operators.

In a general non-Markovian master equation, the invalidity of the above incorrect master equation would be readily apparent as it would likely violate positivity, uncertainty, etc.
Positivity violation will not occur when adding Lindblad terms.
But if one has a non-Markovian Lindblad equation, such as given by the rotating-wave approximation, then the mistake has only become more subtle and therefore more dangerous.
The master equation might be completely positive, but it does not correspond to the dynamics of the physical system considered.

The above example of interacting systems coupled to an environment should not be surprising.
If the environment is thermal, then to zeroth-order in the system-environment interaction, the systems should relax to a Boltzmann state which includes the system-system interaction Hamiltonian.
Obviously this can only happen if the master equation coefficients are aware of the system-system interaction.
However, this same mistake is often applied in more subtle ways.
For instance, if the system is driven by an external force, then due to non-Markovian response the driving terms in the Hamiltonian theory and in the open-system master equation are generally inequivalent~\cite{Xu03,QBM}.
This is to compensate for the difference between the actual nonlocal response of the system and the superficial time-local representation of the master equation.

This issue has also been commented on in the context of cavity QED.
The often-used master equation includes the Hamiltonian for the atom, intracavity field, and atom-field interaction,
but the master equation used is exactly that of an empty cavity with dissipation plus that of an atom spontaneously emitting into empty space. This situation is like that depicted in \ref{eq:combo_master_eq}. 
Indeed, if one begins instead with the atom-cavity system and derives the microscopic master equation using the standard technique \cite{Breuer02},
one finds that the master equation has a different dissipative term \cite{Turku07,Turku07b}.
As explained in \cite{Turku07b} if the spectrum of environmental noise is sufficiently flat then the difference is suppressed,
which explains the success of the standard cavity QED master equation.
But not so otherwise, which is something often overlooked and could lead to mistakes.

\subsection{Complete Positivity}
\label{sec:CP}

\emph{Positive maps} for a system S will translate the physical states of system S (positive-definite, trace-unity density matrices in the product Hilbert space of S) strictly into physical states.
\emph{Completely positive maps} for a system S will additionally provide positive maps for S+A, where A is an arbitrary ancillary system.
Completely positive maps are therefore compatible with entanglement to additional degrees of freedom not included in the system.
Any physical theory of quantum mechanics must take all input states and map them in a completely positive manner into the future.
We will momentarily ignore the subtleties and disagreement~\cite{Pechukas94,Sudarshan05} with this commonly accepted wisdom.

Given an initial-time propagator $\boldsymbol{\mathcal{G}}(t): \boldsymbol{\rho}(0) \to \boldsymbol{\rho}(t)$ which maps initial states into the future,
and which can be represented via some basis in Hilbert space
\begin{align}
\boldsymbol{\rho}(t) &= \boldsymbol{\mathcal{G}}(t) \left\{ \boldsymbol{\rho}(0) \right\} \, , \\
\boldsymbol{\mathcal{G}}_{ij;i'\!j'}(t) &= \bra{i} \boldsymbol{\mathcal{G}}(t) \left\{ \ket{i'}\!\!\bra{j'} \right\} \ket{j} \, ,
\end{align}
preservation of normalization and Hermiticity dictates the relations
\begin{align}
\sum_i \boldsymbol{\mathcal{G}}_{ii;i'\!j'}(t) &= \delta_{i'\!j'} \, , \\
\boldsymbol{\mathcal{G}}_{ij;i'\!j'}(t) &= \boldsymbol{\mathcal{G}}_{ji;j'\!i'}^*(t) \, .
\end{align}
while non-complete positivity requires the positivity of all bi-quadratic forms
\begin{align}
\bra{\psi} \boldsymbol{\mathcal{G}}(t) \left\{ \ket{\phi}\!\!\bra{\phi} \right\} \ket{\psi} & \geq 0 \, , \\
\left( \psi_i \phi_{i'}^* \right)^{\!*} \boldsymbol{\mathcal{G}}_{ij;i'\!j'}(t) \left( \psi_j \phi_{j'}^* \right) & \geq 0 \, ,
\end{align}
where $\boldsymbol{\psi}$ and $\boldsymbol{\phi}$ are Hilbert space vectors.
Determination of non-complete positivity is an NP-Hard problem~\cite{Ling08}.
\emph{Choi's theorem}~\cite{Choi75} proves that completely-positive maps are determined by the positivity of all quadratic forms
\begin{align}
\Psi_{ii'}^* \, \boldsymbol{\mathcal{G}}_{ij;i'\!j'}(t) \, \Psi_{jj'} & \geq 0 \, ,
\end{align}
which is (formally) a simple matter of linear algebra to resolve.
Note that the propagator's indices are interpreted differently from when ordinarily used as a transition matrix.
\begin{align}
\ket{i'}\!\!\bra{j'} \; \to & \ket{i}\!\!\bra{j} &\;\; \mathrm{(Transition \, Matrix)} \, , \\
\ket{j}\!\!\bra{j'} \; \to & \ket{i'}\!\!\bra{i} &\;\; \mathrm{(Choi \, Matrix)} \, .
\end{align}
The eigen-value decomposition of the positive-definite \emph{Choi matrix} is known as the \emph{Kraus representation}~\cite{Kraus83}.

Just as Choi's theorem characterizes the Lie group elements $\boldsymbol{\mathcal{G}}(t)$ of all valid quantum maps,
the theorem due to Lindblad~\cite{Lindblad76} and Gorini, Kossakowski and Sudarshan~\cite{Gorini76},
in its most general application, characterizes the associated Lie algebra.
Specifically, given the exponential representation
\begin{align}
\boldsymbol{\mathcal{G}}(t) &= e^{\boldsymbol{\mathbf{\Phi}}(t)} \, ,
\end{align}
then with the assumption of a semi-group structure,
i.e. $e^{\eta \, \boldsymbol{\mathbf{\Phi}}(t)}$ is also a group element for $\eta > 0$,
the algebraic generator must be of \emph{Lindblad form}:
\begin{equation}
\boldsymbol{\mathbf{\Phi}} \, \boldsymbol{\rho} =
\underbrace{-\imath \left[ \boldsymbol{\Theta} , \boldsymbol{\rho} \right]}_\mathrm{unitary}
+ \underbrace{\sum_{IJ} \Delta_{I\!J} \! \left( \mathbf{e}_I \, \boldsymbol{\rho} \, \mathbf{e}_{\!J}^\dagger -\frac{1}{2} \left\{ \mathbf{e}_{\!J}^\dagger \, \mathbf{e}_I , \boldsymbol{\rho} \right\} \right) }_\mathrm{decoherent} \, , \label{eq:LindbladFormPhi}
\end{equation}
where the capital indices $I$ and $J$ index the space of Hilbert space operators, which can be double indexed by Hilbert space vectors as above,
with $I=i i'$ and $J=j j'$ such that
\begin{equation}
\mathbf{e}_I = \mathbf{e}_{ii'} = \ket{i} \! \bra{i'} \, .
\end{equation}
The operator $\boldsymbol{\Theta}$ is Hermitian and $\boldsymbol{\Delta}$ is a Hermitian and positive-definite coefficient matrix, with $\mathbf{e}_I$ a particular operator (or double Hilbert space) basis of representation.
Such algebraic generators and the dynamics arising when the Liouvillian appearing in the master equation has Lindblad form have been extensively studied~\cite{Kossakowski72,Davies74,Davies76a,Davies77,Alicki07,Accardi02,Attal06,Ingarden97,Lindblad83,Weiss93}.
Due to non-commutativity, the time-local master equation is in general not directly determined by $\dot{\boldsymbol{\mathbf{\Phi}}}(t)$,
\begin{align}
\dot{\boldsymbol{\mathcal{G}}}(t) &= \boldsymbol{\mathcal{L}}(t) \, \boldsymbol{\mathcal{G}}(t) \, , \\
\boldsymbol{\mathcal{L}}(t) &= \int_0^1 \!\! d\eta \, e^{+\eta \boldsymbol{\mathbf{\Phi}}(t)} \, \dot{\boldsymbol{\mathbf{\Phi}}}(t) \, e^{-\eta \boldsymbol{\mathbf{\Phi}}(t)} \, ,
\end{align}
and therefore not all valid master equations must be of (time-dependent) Lindblad form
\begin{equation}
\boldsymbol{\mathcal{L}} \, \boldsymbol{\rho} = -\imath \left[ \mathbf{H} + \mathbf{V}, \boldsymbol{\rho} \right] + \sum_{IJ} \mathcal{D}_{I\!J} \left( \mathbf{e}_I \, \boldsymbol{\rho} \, \mathbf{e}_{\!J}^\dagger -\frac{1}{2} \left\{ \mathbf{e}_{\!J}^\dagger \mathbf{e}_I , \boldsymbol{\rho} \right\} \right) \, , \label{eq:LindbladFormL}
\end{equation}
where the Hermitian operator $\mathbf{V}$ arises with the existence of the non-unitary $\boldsymbol{\mathcal{D}}$ and
therefore $\mathbf{H} + \mathbf{V}$ should not be identified with the system Hamiltonian.
Moreover, we will argue that this unitary generator should also not (generally) be identified with the renormalized system Hamiltonian in Sec.~\ref{sec:renormalization1}.

This class inequivalence between the instantaneous \emph{time-translation generators} $\boldsymbol{\mathcal{L}}(t)$ and the all-time \emph{algebraic generators} $\boldsymbol{\mathbf{\Phi}}(t)$ does not exist for unitary transformations,
where both share the same adjoint symmetry.
However, following the generalization of Choi's theorem on merely Hermitian-preserving maps,
one can prove that any Hermitian and trace-preserving time-local master equation must have a \emph{pseudo-Lindblad form} $\boldsymbol{\mathcal{L}}(t)$ with merely Hermitian $\boldsymbol{\mathcal{D}}$.
The super-operator $\boldsymbol{\mathcal{D}}$ is generally referred to  as the \emph{dissipator} in the mathematics language, very different from the sense used by physicists:
The ``dissipation" generated by the dissipator refers to that of states, not of energy%
\footnote{Later in Sec.~\ref{sec:Damping} it will be shown that the signature of the damping kernel distinguishes between energy dissipation and amplification,
and that energy amplification can accompany state dissipation.}; its signature distinguishes between decoherent and recoherent evolution.
The ``dissipator" and unitary generator $\mathbf{V}$ together contain both dissipation and diffusion in the physics usage. In general, $\mathbf{V}$ should not be considered a renormalization of $\mathbf{H}$.

The state dissipation generated by the dissipator can be given a more precise geometrical meaning.
For any distance $D$ on the space of density operators which is constructed from a \emph{monotonic} metric (e.g. trace distance or Bures distance),
then CP evolution cannot cause any Hilbert-space distances to expand \cite{Petz96}.
\begin{equation}
D\!\left[ \boldsymbol{\mathcal{G}}(t) \, \boldsymbol{\rho}_1 , \boldsymbol{\mathcal{G}}(t) \, \boldsymbol{\rho}_2 \right] \leq D[ \boldsymbol{\rho}_1 , \boldsymbol{\rho}_2 ] \, .
\end{equation}
From this result it is easy to prove that positive-definite dissipators contract the state-space volume,
whereas negative-definite dissipators expand the state-space volume (as they appear to be time-reversed contractions).

Testing for the complete positivity of a time-dependent master equation involves a bit more effort than a determination of $\boldsymbol{\mathcal{D}}(t) > \mathbf{0}$ in the pseudo-Lindblad $\boldsymbol{\mathcal{L}}(t)$.
A Lindblad form master equation is sufficient but not necessary.
In Sec.~\ref{sec:PosProof} we prove that all microscopically-derived second-order master equations are completely positive to second order.
We show that the corresponding \emph{weak test for complete positivity} in an arbitrary (time-local) master equation
is that the pseudo-dissipator is not necessarily positive-definite for all times but for all running averages while in the interaction picture:
\begin{align}
\int_0^t \!\! d\tau \, \underline{\boldsymbol{\mathcal{D}}}(\tau) &\geq \mathbf{0} \, .
\end{align}
A similar analysis exists to check for higher-order consistency.

\subsubsection{Physical Limitations of the Lindblad Master Equation}

Any time-local master equation which preserves the trace and complete positivity for every time translation $t_1 \to t_2$ given arbitrary state $\boldsymbol{\rho}(t_1)$ must be of Lindblad form.
This is not exhaustive of all valid theories.
Quantum open systems provide valid master equations which are generally not of Lindblad form.
In the simplest models of quantum open systems, one begins at some initial time $t_0$ with an uncoupled and uncorrelated (factorized) system S and environment E.
The system and environment are then coupled together via an interaction Hamiltonian and the environmental degrees of freedom are traced over to obtain the open-system dynamics of only S.
It is a trivial matter to show that all initial system states $\boldsymbol{\rho}(t_0)$ will evolve in a completely positive manner.
\begin{equation}
\boldsymbol{\mathcal{G}}(t,t_0): \boldsymbol{\rho}(t_0) \to \boldsymbol{\rho}(t) \, ,
\end{equation}
but the intermediate mappings
\begin{equation}
\boldsymbol{\mathcal{G}}(t_2,t_1) = \boldsymbol{\mathcal{G}}(t_2,t_0) \, \boldsymbol{\mathcal{G}}(t_1,t_0)^{-1} \, ,
\end{equation}
are not necessarily completely positive.
The total volume of possible physical states can shrink such that an arbitrary intermediate state $\boldsymbol{\rho}(t_1)$ can fall into the category of impossible physical states, or states which did not evolve from any physical state at $t_0$.
Such a state can then evolve into the category of unphysical states at a later time $t_2$ and there is nothing fundamentally wrong with this;
it is an irrelevant evolution which has nothing to do with any physical prediction of the theory.
This is a typical feature of non-Markovian master equations; they can generate a less uniform decoherence than Lindblad equations.

In summary, completely-positive maps are much less useful outside of the Markovian regime, as one rarely has the all-time maps.
``All-time'' here meaning that the maps must describe all times wherein there  may not be any correlation to anything.
More precisely, such a map would have to describe the entire universe from its very birth.
Typically and empirically, one usually  has information only pertaining to the two-time maps of some limited set of states.
These maps are not completely positive in the non-Markovian regime and not much is known about them.

\section{Time-Local Master Equations}\label{sec:TLME}
\subsection{Perturbative Master Equations}
\label{sec:PME}
We  consider a fairly general microscopic model of an open quantum system, comprising a quantum system, its  environment, and system-environment interaction,  all separable in the Hamiltonian.
In the Schr\"{o}dinger picture, the combined closed system (system + environment) dynamics may be expressed
\begin{align}
\frac{d}{dt} \boldsymbol{\rho}_\mathrm{C}(t) &= \boldsymbol{\mathcal{L}}_\mathrm{C}(t) \, \boldsymbol{\rho}_\mathrm{C}(t) \, = \, \left[ -\imath \, \mathbf{H}_\mathrm{C}(t) , \boldsymbol{\rho}_\mathrm{C}(t) \right] \, , \label{eq:CSME} \\
\mathbf{H}_\mathrm{C}(t) &= \mathbf{H}_\mathrm{F}(t) + \mathbf{H}_\mathrm{I}(t) \, ,
\end{align}
where a subscript $\mathrm{C}$  refers to quantities of the closed  system (combined with its environment),
$\mathrm{F}$ subscript quantities refer to the free system + environment without interaction,
and $\mathrm{I}$ subscript quantities refer to the system-bath interaction.
S and E subscript quantities will refer to system and environment quantities respectively;
quantities without subscripts will always refer to the system as the vast majority of this work is with regards to the open system.%
\footnote{To be more specific on the terminology, when the degrees of freedom in an \emph{environment} E are overwhelmingly greater than those in the system, we call E a \emph{reservoir}
(In most cases this is assumed to be the case, but for small quantum systems it helps to make such a distinction as the physics could be very different.)
When the reservoir is at a finite temperature we call it a \emph{bath} = thermal reservoir.
It is then that the temperature $T$ appears, yet the system is in no sense at equilibrium.}

We will now proceed with a derivation of the open-system dynamics,
under the assumption that we may treat the effect of the environment upon the system perturbatively.
At least initially we will take the initial state of the system + environment to be uncorrelated, $\boldsymbol{\rho}_\mathrm{C}(0) = \boldsymbol{\rho}_\mathrm{S}(0) \otimes \boldsymbol{\rho}_\mathrm{E}(0)$, and uncoupled for all previous times.
Without abandoning the linear master-equation formalism, the existence of initial-time system-environment correlations are considered in Ref.~\cite{Correlations}.

Rotating Eq.~\eqref{eq:CSME} from the Schr\"{o}dinger picture to the interaction (Dirac) picture, the equivalent integral equation of motion is
\begin{align}
\underline{\boldsymbol{\mathcal{G}}}_\mathrm{C}(t) &= \mathbf{1} + \int_0^t \!\! d\tau \, \underline{\boldsymbol{\mathcal{L}}}_\mathrm{I}(\tau) \, \underline{\boldsymbol{\mathcal{G}}}_\mathrm{C}(\tau) \, ,
\end{align}
where the interaction-picture propagator and super-operators are defined
\begin{align}
\boldsymbol{\mathcal{G}}_\mathrm{C}(t) &= \boldsymbol{\mathcal{G}}_\mathrm{F}(t) \, \underline{\boldsymbol{\mathcal{G}}}_\mathrm{C}(t) \, , \\
\boldsymbol{\mathcal{L}}_\mathrm{I}(t) &= \boldsymbol{\mathcal{G}}_\mathrm{F}(t) \, \underline{\boldsymbol{\mathcal{L}}}_\mathrm{I}(t) \, \boldsymbol{\mathcal{G}}_\mathrm{F}^{-1}(t) \, .
\end{align}
This integral equation is directly amenable to perturbation theory via a Neumann series.
Tracing over the environment we then have
\begin{align}
\underline{\boldsymbol{\mathcal{G}}}(t) =\, & \mathbf{1} + \int_0^t \!\! d\tau_1 \left\langle \underline{\boldsymbol{\mathcal{L}}}_\mathrm{I}(\tau_1) \right\rangle_{\!\mathrm{E}} \label{eq:Neumann} \\
& \phantom{\mathbf{1}} + \int_0^t \!\! d\tau_1 \! \int_0^{\tau_1} \!\!\! d\tau_2 \left\langle \underline{\boldsymbol{\mathcal{L}}}_\mathrm{I}(\tau_1) \, \underline{\boldsymbol{\mathcal{L}}}_\mathrm{I}(\tau_2) \right\rangle_{\!\mathrm{E}} + \cdots \, , \nonumber
\end{align}
for the open-system propagator or transition matrix.
Note that this perturbative series has a time-secular behavior and is of little direct use.
But given some level of approximation to the propagator, we can then extract an approximate open-system Liouvillian via the relation $\boldsymbol{\mathcal{L}}(t) = \, \dot{\boldsymbol{\mathcal{G}}}(t) \, \boldsymbol{\mathcal{G}}^{-1}(t)$ or equivalently
\begin{align}
\boldsymbol{\mathcal{L}}(t) &= \boldsymbol{\mathcal{L}}_0(t) + \boldsymbol{\mathcal{G}}_0(t) \,  \dot{\underline{\boldsymbol{\mathcal{G}}}}(t) \, \underline{\boldsymbol{\mathcal{G}}}^{-1}(t) \, \boldsymbol{\mathcal{G}}_0^{-1}(t)  \, .
\end{align}
Assuming the odd moments of the environment vanish, which is always true for a Gaussian noise distributional (e.g. coupling to an environment of oscillators with linear environment operators), we only require the perturbative expansions
\begin{align}
\dot{\underline{\boldsymbol{\mathcal{G}}}}(t) &= \dot{\underline{\boldsymbol{\mathcal{G}}}}_2(t) + \dot{\underline{\boldsymbol{\mathcal{G}}}}_4(t) + \cdots \, , \\
\underline{\boldsymbol{\mathcal{G}}}^{-1}(t) &= \mathbf{1} - \underline{\boldsymbol{\mathcal{G}}}_2(t) + \cdots \, ,
\end{align}
where the inverse propagator can also be generated via Neumann series, though it will involve a double summation.
Ordinary perturbation in the coupling will then yield the following results expressed succinctly in the interaction picture:
\begin{align}
\underline{\boldsymbol{\mathcal{L}}}_2(t) =& \int_0^t \!\! d\tau \left\langle \underline{\boldsymbol{\mathcal{L}}}_\mathrm{I}(t) \, \underline{\boldsymbol{\mathcal{L}}}_\mathrm{I}(\tau) \right\rangle_{\!\mathrm{E}} \, , \label{eq:WCL} \\
\underline{\boldsymbol{\mathcal{L}}}_4(t) =& \int_0^t \!\! d\tau_1 \! \int_0^{\tau_1} \!\!\! d\tau_2 \! \int_0^{\tau_2} \!\!\! d\tau_3 \left\langle \underline{\boldsymbol{\mathcal{L}}}_\mathrm{I}(t) \, \underline{\boldsymbol{\mathcal{L}}}_\mathrm{I}(\tau_1) \, \underline{\boldsymbol{\mathcal{L}}}_\mathrm{I}(\tau_2) \, \underline{\boldsymbol{\mathcal{L}}}_\mathrm{I}(\tau_3) \right\rangle_{\!\mathrm{E}} \nonumber \\
& - \underline{\boldsymbol{\mathcal{L}}}_2(t) \! \int_0^t \! d\tau \, \underline{\boldsymbol{\mathcal{L}}}_2(\tau) \, , \label{eq:L4}
\end{align}
and for stationary environments the Liouville operators can be easily expressed in the Schr\"{o}dinger picture as
\begin{equation}
\boldsymbol{\mathcal{L}}_2(t) = \int_0^t \!\! d\tau \left\langle \boldsymbol{\mathcal{L}}_\mathrm{I}(t) \, \underline{\boldsymbol{\mathcal{L}}}_\mathrm{I}(\tau,t) \right\rangle_{\!\mathrm{E}} \, ,
\end{equation}
for the second-order master equation and
\begin{align}
& \boldsymbol{\mathcal{L}}_4(t) = - \boldsymbol{\mathcal{L}}_2(t) \! \int_0^t \! d\tau \, \underline{\boldsymbol{\mathcal{L}}}_2(\tau,t) \\
& +\! \int_0^t \!\! d\tau_1 \! \int_0^{\tau_1} \!\!\!\! d\tau_2 \! \int_0^{\tau_2} \!\!\!\! d\tau_3 \left\langle \boldsymbol{\mathcal{L}}_\mathrm{I}(t) \, \underline{\boldsymbol{\mathcal{L}}}_\mathrm{I}(\tau_1,t) \, \underline{\boldsymbol{\mathcal{L}}}_\mathrm{I}(\tau_2,t) \, \underline{\boldsymbol{\mathcal{L}}}_\mathrm{I}(\tau_3,t) \right\rangle_{\!\mathrm{E}} \, , \nonumber
\end{align}
for the fourth-order master equation, where two-time interaction-picture operators are given by
\begin{equation}
\underline{\boldsymbol{\mathcal{L}}}(t_1,t_2) \equiv \boldsymbol{\mathcal{G}}_\mathrm{F}(t_1,t_2)^{-1} \, \boldsymbol{\mathcal{L}}(t_1) \, \boldsymbol{\mathcal{G}}_\mathrm{F}(t_1,t_2) \, ,
\end{equation}
and the two-time propagator is given by
\begin{equation}
\boldsymbol{\mathcal{G}}(t_1,t_2) = \boldsymbol{\mathcal{G}}(t_1) \, \boldsymbol{\mathcal{G}}^{-1}(t_2) \, .
\end{equation}

Equivalent perturbative expansions of the open-system master equation have been derived by
projection operator techniques~\cite{Kampen97},
series expansion of the influence functional~\cite{Breuer03},
and coherent unraveling of the influence functional~\cite{Strunz04}.

\subsubsection{The Second-Order Master Equation}
To specify the second-order master equation in Eq.~\eqref{eq:WCL}, we will first expand the interaction Hamiltonian into a sum of separable couplings.
\begin{align}
\mathbf{H}_\mathrm{I}(t) &= \sum_{n} \mathbf{L}_n(t) \otimes \mathbf{l}_n(t) \, , \label{eq:sscouple}
\end{align}
with Hermitian system coupling variables $\mathbf{L}_n(t)$ and collective environment coupling variables $\mathbf{l}_n(t)$.
The environmental variables should be assumed to be completely non-stationary, in that they are off-diagonal in the free energy basis of the environment.
Any stationary environmental coupling would commute with the free environment Hamiltonian and could be effectively absorbed into the free system Hamiltonian at this order.
The second-order master equation can be expressed
\begin{equation}
\boldsymbol{\mathcal{L}}_2 \, \boldsymbol{\rho} = \sum_{nm} \left[ \mathbf{L}_n, \boldsymbol{\rho} \, (\mathbf{A}_{nm}\! \diamond \mathbf{L}_m)^\dagger - (\mathbf{A}_{nm}\! \diamond \mathbf{L}_m) \, \boldsymbol{\rho} \right] \, , \label{eq:WCGME}
\end{equation}
where the $\mathbf{A}$ operators and $\diamond$ product define the second-order operators
\begin{equation}
(\mathbf{A}_{nm}\! \diamond \mathbf{L}_m)(t) \equiv \int_0^t \!\! d\tau \, \alpha_{nm}(t,\tau) \, \left\{ \boldsymbol{\mathcal{G}}_0(t,\tau) \, \mathbf{L}_m(\tau) \right\} \, , \label{eq:WCOG}
\end{equation}
in terms of the (multivariate) environmental correlation function
\begin{align}
\alpha_{nm}(t,\tau) & \equiv \left\langle \underline{\mathbf{l}}_n(t) \, \underline{\mathbf{l}}_m(\tau) \right\rangle_{\mathrm{E}} \, , \label{eq:alpha}
\end{align}
which will be discussed more thoroughly in Sec.~\ref{sec:QNC}.
For now simply note that the correlation function is Hermitian and positive definite.
The pseudo-Lindblad coefficients of this master equation can then be expressed
\begin{equation}
\mathbf{V} \equiv \frac{1}{2\imath} \sum_{nm} \left[ \mathbf{L}_n \, (\mathbf{A}_{nm}\! \diamond \mathbf{L}_m) - (\mathbf{A}_{nm}\! \diamond \mathbf{L}_m)^\dagger \, \mathbf{L}_n \right] \, , \label{eq:V}
\end{equation}
for the unitary generator and for the dissipator
\begin{align}
\mathcal{D}_{ii';jj'} =& +  \sum_{nm} \bra{i} \mathbf{L}_n \ket{i'} \overline{\bra{j} (\mathbf{A}_{nm}\! \diamond \mathbf{L}_m) \ket{j'}} \label{eq:2OLindblad}  \\
& + \sum_{nm} \bra{i} (\mathbf{A}_{nm}\! \diamond \mathbf{L}_m) \ket{i'} \overline{\bra{j} \mathbf{L}_n \ket{j'}} \, , \nonumber
\end{align}
where the dissipator has been evaluated in some basis $\mathbf{e}_{ii'}=\ket{i}\!\!\bra{i'}$ given representation \eqref{eq:LindbladFormL}.
Though the second-order master equation is not of Lindblad form, in Sec.~\ref{sec:PosProof} we prove that it generates dynamics which are completely positive to second order.

\subsubsection{Validity of the Late-Time Limit}\label{sec:LateTime}
The second-order Liouvillian $\boldsymbol{\mathcal{L}}_2(t)$ can have a well defined late-time limit for many environments.
But because of the convergence of $\boldsymbol{\mathcal{L}}_2(t)$, this necessarily implies that the $\boldsymbol{\mathcal{L}}_2^2$ contribution to $\boldsymbol{\mathcal{L}}_4(t)$ in Eq.~\eqref{eq:L4} has the potential to give rise to an $\mathcal{O}(t)$ secular term.
Even assuming an asymptotic limit for the lower-order perturbative master-equation coefficients, secularly-evolving higher-order terms could invalidate taking the late-time limit after the perturbative expansion.
We will argue that the second-order master equation is exempt from this worry, and that higher-order perturbative master equations can also be acceptable if (1) the environment is Gaussian (in its cumulants) to that order of perturbation and (2) the correlation function is sufficiently localized.

First we must express the fourth-order Liouvillian strictly in terms of time-ordered integrals.
\begin{align}
\underline{\boldsymbol{\mathcal{L}}}_4(t) = & \int_0^t \!\! d\tau_1 \! \int_0^{\tau_1} \!\!\!\! d\tau_2 \! \int_0^{\tau_2} \!\!\!\! d\tau_3 \left[ \left\langle \underline{\boldsymbol{\mathcal{L}}}_\mathrm{I}(t) \, \underline{\boldsymbol{\mathcal{L}}}_\mathrm{I}(\tau_1) \, \underline{\boldsymbol{\mathcal{L}}}_\mathrm{I}(\tau_2) \, \underline{\boldsymbol{\mathcal{L}}}_\mathrm{I}(\tau_3) \right\rangle_{\!\mathrm{E}} \right. \nonumber \\
& - \left\langle \underline{\boldsymbol{\mathcal{L}}}_\mathrm{I}(t) \, \underline{\boldsymbol{\mathcal{L}}}_\mathrm{I}(\tau_1) \right\rangle_{\!\mathrm{E}} \left\langle \underline{\boldsymbol{\mathcal{L}}}_\mathrm{I}(\tau_2) \, \underline{\boldsymbol{\mathcal{L}}}_\mathrm{I}(\tau_3) \right\rangle_{\!\mathrm{E}} \nonumber \\
& - \left\langle \underline{\boldsymbol{\mathcal{L}}}_\mathrm{I}(t) \, \underline{\boldsymbol{\mathcal{L}}}_\mathrm{I}(\tau_2) \right\rangle_{\!\mathrm{E}} \left\langle \underline{\boldsymbol{\mathcal{L}}}_\mathrm{I}(\tau_1) \, \underline{\boldsymbol{\mathcal{L}}}_\mathrm{I}(\tau_3) \right\rangle_{\!\mathrm{E}} \nonumber \\
& - \left. \left\langle \underline{\boldsymbol{\mathcal{L}}}_\mathrm{I}(t) \, \underline{\boldsymbol{\mathcal{L}}}_\mathrm{I}(\tau_3) \right\rangle_{\!\mathrm{E}} \left\langle \underline{\boldsymbol{\mathcal{L}}}_\mathrm{I}(\tau_1) \, \underline{\boldsymbol{\mathcal{L}}}_\mathrm{I}(\tau_2) \right\rangle_{\!\mathrm{E}} \, \right] \, . \label{eq:L4to}
\end{align}
The previously discussed secular behavior would appear to be now located in the second term of this equation.
In the two following terms, the arguments of the 2-time correlations are intertwined between correlations with respect to their order of integration.
For a Gaussian environment, or an environment which is at least Gaussian to this order, the 4-time correlation can then be decomposed into a sum of 3 products of 2 2-time correlations.
The result is difficult to express in standard notation because the super-operators do not commute.
The integrand of the 4-time correlation becomes
\begin{align}
& \underbrace{\underline{\boldsymbol{\mathcal{L}}}_\mathrm{I}(t) \, \underline{\boldsymbol{\mathcal{L}}}_\mathrm{I}(\tau_1)} \, \overbrace{\underline{\boldsymbol{\mathcal{L}}}_\mathrm{I}(\tau_2) \, \underline{\boldsymbol{\mathcal{L}}}_\mathrm{I}(\tau_3)} + \underline{\boldsymbol{\mathcal{L}}}_\mathrm{I}(t) \, \overbrace{ \underline{\boldsymbol{\mathcal{L}}}_\mathrm{I}(\tau_1) \, \underline{\boldsymbol{\mathcal{L}}}_\mathrm{I}(\tau_2) \makebox[0pt][r]{$\underbrace{\phantom{ \underline{\boldsymbol{\mathcal{L}}}_\mathrm{I}(t) \, \underline{\boldsymbol{\mathcal{L}}}_\mathrm{I}(\tau_1) \, \underline{\boldsymbol{\mathcal{L}}}_\mathrm{I}(\tau_2) }}$} \, \underline{\boldsymbol{\mathcal{L}}}_\mathrm{I}(\tau_3) } \nonumber \\
& + \underbrace{\underline{\boldsymbol{\mathcal{L}}}_\mathrm{I}(t) \, \overbrace{ \underline{\boldsymbol{\mathcal{L}}}_\mathrm{I}(\tau_1) \, \underline{\boldsymbol{\mathcal{L}}}_\mathrm{I}(\tau_2) } \, \underline{\boldsymbol{\mathcal{L}}}_\mathrm{I}(\tau_3)} \, ,
\end{align}
where the brackets denote the pairs of operators being traced over with respect to the environment, and we would refer the reader to Ref.~\cite{Breuer03} for a more thorough examination of arbitrary orders.
Non-commutativity is not a problem with the first term of the decomposition, which precisely cancels the secular term in Eq.~\eqref{eq:L4to}.
This cancelation of secular terms will continue at higher orders of perturbation theory, as long as one can apply a Gaussian decomposition to the moments of the environment.
Therefore Gaussian environments, and possibly other environments which admit analogous cumulant decomposition, can be effectively described by perturbative master equations even in the late-time limit.

The stability of Gaussian environments bodes well for the second-order master equation even when the environment correlations are not exactly Gaussian.
We can make a Gaussian approximation to the environmental influence~\cite{Feynman63} by truncating the influence phase to quadratic order.
Because this perturbation is done in the influence phase as opposed to the influence functional, it should be good for long times as higher-order terms in the influence phase should be well controlled.
This is essentially the saddle-point approximation for path integrals.
From the Gaussian approximation we may then take a weak-coupling approximation, which as we have shown is also justified for long times,
and we will arrive at the same second-order master equation that we would have gotten if we had never taken the Gaussian approximation.
Therefore the second-order master equation can be safe in the late-time limit because it is also an effectively Gaussian approximation.

\subsubsection{Complete Positivity}
\label{sec:PosProof}
As explained in Sec.~\ref{sec:CP}, application of the Lindblad-GKS theorem to test for complete positivity requires,
not the time-translation generator, but the algebraic generator.
Consider the Neumann series, Eq.~\eqref{eq:Neumann}, whose terms are given by
\begin{align}
\underline{\boldsymbol{\mathcal{G}}}_k(t) &= \left\langle \prod_{i=1}^k \int_0^{\tau_{i-1}} \!\! d\tau_{i} \, \underline{\boldsymbol{\mathcal{L}}}_\mathrm{I}(\tau_{i}) \right\rangle_{\!\!\!\mathrm{E}} \, ,
\end{align}
where $\tau_0=t$.
This series can be contracted into the single exponential
\begin{align}
\underline{\boldsymbol{\mathcal{G}}}(t) &= e^{\underline{\boldsymbol{\Phi}}(t)} \, , \\
\boldsymbol{\mathcal{G}}(t) &= \boldsymbol{\mathcal{G}}_0(t) \, e^{\underline{\boldsymbol{\Phi}}(t)} \, ,
\end{align}
where for symmetric noise, e.g. Gaussian, the perturbative generators can then be found to be
\begin{align}
\underline{\boldsymbol{\Phi}}_{2}(t) &= \underline{\boldsymbol{\mathcal{G}}}_2(t) \, , \\
\underline{\boldsymbol{\Phi}}_{4}(t) &= \underline{\boldsymbol{\mathcal{G}}}_4(t) - \frac{1}{2} \underline{\boldsymbol{\mathcal{G}}}_2^2(t) \, ,
\end{align}
This is equivalent to solving the master equation via \emph{Magnus series}~\cite{Magnus54} in the interaction picture.
It should be noted that Magnus-series solutions are slightly secular in time, since in general the Magnus series has a finite radius of convergence \cite{Blanes98}.
In this context the second-order Magnus-series solution will accurately match the correct solution to second order at early times, and then only match the correct solution to zeroth order at later times, wherein it converges to the RWA solution.
For some aspects of the solution this accuracy can be improved with a less-secular integrator.
However, a careful analysis of the master equation and its solutions shows that, due to unavoidable degeneracy, the second-order master equation is fundamentally incapable of providing the full second-order solutions (see Sec.~\ref{sec:StatSol} and Ref.~\cite{Accuracy}).
Therefore we would not promote these solutions as \emph{the} second-order solutions,
but they contain the most information pertaining to the non-Markovian dissipator that one can extract from the second-order master equation.

The Magnus-series solution to the second-order master equation gives rise to the second-order algebraic generator
\begin{equation}
\underline{\boldsymbol{\Phi}}(t) = \int_0^t \!\! d\tau \! \int_0^\tau \!\! d\tau' \left\langle \underline{\boldsymbol{\mathcal{L}}}_\mathrm{I}(\tau) \, \underline{\boldsymbol{\mathcal{L}}}_\mathrm{I}(\tau') \right\rangle_{\!\mathrm{E}} + \mathcal{O}(\boldsymbol{\mathcal{L}}_\mathrm{I}^4) \, ,
\end{equation}
in the interaction picture.
In terms of the interaction Hamiltonian, the Lindblad coefficients of the algebraic generator $\underline{\boldsymbol{\Phi}}(t)$ are then
\begin{equation}
\underline{\Delta}_{ii';jj'}(t) = \int_0^t \!\! d\tau \! \int_0^t \!\! d\tau' \bigl\langle \bra{i} \underline{\mathbf{H}}_\mathrm{I}(\tau) \ket{i'} \bra{j'} \underline{\mathbf{H}}_\mathrm{I}(\tau') \ket{j} \bigr\rangle_{\!\mathrm{E}} \, ,
\end{equation}
given representation \eqref{eq:LindbladFormL}.
With the interaction Hamiltonian expanded as a sum of separable operators, as per Eq.~\eqref{eq:sscouple}, the coefficients evaluate to
\begin{equation}
\sum_{nm} \int_0^t \!\! d\tau \!\int_0^t \!\! d\tau' \bra{i} \underline{\mathbf{L}}_m(\tau) \ket{i'} \alpha_{nm}(\tau'\!,\tau) \, \overline{\bra{j} \underline{\mathbf{L}}_n(\tau') \ket{j'}} \, ,
\end{equation}
in terms of the environmental correlation function.
Both forms are positive-definite quadratic forms,
therefore the second-order master equation must generate completely-positive maps to second order and
the second-order Magnus-series solution $\boldsymbol{\mathcal{G}}(t) = \boldsymbol{\mathcal{G}}_0(t) \, e^{\underline{\boldsymbol{\Phi}}_2\!(t)}$ happens to be \emph{exactly} completely positive.

The algebraic generator for intermediate transitions is not generally of Lindblad form.
For $t_1 < t_2$, the integration kernel for $\underline{\boldsymbol{\Phi}}_2(t_1)$ in the quadratic form is effectively a leading principal minor of the corresponding integration kernel of $\underline{\boldsymbol{\Phi}}_2(t_2)$.
Therefore, the intermediate transitions can only be ensured completely positive (for arbitrary states and couplings) given delta correlations or white noise.
One can also have Lindblad master equations for specific system-environment couplings, such as the RWA-interaction, but this is a coupling dependent result.

Finally note that $\underline{\boldsymbol{\Phi}}_2(t) = \int_0^t \! d\tau \, \underline{\boldsymbol{\mathcal{L}}}_2(\tau)$ and
therefore the \emph{weak test for complete positivity} in an arbitrary time-local master equation is
\begin{align}
\int_0^t \!\! d\tau \, \underline{\boldsymbol{\mathcal{D}}}(\tau) &\geq \mathbf{0} \, , \label{eq:WeakTestCP}
\end{align}
that the dissipator is on-average positive definite in the interaction picture.
Higher-order tests for complete positivity would rely upon higher-order terms of the Magnus series.


\subsubsection{Exact Second-Order Master Equations}
\label{sec:E2OME}
In Sec.~\ref{sec:LateTime}, we have shown that a Gaussian noise cumulant decomposition causes the cancelation of potentially secular terms which would arise at fourth order and higher orders in the master equation.
The remaining non-secular terms do not cancel because their operator, time, and trace ordering are not generally the same.
However, there are situations where these higher-order terms do cancel, specifically
when the system coupling operators are constant or at least effectively constant when integrated alongside environmental correlations.
Such is the case when the noise is delta correlated, in which case the system operators do not have time to evolve under integration.
Also if the system coupling operators commute with the system Hamiltonian, then they will be stationary in the interaction picture and will not evolve regardless of environment timescales.
Therefore, \emph{given Gaussian noise, the second-order master equation is exact for either a Markovian process or stationary coupling}. 

For the Markovian process this property can also be seen most easily in the influence functional.
The influence phase trivially resolves into a much more mundane second-order term.
For stationary system couplings this property can also be seen in the stochastic Schr\"{o}dinger equation and corresponding convolutionless master-equation formalism~\cite{Strunz04}.
In this case the effective noise derivative can be exactly solved for, as it is completely lacking in dynamics.

\subsubsection{The Born-Markov Approximation}
The second-order master equation is both consistent with second-order perturbation in the system-environment coupling
and consistent with the Markovian limit, which can be viewed as the zeroth-order limit of weak system-energy perturbation.
In general, these are two radically different regimes.
Related to these two regimes of validity, the second-order master equation is in agreement with the \emph{Born-Markov approximation} and \emph{Redfield equation} after time localization.
The \emph{Born approximation} is to assume the system and environment, $\boldsymbol{\rho}_\mathrm{C}(t)$, to remain factorized in time
\begin{align}
\boldsymbol{\rho}_\mathrm{C}(t) &\approx \boldsymbol{\rho}_\mathrm{S}(t) \otimes \boldsymbol{\rho}_\mathrm{E}(0)
\end{align}
with the environment insignificantly influenced by the system.
This is a reasonable approximation in the weak-coupling regime.
The \emph{Markov Approximation} is then said to approximate environment correlations as delta correlations when integrated alongside system variables.
When used together in deriving the open-system master equation these two approximations constitute the \emph{Born-Markov Approximation}~\cite{Breuer02}.
The often stated Markov approximation in the interaction picture is
\begin{align}
\dot{\underline{\boldsymbol{\rho}}}(t) &= \int_0^t \!\! d\tau \left\langle \underline{\boldsymbol{\mathcal{L}}}_\mathrm{I}(t) \, \underline{\boldsymbol{\mathcal{L}}}_\mathrm{I}(\tau) \, \underline{\boldsymbol{\rho}}(\tau) \right\rangle_\mathrm{E} \, , \\
&\approx \int_0^t \!\! d\tau \left\langle \underline{\boldsymbol{\mathcal{L}}}_\mathrm{I}(t) \, \underline{\boldsymbol{\mathcal{L}}}_\mathrm{I}(\tau) \right\rangle_\mathrm{E} \underline{\boldsymbol{\rho}}(t) \, . \label{eq:BornMarkov}
\end{align}
Note that the density matrix can be pulled out of the integral if the remaining kernel is a delta correlation.
Finally, the stationary limit of the master equation coefficients often accompanies this approximation.

It may seem curious that the Born-Markov approximation is more generally consistent with the second-order (in the system-environment interaction) master equation,
even when far from the Markovian regime such as for zero-temperature reservoirs with moderate cutoff frequencies.
In general, a Markovian approximation is not reasonable outside of the Markovian or near-Markovian regime.
Indeed if one were to apply the very same Markov approximation to remaining system variables still inside the integral of Eq.~\eqref{eq:BornMarkov},
then the result would be incorrect outside of the Markovian regime.
It is specifically in the Born-Markov approximation,
that the Markov approximation is consistent with lowest-order perturbation theory and does not require highly localized correlations.
The nonlocal effects can only arrive at higher-order perturbation regardless of environment correlation timescales.
This is a typical occurrence in integro-differential equations which are only perturbatively nonlocal.

\subsection{Stationary Master Equations}
In this section we consider the case where all Hamiltonian terms are constant in time.
The second-order operator $(\mathbf{A}_{nm} \diamond \mathbf{L}_m)$ in Eq.~\eqref{eq:WCOG} is reduced to quadrature in the energy basis where it is given by a Hadamard product.
\begin{equation}
\bra{\omega_i} \mathbf{A}_{nm}\! \diamond \mathbf{L}_m \ket{\omega_{i'}} = \bra{\omega_i} \mathbf{A}_{nm} \ket{\omega_{i'}} \, \bra{\omega_i} \mathbf{L}_m \ket{\omega_{i'}} \, , \label{eq:Hadamard1}
\end{equation}
with the $\mathbf{A}_{nm}$ operator defined
\begin{align}
\bra{\omega_i} \mathbf{A}_{nm} \ket{\omega_{i'}} &\equiv A_{nm}\!\left( \omega_{ii'} \right) \, , \label{eq:Hadamard2} \\
A_{nm}(t;\omega) &= \int_0^t \!\! d\tau \, \alpha_{nm}(t,\tau) \, e^{-\imath \omega (t-\tau)} \label{eq:beta} \, ,
\end{align}
where $\omega_{ii'}=\omega_i\!-\!\omega_{i'}$.
If the correlation function is sufficiently localized in time, then these coefficients will have a stationary limit.

It will be useful to decompose $A_{nm}$ into its Hermitian and anti-Hermitian parts in the noise index
\begin{align}
\mathrm{He}[A_{nm}(t;\omega)] &= \frac{1}{2} \left[ A_{nm}(t;\omega) + A^*_{mn}(t;\omega) \right] \, , \\
\mathrm{An}[A_{nm}(t;\omega)] &= \frac{1}{2\imath} \left[ A_{nm}(t;\omega) - A^*_{mn}(t;\omega) \right] \, ,
\end{align}
where for a univariate correlation function these will reduce to the real and imaginary parts respectively.
Next we define the pseudo-stationary correlation
\begin{align}
\boldsymbol{\alpha}_t(\Delta t) &= \left\{ \begin{array}{cccc} \boldsymbol{\alpha}(t\!+\!\Delta t,t) & \Delta t & < & 0 \\ \boldsymbol{\alpha}(t,t\!-\!\Delta t) & 0 & < & \Delta t \end{array} \right. \, ,
\end{align}
with Hermiticity relation $\boldsymbol{\alpha}_t(\Delta t) = \boldsymbol{\alpha}_t^\dagger(-\Delta t)$ and
where for truly stationary noise one simply has $\boldsymbol{\alpha}_t(\Delta t) = \boldsymbol{\alpha}(\Delta t)$.
[By the emboldened $\boldsymbol{\alpha}$, we mean to imply the noise-indexed matrix with elements $\alpha_{nm}$.]
All of the following calculations may then proceed as if we have a stationary correlation.
E.g. the characteristic function, or power spectral density, is defined to be
\begin{align}
\tilde{\boldsymbol{\alpha}}_t(\omega) &= \int_{\!-\infty}^{+\infty} \!\!\! d\tau \, e^{-\imath \omega \tau} \, \boldsymbol{\alpha}_t(\tau) \, .
\end{align}
The full-time master equation coefficients can now be represented
\begin{align}
\mathrm{He}\!\left[ A_{nm}(t;\omega) \right] &= \frac{1}{2} \, \delta_t(\omega) * \tilde{\alpha}_{nm}(\omega) \, , \label{eq:fulltReA} \\
\mathrm{An}\!\left[ A_{nm}(t;\omega) \right] &= - \mathcal{P}_t\!\left[ \frac{1}{\omega} \right] \! * \delta_t(\omega) * \tilde{\alpha}_{nm}(\omega) \label{eq:fulltRImA} \, ,
\end{align}
with $*$ the appropriate Fourier convolution in frequency space
and the late-time limiting delta function and Cauchy principal value defined
\begin{align}
\delta_t(\omega) & \equiv  \frac{\sin(\omega t)}{\pi \omega} \, , \\
\mathcal{P}_t\!\left[\frac{1}{\omega}\right] & \equiv  \frac{1-\cos(\omega t)}{\omega} \, .
\end{align}
Assuming a sufficiently smooth and localized correlation function,
the late-time coefficients will then be
\begin{align}
\mathrm{He}\!\left[ A_{nm}(\omega) \right] &= \frac{1}{2} \tilde{\alpha}_{nm}(\omega) \, , \label{eq:MNC1} \\
\mathrm{An}\!\left[ A_{nm}(\omega) \right] &= - \mathcal{P}\!\left[ \frac{1}{\omega} \right] \! * \tilde{\alpha}_{nm}(\omega) \, , \label{eq:MNC2}
\end{align}
and thus the late-time coefficients obey a Kramers \!\!- \!\!Kronig relation and are causal response functions. 
Bochner's theorem dictates that positive-definite correlation functions in the frequency domain will arise from stationary (and positive-definite) correlation functions in the time domain.
Therefore, for stationary correlations and with some assumptions of continuity, the Hermitian coefficients will comprise a positive-definite matrix in the late-time limit.

Perhaps more useful computationally is the Laplace representation
\begin{align}
\hat{\boldsymbol{\alpha}}_t(s) = \int_{0}^{\infty} \!\! d\tau \, e^{-s \tau} \, \boldsymbol{\alpha}_t(\tau) \, .
\end{align}
In the Laplace domain, Eq.~\eqref{eq:beta} is merely a frequency shift of the correlation function before the time integration.
\begin{align}
\hat{A}_{nm}(s;\omega) &= \frac{1}{s} \hat{\alpha}_{nm}(s\!+\!\imath \omega) \, , \label{eq:Alaplace}
\end{align}
and from the final value theorem we have the late-time coefficients
\begin{align}
A_{nm}(\omega) &= \hat{\alpha}_{nm}(\imath \omega) \, . \label{eq:aAlate}
\end{align}
This is generally the fastest method for obtaining the late-time coefficients, assuming one has obtained functions with analytic continuation into the complex plane.
From the Kramers \!\!- \!\!Kronig relation, the late-time coefficients should be analytic in the upper half of the complex plane.

Finally, the pseudo-Lindblad master-equation coefficients can be represented
\begin{equation}
\mathcal{D}_{ii';jj'} = \sum_{nm} \bra{\omega_{i}} \mathbf{L}_m \ket{\omega_{i'}} \mathcal{A}_{ii';jj'} \overline{\bra{\omega_j } \mathbf{L}_n \ket{\omega_{j'}}} \, , \label{eq:p-Lindblad}
\end{equation}
in terms of the kernel
\begin{align}
\mathcal{A}_{ii';jj'} & \equiv A_{nm}(\omega_{ii'}) + A^*_{mn}(\omega_{jj'}) \, ,
\end{align}
all evaluated in the energy basis $\mathbf{e}_{ii'}=\ket{\omega_i}\!\!\bra{\omega_{i'}}$.
In the \emph{rotating-wave approximation} one only keeps the terms stationary in the interaction picture of the free system:
\begin{equation}
\mathcal{D}_{ii';ii'} = \sum_{nm} \bra{\omega_i} \mathbf{L}_m \ket{\omega_{i'}} \,2 \,\mathrm{He}\!\left[ A_{nm}(\omega_{ii'}) \right] \overline{\bra{\omega_i } \mathbf{L}_n \ket{\omega_{i'}}} \, , \label{eq:DRWA}
\end{equation}
and
\begin{equation}
\mathcal{D}_{ii;jj} = \sum_{nm} \bra{\omega_{i}} \mathbf{L}_m \ket{\omega_i} \,2 \,\mathrm{He}\!\left[ A_{nm}(0) \right] \overline{\bra{\omega_j } \mathbf{L}_n \ket{\omega_j}} \, . \label{eq:DRWA2}
\end{equation}
as well as the corresponding stationary contributions of $\mathbf{V}$.
These terms are a quadratic form on the correlation function.
If one has a correlation function which is stationary, at least in the late-time limit, then the RWA coefficients will take on a Lindblad form.
Therefore the rotating-wave approximation constitutes a Lindblad-projection of the master equation.
Correspondence between the RWA-Lindblad master equation and perturbation theory is discussed in Ref.~\cite{RWA}.

\subsubsection{Second-Order Solutions}
\label{sec:StatSol}
Given sufficiently-localized environment correlations, the master-equation coefficients asymptote primarily within the environment timescales
and secondarily within the free-system timescales.
Some exact results for linear systems are given in Ref.~\cite{QBM} where this behavior can be seen explicitly.
The short-time relaxation, which occurs on the order of the environment cutoff frequency, can be particularly violent and the master equation coefficients are said to \emph{jolt}.
Such jolts are result of specially chosen initial conditions where the system environment are initially factorized, yet subsequent evolution occurs with finite interaction strength.
In the context of linear master equations, avoidance of jolts by the preparation of correlated initial states or by turning on the interaction gradually are discussed in Ref.~\cite{Correlations}.

Therefore given some properly correlated or coupled environment,
the master equation should asymptote to its stationary value smoothly and (for weak coupling) within timescales which are much shorter than those of its effects.%
\footnote{Multipartite systems in relativistic fields also experience retarded jolts. See, e.g. Ref.~\cite{Dipole,ADL}.
These are also peculiarities associated with factorized initial states.}
Upon this initial relaxation, the system evolves in a manner invariant to time translations
\begin{align}
\lim_{t > \tau \gg 0} \boldsymbol{\mathcal{G}}(t,\tau) &= e^{(t-\tau) \boldsymbol{\mathcal{L}}(\infty)} \, .
\end{align}
Therefore in the weak-coupling regime it typically suffices to consider the dynamics generated by the stationary limit of the master equation.
As with the time-independent Schr\"{o}dinger equation, one can use canonical perturbation theory to compute the stationary propagator $e^{t \boldsymbol{\mathcal{L}}}$.
Full-time solutions can also be calculated analytically, and to within some perturbative order, via Fer expansion \cite{Fer58}.

Canonical perturbation theory is applied specifically to the eigen-value problem
\begin{align}
\boldsymbol{\mathcal{L}} \, \boldsymbol{\sigma}_{\!ij} &= f_{ij} \, \boldsymbol{\sigma}_{\!ij} \, ,
\end{align}
where the eigen-values $f_{ij}$ and corresponding right eigen-operators $\boldsymbol{\sigma}_{\!ij}$ of $\boldsymbol{\mathcal{L}}$ can be expanded perturbatively in powers of the coupling as
\begin{align}
\boldsymbol{\sigma}_{\!ij} &= \ket{\omega_i}\!\!\bra{\omega_j} + \boldsymbol{\delta\sigma}_{\!ij} + \cdots \, , \\
f_{ij} &= -\imath \, \omega_{ij} + \delta\! f_{ij} + \cdots \, ,
\end{align}
here to second order, where $\omega_{ij}=\omega_i-\omega_j$.
We will assume no resonance or near resonance in the energy-level splittings,
though it will be more-or-less clear how to apply degenerate perturbation theory to these cases.
By construction, the zeroth-order terms are set correctly.
The second-order terms are set by the corresponding order of terms in the master equation:
\begin{align}
&\bra{\omega_{i'}} \boldsymbol{\mathcal{L}}_2\!\left\{ \ket{\omega_i}\!\!\bra{\omega_j} \right\} \ket{\omega_{j'}} = \\
& -\imath(\omega_{ij} \!-\! \omega_{i'\!j'}) \bra{\omega_{i'}} \boldsymbol{\delta\sigma}_{\!ij} \ket{\omega_{j'}} + \delta\! f_{ij} \, \delta_{ij;i'\!j'} \, . \nonumber
\end{align}
Evaluating the components of this equation yields the non-degenerate corrections
\begin{align}
\bra{\omega_{i'}} \boldsymbol{\delta\sigma}_{\!ij} \ket{\omega_{j'}} &= \frac{ \bra{\omega_{i'}} \boldsymbol{\mathcal{L}}_2\!\left\{ \ket{\omega_i}\!\!\bra{\omega_j} \right\} \ket{\omega_{j'}} }{-\imath(\omega_{ij}\!-\!\omega_{i'\!j'})} \, , \label{eq:per_s} \\
\delta\! f_{ij} &= \bra{\omega_i} \boldsymbol{\mathcal{L}}_2\!\left\{ \ket{\omega_i}\!\!\bra{\omega_j} \right\} \ket{\omega_j} \, , \label{eq:per_f}
\end{align}
where $\omega_{ij} \neq \omega_{i'\!j'}$.
The second-order timescales $f_{ij}$ are determined strictly by the RWA coefficients, Eq.~\eqref{eq:DRWA}-\eqref{eq:DRWA2}.
The non-degenerate perturbative frequency shifts are found to be
\begin{align}
\delta\! f_{ij} &= \sum_{nm} \bra{\omega_i} \mathbf{L}_m \ket{\omega_i} 2\, \mathrm{He}\!\left[ A_{nm}(0) \right] \overline{\bra{\omega_j } \mathbf{L}_n \ket{\omega_j}} \nonumber \\
& - \sum_{nmk} \bra{\omega_k} \mathbf{L}_m \ket{\omega_i} A_{nm}(\omega_{ki}) \overline{\bra{\omega_k} \mathbf{L}_n \ket{\omega_i}} \nonumber \\
& - \sum_{nmk} \bra{\omega_k} \mathbf{L}_m \ket{\omega_j} A^*_{mn}(\omega_{kj}) \overline{\bra{\omega_k} \mathbf{L}_n \ket{\omega_j}} \, . \label{eq:off-freq}
\end{align}
This relation reveals that the Hermitian part of $A_{nm}(\omega)$ only gives rise to the real timescales, i.e. growth and decay, while the anti-Hermitian part only gives rise to imaginary timescales.
Upon completing the square with the first term, as we shall do in Eq.~\eqref{eq:pdissf}, one can see that He$[A_{nm}(\omega)]$ (and thus $\tilde{\boldsymbol{\alpha}}(\omega)$ at late time) being positive definite will force the real timescales to be negative.
This decay corresponds to the decoherence of (perturbatively) off-diagonal components of the density matrix while in the energy basis.

Even when far from resonance, the unperturbed eigen-operators are always degenerate along the diagonal where $\omega_{ii}=0$
and the previous perturbation theory only applies to the off-diagonal entries.
For the diagonal evolution, we need the linear combinations of diagonal entries that branch under perturbation.
The correct diagonal combination
\begin{align}
\mathbf{p} &= \sum_i p_i \ket{\omega_i}\!\!\bra{\omega_i} \, ,
\end{align}
with second-order eigen-frequency $\delta\!f$ must satisfy the characteristic equation
\begin{align}
\bra{\omega_i} \boldsymbol{\mathcal{L}}_2\!\left\{ \mathbf{p} \right\} \ket{\omega_i} &= \delta\! f \bra{\omega_i} \mathbf{p} \ket{\omega_i} \, ,
\end{align}
This is also a well defined eigen-value problem.
Written more conveniently in matrix notation we have
\begin{align}
\mathbf{W} \, \vec{\mathbf{p}} &= \delta\! f \, \vec{\mathbf{p}} \, , \label{eq:PauliME}
\end{align}
with the characteristic matrix and diagonal vector defined
\begin{align}
{[\![ \mathbf{W} ]\!]}_{ij} &= \bra{\omega_i} \boldsymbol{\mathcal{L}}_2\!\left\{ \ket{\omega_j}\!\!\bra{\omega_j} \right\} \ket{\omega_i} \, , \\
{[\![ \vec{\mathbf{p}} ]\!]}_{i} & \equiv \bra{\omega_i} \mathbf{p} \ket{\omega_i} \, ,
\end{align}
and where the characteristic matrix evaluates to
\begin{equation}
{[\![ \mathbf{W} ]\!]}_{ij} = \sum_{nm} \bra{\omega_i} \mathbf{L}_m \ket{\omega_j} 2 \, \mathrm{He}\!\left[ A_{nm}(\omega_{ij}) \right] \overline{\bra{\omega_i} \mathbf{L}_n \ket{\omega_j}} \, , \label{eq:C}
\end{equation}
for $i \neq j$ and for the diagonals
\begin{align}
{[\![ \mathbf{W} ]\!]}_{ii} &= -\sum_{k \neq i} \bra{\omega_k} \mathbf{W} \ket{\omega_i} \, .
\end{align}
Note that away from resonance $\mathbf{W}$ receives no contribution from the anti-Hermitian coefficients and is encapsulated by the RWA coefficients.
Associated with this characteristic equation is the Pauli master equation
\begin{align}
\frac{d}{dt} \vec{\mathbf{p}} &= \mathbf{W} \, \vec{\mathbf{p}} + \mathcal{O}(\boldsymbol{\mathcal{L}}_\mathrm{I}^2) \, ,
\end{align}
though it should be noted that this neglects the second-order corrections to the eigen-operators, of which the second-order off-diagonal perturbations are obtainable directly from the second-order master equation.

For $n$ energy levels, then there is an order $n$ characteristic polynomial that needs to be factored.
Directly solving the master equation would have been order $n^2$.
The zero temperature limit is a particularly simple case to solve for.
$\mathbf{W}$ is upper triangular in accord with the lack of thermal activation;
eigen-values are determined by the diagonal entries and eigen-vectors by simple matrix inversion.

The diagonal entries ${[\![ \mathbf{W} ]\!]}_{ii}$ are the maximal instantaneous relaxation rates for the perturbatively diagonal entries of the density matrix in the energy basis.
The perturbatively off-diagonal relaxation (decoherence) rates can be expressed
\begin{equation}
\mathrm{Re}\!\left[\delta\! f_{ij}\right] = \frac{{[\![ \mathbf{W} ]\!]}_{ii} + {[\![ \mathbf{W} ]\!]}_{jj}}{2} - \sum_{nm} d\ell_m^{[ij]} \, \mathrm{He}\!\left[ A_{nm}(0) \right] \overline{d\ell}_n^{[ij]} \, , \label{eq:pdissf}
\end{equation}
given the matrix elements
\begin{align}
d\ell_n^{[ij]} & \equiv \bra{\omega_i} \mathbf{L}_n \ket{\omega_i} - \bra{\omega_j } \mathbf{L}_n \ket{\omega_j} \, ,
\end{align}
and therefore the decoherence rates are strictly larger (more negative) than the (average) maximal diagonal decay rates.
Aside from an overall Markovian limit, only the decoherence rates reference $A_{nm}(0)$ or the Markovian coefficients.
This gives the decoherence rates additional sensitivities to the infrared regime (e.g. $1/f$ noise) which are not present among the stationary relaxation rates.
\emph{Decoherence can occur on timescales much more rapid than thermalization.}

It is not difficult to see from \eqref{eq:C} that the columns of $\mathbf{W}$ are not independent, as they all add up to zero.
In fact they must do so to preserve the trace of the density matrix.
Therefore there is always at least one state within the null space of $\mathbf{W}$.
This is a stationary state with characteristic frequency $\delta\! f=0$.
Any stationary master equation must have a stationary state.
Depending on the details of the model, there may be additional stationary states.
These will also be energy states (to lowest order in the coupling) in accord with the ``quantum limit of einselection''~\cite{Paz99}, but here as a simple consequence of perturbation theory.
$\mathbf{W}$ is weakly column diagonal dominant given a positive definite He$\left[A_{nm}(\omega)\right]$, as will be the case for stationary environment in the late-time limit where He$\left[A_{nm}(\omega)\right]=\frac{1}{2}\tilde{\alpha}_{nm}(\omega)$.
The Gershgorin circle theorem immediately implies that there can only be damped oscillations and that the damping rates strictly outpace the oscillation rates.

Finally we note that second-order diagonal perturbations to the operators $\mathbf{p}$ in the degenerate subspace require the fourth-order Pauli master equation,
as there will be corrections of the form $\mathbf{W}_{\!4} / f_2$ because of degeneracy~\cite{Accuracy}.
\emph{There is often a mistaken expectation that the second-order master equation should produce second-order solutions},
e.g. complete positivity to second order, but this is not the case.
The range of approximations between zeroth and second-order perturbation can be organized as follows: zeroth-order limiting solutions,
RWA solutions,
solutions to the second-order master equation,
and finally the second-order solutions.
Each successive approximation contains more information than the last~\cite{RWA},
however there can be a trade-off in positivity between the RWA and second-order master equation.
Solutions to the second-order master equation without fourth-order Pauli equation can easily violate positivity at second order.
This is particularly exacerbated by low-temperature environments where the asymptotic state has a largely vanishing diagonal to zeroth-order
and one lacks the proper second-order correction to the diagonals which are needed to accommodate off-diagonal perturbations.

\subsubsection{The Damping Basis}
\label{sec:dampingbasis}
The damping basis~\cite{Briegel93} is simply the basis which diagonalizes $\boldsymbol{\mathcal{L}}$.
It is a basis of operators or matrices and not necessarily physical states or vectors, as in general the eigen-operators $\boldsymbol{\sigma}_{\!ij}$ cannot be expressed as an outer product of state vectors.
$\boldsymbol{\mathcal{L}}$ is not necessarily Hermitian or normal in the ordinary sense of linear algebra,
and so the left eigen-operators are not determined by the super-adjoint of the right eigen-operators.
This is not to be confused with the fact that $\boldsymbol{\mathcal{L}}$ is Hermitian and normal in the sense of preserving the Hermiticity and trace of the density matrix.
The master equation sense of Hermiticity implies that the eigen-system has an adjoint symmetry, $\boldsymbol{\mathcal{L}} \, \boldsymbol{\sigma}_{\!ij}^\dagger = f_{ij}^* \, \boldsymbol{\sigma}_{\!ij}^\dagger$.
The master equation sense of normality implies that the identity matrix is always a left eigen-matrix with eigen-value zero as
\begin{equation}
\frac{d}{dt} \mathrm{Tr}[ \boldsymbol{\rho} ] = \mathrm{Tr}[ \mathbf{1} \, \boldsymbol{\mathcal{L}} \, \boldsymbol{\rho} ] = 0 \, ,
\end{equation}
for all $\boldsymbol{\rho}$; the corresponding right eigen-state to $\mathbf{1}$ being a stationary state.

In dissipative quantum mechanics we must resort to calculating the dual, left eigen-operators (super-vectors) $\boldsymbol{\sigma}^\star_{\!ij}$ such that $\boldsymbol{\sigma}^\star_{\!ij} \, \boldsymbol{\mathcal{L}} = \boldsymbol{\sigma}^\star_{\!ij} \, f_{ij}$
where $f_{ij}$ is also the eigen-value of the corresponding right eigen-operator $\boldsymbol{\sigma}_{\!ij}$.
To clarify, left and right super-multiplication with super-vectors is defined as
\begin{equation}
\boldsymbol{\rho}^\star_\mathrm{L} \, \boldsymbol{\mathcal{L}} \, \boldsymbol{\rho}_\mathrm{R} = \sum_{ij;i'\!j'} \bra{i} \boldsymbol{\rho}_\mathrm{L} \ket{j} \, \bra{i} \boldsymbol{\mathcal{L}}\!\left\{ \ket{i'}\!\!\bra{j'} \right\} \ket{j} \, \bra{i'} \boldsymbol{\rho}_\mathrm{R} \ket{j'} \, ,
\end{equation}
and the spectral decomposition of our open-system propagator can be represented by the sum of outer-products
\begin{align}
e^{t\, \boldsymbol{\mathcal{L}}} &= \sum_{ij} e^{f_{ij} t} \, \boldsymbol{\sigma}_{\!ij} \, \boldsymbol{\sigma}^\star_{\!ij} \, ,
\end{align}
which acts upon states as
\begin{equation}
e^{t\, \boldsymbol{\mathcal{L}}} \, \boldsymbol{\rho} = \sum_{ij} e^{f_{ij} t} \left(\sum_{i'\!j'} \bra{i'} \boldsymbol{\sigma}^\star_{\!ij} \ket{j'} \, \bra{i'} \boldsymbol{\rho} \ket{j'} \right) \boldsymbol{\sigma}_{\!ij} \, .
\end{equation}
The left eigen-operator $\boldsymbol{\sigma}^\star_{\!ij}$ projects out the $\boldsymbol{\sigma}_{\!ij}$ component from $\boldsymbol{\rho}$.
Perturbatively, the dual vectors can be written $\boldsymbol{\sigma}^\star_{\!ij} = \ket{\omega_i}\!\!\bra{\omega_j} + \boldsymbol{\delta}\boldsymbol{\sigma}^\star_{\!ij} + \mathcal{O}(\boldsymbol{\mathcal{L}}_\mathrm{I}^4)$,
and the second-order terms of the left eigen-value equation are then
\begin{align}
& \bra{\omega_i} \boldsymbol{\mathcal{L}}_2\!\left\{ \ket{\omega_{i'}}\!\!\bra{\omega_{j'}} \right\} \ket{\omega_j} = \\
& -\imath(\omega_{ij} \!-\! \omega_{i'\!j'}) \bra{\omega_{i'}} \boldsymbol{\delta}\boldsymbol{\sigma}^\star_{\!ij} \ket{\omega_{j'}} + \delta\! f_{ij} \, \delta_{ij;i'\!j'} \, . \nonumber
\end{align}
As must be the case, the corresponding eigen-values are the same while the left eigen-operators become
\begin{equation}
\bra{\omega_{i'}} \boldsymbol{\delta}\boldsymbol{\sigma}^\star_{\!ij} \ket{\omega_{j'}} = \frac{ \bra{\omega_i} \boldsymbol{\mathcal{L}}_2\!\left\{ \ket{\omega_{i'}}\!\!\bra{\omega_{j'}} \right\} \ket{\omega_j} }{-\imath(\omega_{ij}\!-\!\omega_{i'\!j'})} \, ,
\end{equation}
where $\omega_{ij} \neq \omega_{i'\!j'}$.
Due to degeneracy, the left eigen-vectors of $\mathbf{W}$ must be solved non-perturbatively.
As for the non-degenerate eigen-vectors, their orthogonality relation perturbatively evaluates to
\begin{align}
\boldsymbol{\sigma}^\star_{\!ij} \, \boldsymbol{\sigma}_{i'\!j'} &= \delta_{ij;i'\!j'} + \mathcal{O}(\boldsymbol{\mathcal{L}}_\mathrm{I}^4) \, ,
\end{align}
and therefore the eigen-basis is not only perturbatively orthogonal but also normalized to second order.

\subsection{Cyclo-Stationary Master Equations}
Here we will briefly discuss the applicability of second-order master equations to time-dependent problems by a simple analysis of asymptotically cyclo-stationary master equations.
Such master equations can arise when stationary free-system dynamics are under the influence of cyclo-stationary correlations (defined in Sec.~\ref{sec:NoiseStability}) or when periodic free-system dynamics are under the influence of stationary correlations.

As a brief review of periodic systems,
recall that Floquet's theory asserts that given a periodic Hamiltonian $\mathbf{H}(t)$, the free solutions take the form
\begin{align}
\boldsymbol{\psi}(t) &= \mathbf{P}_0(t) \, e^{-\imath \boldsymbol{\mathcal{H}} t} \, \boldsymbol{\psi}(0) \, ,
\end{align}
for pure states in the Schr\"{o}dinger picture,
where $\mathbf{P}_0(t)$ is a unitary operator with the same period as $\mathbf{H}(t)$ and $\boldsymbol{\mathcal{H}}$ is the time-homogeneous pseudo-Hamiltonian whose eigen-values are the pseudo-energy levels.
Effectively, $\mathbf{P}_0(t)$ acts as a change of basis between the time-dependent Hamiltonian motion and the time-independent pseudo-Hamiltonian motion.
$\mathbf{P}_0(t)$ has its own Fourier decomposition in terms of the Hamiltonian frequency $\Omega_\mathrm{H}$
and induces a Fourier decomposition upon transformed operators
\begin{align}
\mathbf{P}_0^\dagger(t) \, \mathbf{X} \, \mathbf{P}_0(t) &= \sum_{u} \mathbf{X}_{[u]} \, e^{+\imath u \Omega_\mathrm{H} t} \, ,
\end{align}
with which one can calculate the full spectral decomposition of Heisenberg picture observables
\begin{equation}
\bra{\omega_i} \underline{\mathbf{X}}(t) \ket{\omega_j} = \sum_{u} \bra{\omega_i} \mathbf{X}_{[u]} \ket{\omega_j} e^{+\imath \omega_{ij} t} \, e^{+\imath u \Omega_\mathrm{H} t} \, , \label{eq:Xspec}
\end{equation}
where $\ket{\omega}$ here denotes the pseudo-energy basis.
This spectral decomposition, whether for the system or environment, will allow for calculation of the second-order master equation coefficients in Eq.~\eqref{eq:WCOG} in terms of the more mundane stationary coefficients in Eq.~\eqref{eq:beta}.

Given stationary environment correlations and periodic system dynamics, the second-order operator
\begin{equation}
\bra{\omega_i} \mathbf{P}_0^\dagger(t) \, (\mathbf{A}_{nm} \diamond \mathbf{L}_m) \, \mathbf{P}_0(t) \ket{\omega_j} \, ,
\end{equation}
can be decomposed into
\begin{equation}
\sum_u A_{nm}(\omega_{ij}\!+\!u\Omega_\mathrm{H}) \, e^{+\imath u \Omega_\mathrm{H} t} \bra{\omega_i} \mathbf{L}_m^{[u]} \ket{\omega_j} \, ,
\end{equation}
here evaluated in the free concyclic picture of the system, which will turn out to be more useful than the Schr\"{o}dinger picture.

Given stationary system dynamics yet cyclo-stationary environment correlations as resolved in Eq.~\eqref{eq:CycloAlpha},
the second-order coefficients can be expressed by the Hadamard product formulas, Eq.~\eqref{eq:Hadamard1}-\eqref{eq:Hadamard2}, with the cyclo-stationary coefficients
\begin{align}
A_{nm}(\omega) &= \sum_{uv} A_{nm}^{[uv]}(\omega\!-\!v\Omega_\mathrm{H}) \, e^{\imath(u-v)\Omega_\mathrm{H} t} \, , \label{eq:WCCcyc1}
\end{align}
where the coefficients $A_{nm}^{[uv]}(\omega)$ are calculated also according to Eq.~\eqref{eq:beta}, but with the stationary kernels $\boldsymbol{\alpha}_{[uv]}(t)$ given by Eq.~\eqref{eq:CycloStationary}.

In either case the mundane stationary coefficients $A(\omega\!+\!u\Omega_\mathrm{H})$ remain asymptotically stationary and therefore the full coefficients are asymptotically periodic.
The cyclo-stationary master equation can therefore be represented with modulated stationary master equation coefficients.

\subsubsection{Second-Order Solutions}
Let us assume that we have an asymptotically and perturbatively cyclo-stationary master equation of the general form
\begin{align}
\dot{\boldsymbol{\rho}}(t) &= -\imath \left[ \boldsymbol{\mathcal{H}} , \boldsymbol{\rho}(t) \right] + \boldsymbol{\mathcal{L}}_2(t) \! \left\{ \boldsymbol{\rho}(t) \right\} \, , \\
\boldsymbol{\mathcal{L}}_2(t) &= \boldsymbol{\mathcal{L}}_2(t \!+\! \mathcal{T}) \, ,
\end{align}
where if our system had a periodic Hamiltonian, then we would have transformed into the free concyclic picture via $\mathbf{P}_0(t)$
\begin{align}
\boldsymbol{\rho}(t) &\to \mathbf{P}_0^\dagger(t) \, \boldsymbol{\rho}(t) \, \mathbf{P}_0(t) \, ,
\end{align}
so that the dynamics are zeroth-order stationary and we now work with the pseudo-Hamiltonian and concyclic dissipator.
If our system has stationary dynamics and the environment has cyclo-stationary correlations then we simply remain in the Schr\"{o}dinger picture.

Floquet's theorem then asserts that there is a periodic transformation of the density matrix into a basis wherein the dynamics are generated by a time-independent pseudo-Liouville operator $\boldsymbol{\mathfrak{L}}$ where
\begin{align}
\boldsymbol{\mathfrak{L}} &= \boldsymbol{\mathfrak{L}}_{\!0} + \boldsymbol{\mathfrak{L}}_{\!2} + \mathcal{O}(\boldsymbol{\mathcal{L}}_\mathrm{I}^4) \, , \\
\boldsymbol{\mathfrak{L}}_{\!0} \, \boldsymbol{\rho} &= -\imath \left[ \boldsymbol{\mathcal{H}} , \boldsymbol{\rho} \right] \, .
\end{align}
Determination of these perturbative corrections can be made by Magnus-Floquet series, to be detailed in Ref.~\cite{Floquet}.
The easiest terms to calculate are those stationary in the interaction picture, or the RWA coefficients.
These terms are simply given by the time average of the Liouvillian $\langle \boldsymbol{\mathcal{L}}(t) \rangle$
and are sufficient to determine the timescale perturbations and degenerate evolution as in the stationary master equation.
For instance, the RWA coefficients corresponding to Eq.~\eqref{eq:WCCcyc1} are given by
\begin{align}
\langle A_{nm}(\omega) \rangle &= \sum_{u} A_{nm}^{[uu]}(\omega\!-\!u\Omega) \, .
\end{align}
As with the stationary master equation, the RWA coefficients lack some information.
Here, the pseudo-energy basis is perturbed into a pseudo damping basis and there is also some (perturbatively) periodic transformation between the cyclo-stationary Liouvillian and pseudo-Liouvillian.

For finite-temperature reservoirs, one can see that cyclo-thermalization will not ensue (non-trivially at least) as detailed balance (discussed in Sec.~\ref{sec:DetailedBalance}) is lost with the frequency shifts.
The system will still evolve towards a cyclo-stationary state, only it will not generally be cyclo-thermal.

\section{Time-Nonlocal Master Equations}
\label{sec:QOSn}
Historically, the most general open-system master equations were, as first derived, of nonlocal form such as in the projection-operator formalism of Nakajima \cite{Nakajima58} and Zwanzig \cite{Zwanzig60}.
These nonlocal master equations would then be localized via Born-Markov approximation or a more careful perturbative analysis.
Our simple derivation of the convolutionless master equation in Sec.~\ref{sec:PME} contained no appeal to a nonlocal form.
Here we will demonstrate the dual non-secular perturbation theory can naturally produce a nonlocal master equation.
We will also discuss similarities between the two \emph{representations}.

\subsection{Perturbative Master Equations}
\label{sec:PMEn}
We will consider the same sort of system-environment-interaction decomposition,
but instead of time-local master equations we consider time-nonlocal master equations of the form
\begin{align}
\frac{d}{dt} \boldsymbol{\rho}_\mathrm{C}(t) &= \boldsymbol{\mathcal{K}}_\mathrm{C}(t) * \boldsymbol{\rho}_\mathrm{C}(t) \, , \\
\boldsymbol{\mathcal{K}}_\mathrm{C}(t) &= \boldsymbol{\mathcal{K}}_\mathrm{F}(t) + \boldsymbol{\mathcal{K}}_\mathrm{I}(t) \, ,
\end{align}
where $*$ denotes the Laplace convolution.
Therefore we should expect a perturbative level of agreement between the two analyses for closed (system + environment) Liouvillians which are constant in time.
In the Laplace frequency domain, we have an algebraic equation for the closed (system + environment) propagator
\begin{align}
s \, \hat{\boldsymbol{\mathcal{G}}}_\mathrm{C}(s) - \mathbf{1} &= \hat{\boldsymbol{\mathcal{K}}}_\mathrm{C}(s) \, \hat{\boldsymbol{\mathcal{G}}}_\mathrm{C}(s) \, ,
\end{align}
which readily admits perturbative solutions via the Neumann series
\begin{align}
\hat{\boldsymbol{\mathcal{G}}}(s) =& \, \hat{\boldsymbol{\mathcal{G}}}_0(s) + \left\langle \hat{\boldsymbol{\mathcal{G}}}_\mathrm{F}(s) \, \hat{\boldsymbol{\mathcal{K}}}_\mathrm{I}(s) \, \hat{\boldsymbol{\mathcal{G}}}_\mathrm{F}(s) \right\rangle_\mathrm{E}  \\
& + \left\langle \hat{\boldsymbol{\mathcal{G}}}_\mathrm{F}(s) \left[ \hat{\boldsymbol{\mathcal{K}}}_\mathrm{I}(s) \, \hat{\boldsymbol{\mathcal{G}}}_\mathrm{F}(s) \right]^2 \right\rangle_\mathrm{E} + \cdots \, , \nonumber \\
\hat{\boldsymbol{\mathcal{G}}}_\mathrm{F}(s) \equiv& \left[ s - \hat{\boldsymbol{\mathcal{K}}}_\mathrm{F}(s) \right]^{-1} \, ,
\end{align}
after tracing over the environment.
Again, the perturbative propagator is secular in nature, but we will only use it as an intermediate in the perturbative approximation of the open-system Liouville kernel.
\begin{align}
\hat{\boldsymbol{\mathcal{K}}}(s) &= s - \hat{\boldsymbol{\mathcal{G}}}(s)^{-1} \, .
\end{align}
For Gaussian noise we find the second and fourth-order kernels to be
\begin{align}
\hat{\boldsymbol{\mathcal{K}}}_2(s) =&\, \hat{\boldsymbol{\mathcal{G}}}_0(s)^{-1} \left\langle \hat{\boldsymbol{\mathcal{G}}}_\mathrm{F}(s) \left[ \hat{\boldsymbol{\mathcal{K}}}_\mathrm{I}(s) \, \hat{\boldsymbol{\mathcal{G}}}_\mathrm{F}(s) \right]^2 \right\rangle_{\!\!\mathrm{E}} \hat{\boldsymbol{\mathcal{G}}}_0(s)^{-1} \, , \label{eq:2OMEn} \\
\hat{\boldsymbol{\mathcal{K}}}_4(s) =&\, \hat{\boldsymbol{\mathcal{G}}}_0(s)^{-1} \left\langle \hat{\boldsymbol{\mathcal{G}}}_\mathrm{F}(s) \left[ \hat{\boldsymbol{\mathcal{K}}}_\mathrm{I}(s) \, \hat{\boldsymbol{\mathcal{G}}}_\mathrm{F}(s) \right]^4 \right\rangle_{\!\!\mathrm{E}} \hat{\boldsymbol{\mathcal{G}}}_0(s)^{-1} \nonumber \\
& - \hat{\boldsymbol{\mathcal{K}}}_2(s) \, \hat{\boldsymbol{\mathcal{G}}}_0(s) \, \hat{\boldsymbol{\mathcal{K}}}_2(s) \, .
\end{align}
For a stationary environment this simplifies to
\begin{align}
\hat{\boldsymbol{\mathcal{K}}}_2(s) =& \left\langle \hat{\boldsymbol{\mathcal{K}}}_\mathrm{I}(s) \, \hat{\boldsymbol{\mathcal{G}}}_\mathrm{F}(s) \, \hat{\boldsymbol{\mathcal{K}}}_\mathrm{I}(s) \right\rangle_{\!\mathrm{E}} \, , \\
\hat{\boldsymbol{\mathcal{K}}}_4(s) =& \left\langle \hat{\boldsymbol{\mathcal{K}}}_\mathrm{I}(s) \left[ \hat{\boldsymbol{\mathcal{G}}}_\mathrm{F}(s) \, \hat{\boldsymbol{\mathcal{K}}}_\mathrm{I}(s) \right]^3 \right\rangle_{\!\!\mathrm{E}}  - \hat{\boldsymbol{\mathcal{K}}}_2(s) \, \hat{\boldsymbol{\mathcal{G}}}_0(s) \, \hat{\boldsymbol{\mathcal{K}}}_2(s) \, .
\end{align}

\subsection{The Second-Order Master Equation}
\label{sec:WCMEn}
For a stationary environment with time-local interaction,
the second-order Liouville kernel can be expressed in terms of the time-local Liouville operator as
\begin{align}
\boldsymbol{\mathcal{K}}_2(t) &= \dot{\boldsymbol{\mathcal{L}}}_2(t) \, \boldsymbol{\mathcal{G}}_0(t) \, ,
\end{align}
and for a time-local system Hamiltonian we can additionally express
\begin{equation}
\hat{\boldsymbol{\mathcal{K}}}_2(s) \!\left\{ \ket{\omega_i}\!\!\bra{\omega_j} \right\} = \left. s' \, \hat{\boldsymbol{\mathcal{L}}}_2(s') \! \left\{ \ket{\omega_i}\!\!\bra{\omega_j} \right\} \right|_{s'=s+\imath\, \omega_{ij}} \, , \label{eq:K2-1}
\end{equation}
in Laplace space.
Using Eq.~\eqref{eq:Alaplace}, we have the relation
\begin{align}
s \, \hat{A}_{nm}(s;\omega) &= \hat{\alpha}_{nm}(s\!+\!\imath\, \omega) \, ,
\end{align}
which reveals a particular shift-scale symmetry of the nonlocal master equation.
The Liouville kernel can therefore be expressed
\begin{equation}
\hat{\boldsymbol{\mathcal{K}}}_2(s) \!\left\{ \ket{\omega_i}\!\!\bra{\omega_j} \right\} = s \, \hat{\boldsymbol{\mathcal{L}}}_2(s\!+\!2\,\imath\,\omega_{ij}) \! \left\{ \ket{\omega_i}\!\!\bra{\omega_j} \right\} \, . \label{eq:K2-2}
\end{equation}
In terms of the correlation function $\hat{\boldsymbol{\alpha}}(s)$, it is more obvious that
while the time-local master equation always has Markovian representation,
the time-nonlocal master equation only has Markovian representation for the Markovian process,
or at least when the process appears Markovian with respect to the freely-evolving system coupling operators.

Though the Liouville kernel is slightly more difficult to express than the time-local analog, the full-time solution is simply
\begin{align}
\hat{\boldsymbol{\mathcal{G}}}(s) &= \left[ s - \hat{\boldsymbol{\mathcal{K}}}(s) \right]^{-1} \, ,
\end{align}
in the Laplace domain.
Using the final-value theorem, the asymptotic state is
\begin{align}
\boldsymbol{\rho}(\infty) &= \lim_{s \to 0} s \left[ s - \hat{\boldsymbol{\mathcal{K}}}(s) \right]^{-1} \boldsymbol{\rho}(0) \, ,
\end{align}
which, under some assumptions of convergence, is equivalent to
\begin{align}
\boldsymbol{\rho}(\infty) &= \lim_{t \to \infty} e^{t \, \hat{\boldsymbol{\mathcal{K}}}(0)} \boldsymbol{\rho}(0) \, .
\end{align}
Therefore the stationary operator $\hat{\boldsymbol{\mathcal{K}}}(0)$ must determine the asymptotic state just as the stationary operator $\boldsymbol{\mathcal{L}}(\infty)$ does.
By comparison with either \eqref{eq:K2-1} or \eqref{eq:K2-2} we have
\begin{align}
\hat{\boldsymbol{\mathcal{K}}}_2(s)\!\left\{ \ket{\omega_i}\!\!\bra{\omega_i} \right\} &= s\, \hat{\boldsymbol{\mathcal{L}}}_2(s)\!\left\{ \ket{\omega_i}\!\!\bra{\omega_i} \right\} \, , \label{eq:KLsim}
\end{align}
which is sufficient to ensure that the two operators share the same zeroth-order stationary states,
and second-order corrections to any zeroth-order stationary state.

The late-time dynamics are slightly more involved to compare.
Ideally, one would decompose the full-time propagator into its early and late-time behavior
\begin{align}
\hat{\boldsymbol{\mathcal{G}}}(s) &= \hat{\boldsymbol{\mathcal{G}}}_\infty(s) + \boldsymbol{\delta}\hat{\boldsymbol{\mathcal{G}}}_\mathrm{i}(s) \, ,
\end{align}
where $\hat{\boldsymbol{\mathcal{G}}}_\infty(s)$ denotes the late-time local evolution which decays slowly with the perturbation,
and $\boldsymbol{\delta}\hat{\boldsymbol{\mathcal{G}}}_\mathrm{i}(s)$ denotes the short-time nonlocal evolution which decays more rapidly with system and environment timescales.
The asymptotically dominant $\hat{\boldsymbol{\mathcal{G}}}_\infty(s)$ would then correspond to $e^{t \, \boldsymbol{\mathcal{L}}(\infty)}$, possibly modulo some amplitude and phase differences.
This relation will be demonstrated once we have obtained solutions.

\subsection{Second-Order Solutions}
\label{sec:perturbationn}
Our perturbative solutions will follow the spirit of canonical perturbation theory.
First we inspect the structure of the exact solution, which takes the form
\begin{align}
\hat{\boldsymbol{\mathcal{G}}}(s) &= \left[ s - \hat{\boldsymbol{\mathcal{K}}}(s) \right]^{-1} \, ,
\end{align}
and is a rational function of the Liouville kernel.
Given the eigen-system $\hat{k}_{ij}(s)$, $\hat{\mathbf{o}}_{ij}(s)$ of $\hat{\boldsymbol{\mathcal{K}}}(s)$ such that
\begin{align}
\hat{\boldsymbol{\mathcal{K}}}(s) \, \hat{\mathbf{o}}_{ij}(s) &= \hat{k}_{ij}(s) \, \hat{\mathbf{o}}_{ij}(s) \, ,
\end{align}
then the exact solutions can also be expressed
\begin{align}
\hat{\boldsymbol{\mathcal{G}}}(s) &= \sum_{ij} \frac{1}{ s - \hat{k}_{ij}(s) } \hat{\mathbf{o}}_{ij}(s) \, \hat{\mathbf{o}}_{ij}^\star(s) \, , \label{eq:Gasymp}
\end{align}
where $\hat{\mathbf{o}}_{ij}^\star(s)$ is the left eigen-matrix dual to $\hat{\mathbf{o}}_{ij}(s)$ as described in Sec.~\eqref{sec:dampingbasis}.
We will construct a perturbative solution from the perturbative eigen-system, which will constitute a rational approximation of the exact solution akin to a Pad\'{e} approximant.
Further note that, with $\hat{\boldsymbol{\mathcal{K}}}(s)$ perturbatively truncated,
exact solutions to the perturbative master equation are already rational approximants of the exact solutions, so there is no new kind of mangling in doing this.

The nature of such a perturbative solution is slightly different from that of  the time-local perturbation.
The perturbative evaluation of $e^{dt \, \boldsymbol{\mathcal{L}}(t)}$ is non-secular, but may contain slight amplitude and phase discrepancy
and can only be applied for $\boldsymbol{\mathcal{L}}(t)$ commutating or for small $dt$.
The nonlocal perturbative solutions are effectively full time, but being a rational approximation in Laplace space they can introduce some slightly secular behavior.
For instance, exact timescales can become duplicated with small (here fourth-order) error, giving rise to a very slow beat frequency.
Regardless, we will show these perturbative solutions to be convergent with the correct asymptotics.

The second-order eigen-value problem constrains the non-degenerate perturbative corrections to be
\begin{equation}
\bra{\omega_{i'}} \boldsymbol{\delta}\hat{\mathbf{o}}_{ij}(s) \ket{\omega_{j'}} = \frac{ \bra{\omega_{i'}} \hat{\boldsymbol{\mathcal{K}}}_2(s)\!\left\{ \ket{\omega_i}\!\!\bra{\omega_j} \right\} \ket{\omega_{j'}} }{-\imath(\omega_{ij}\!-\!\omega_{i'\!j'})} \, , \label{eq:per_sn}
\end{equation}
for the operator corrections, where $\omega_{ij} \neq \omega_{i'\!j'}$,
and
\begin{align}
\delta \hat{k}_{ij}(s) &= \bra{\omega_i} \hat{\boldsymbol{\mathcal{K}}}_2(s)\!\left\{ \ket{\omega_i}\!\!\bra{\omega_j} \right\} \ket{\omega_j} \, . \label{eq:per_fn}
\end{align}
for the eigen-value corrections.
Given the similarity mentioned in Eq.~\eqref{eq:KLsim},
the ``frequency'' will be perturbed by a term analogous to the time-local formula.
The eigen-value perturbation is particularly important to analyze,
as the stability of our perturbative solution is dictated by its poles, or equivalently the roots of
\begin{align}
s - \hat{k}_{ij}(s) &= s + \imath \, \omega_{ij} - \delta \hat{k}_{ij}(s) + \cdots \, ,
\end{align}
having negative real part.
Assuming the correlation function to be well regulated, these poles should occur as perturbations of the system frequencies and perturbations of scales which regulate the correlations, such as the cutoff.
For well-regulated correlations our attention turns to the near-resonance poles, $s \approx -\imath \, \omega_{ij}$.
We can actually solve for these poles perturbatively, they are simply
\begin{align}
s_{ij} &= -\imath \, \omega_{ij} + \delta \hat{k}_{ij}(-\imath\omega_{ij}) \, + \cdots ,
\end{align}
and are precisely the late-time eigen-values of the time-local master equation given the relation
\begin{align}
\hat{\boldsymbol{\mathcal{K}}}_2(-\imath\omega_{ij})\!\left\{ \ket{\omega_i}\!\!\bra{\omega_j} \right\} &= \boldsymbol{\mathcal{L}}_2(\infty)\!\left\{ \ket{\omega_i}\!\!\bra{\omega_j} \right\}  \, , \label{eq:KLsim2} \\
\hat{\boldsymbol{\mathcal{K}}}_2 (\boldsymbol{\mathcal{L}}_0) &= \boldsymbol{\mathcal{L}}_2(\infty) \, ,
\end{align}
where the operator-Laplace transform in the second equation is interpreted as resulting in the first equation.
This general relation also reveals the asymptotic propagator from Eq.~\eqref{eq:Gasymp} as correctly being the rational approximant
\begin{align}
\hat{\boldsymbol{\mathcal{G}}}_\infty(s) &= \left[ s - \hat{\boldsymbol{\mathcal{K}}} (\boldsymbol{\mathcal{L}}_0) \right]^{-1} \, ,
\end{align}
to second order,
assuming the degenerate dynamics also work out correctly.

\subsubsection{The Pauli Master Equation}
Just as for the time-local master equation, the nonlocal master equation is confronted with degeneracy in the stationary states.
Here we must find the correct linear combination of diagonal matrices $\hat{\mathbf{q}}$ , and associated vectors $\vec{\mathbf{q}}_i = \bra{\omega_i} \hat{\mathbf{q}} \ket{\omega_i}$, which branch under perturbation
\begin{align}
\hat{\boldsymbol{\mathcal{V}}}(s) \, \vec{\mathbf{q}}(s) &= \delta\hat{k}(s) \, \vec{\mathbf{q}}(s) \, , \\
\bra{\omega_i} \hat{\boldsymbol{\mathcal{V}}}(s) \ket{\omega_j} &= \bra{\omega_i} \hat{\boldsymbol{\mathcal{K}}}_2(s)\{ \ket{\omega_j} \! \bra{\omega_j} \} \ket{\omega_i} \, ,
\end{align}
where from similarity \eqref{eq:KLsim} we have the correspondence
\begin{align}
\hat{\boldsymbol{\mathcal{V}}}(s) &= s \, \hat{\mathbf{W}}(s) \, ,
\end{align}
to the time-local Pauli master equation, Eq.~\eqref{eq:PauliME}.
The asymptotic dynamics follow in accord with the time-local master equation but the full-time evolution is, in general, exceedingly complicated from this perspective.
Full-time solutions may only be feasible in either the Markovian limit, where it is trivial,
or in the zero temperature limit, where $\mathbf{W}$ is upper triangular in accord with the lack of thermal activation.

\section{Non-Markovian Quantum Regression Theorem}\label{sec:QRT}
An open-system master equation provides a propagator $\boldsymbol{\mathcal{G}}(t)$ and super-adjoint propagator $\boldsymbol{\mathcal{G}}^\dagger(t)$ such that
\begin{align}
\mathrm{Tr} \left[ \mathbf{X} \, \boldsymbol{\mathcal{G}}(t) \{ \boldsymbol{\rho} \} \right] &= \mathrm{Tr} \left[ \boldsymbol{\mathcal{G}}^\dagger(t) \{ \mathbf{X} \} \boldsymbol{\rho} \right] \, .
\end{align}
Given this, one could construct a ``Heisenberg picture'' via $\mathbf{X}(t) \equiv \boldsymbol{\mathcal{G}}^\dagger(t) \{ \mathbf{X} \}$.
However, for multi-time correlations this ``Heisenberg picture'' does not necessarily have any relation to the Heisenberg picture of the microscopic model.
This is very easy to see, as in general
\begin{align}
\left\langle \mathbf{X}_1(t_1) \, \mathbf{X}_2(t_2) \right\rangle_{\!\mathrm{E}} & \neq  \left\langle \mathbf{X}_1(t_1) \right\rangle_{\!\mathrm{E}} \left\langle \mathbf{X}_2(t_2) \right\rangle_{\!\mathrm{E}} \, .
\end{align}
For the closed system or an open system driven by Markovian processes,
the Quantum Regression Theorem asserts that the dynamics of multi-time correlations can be generated via $\boldsymbol{\mathcal{L}}(t)$ or more specifically its super-adjoint $\boldsymbol{\mathcal{L}}^\dagger(t)$.
However this is a defining property of the Markovian process and will not hold true in the non-Markovian regime.

In this section we approach the category of non-Markovian dynamics which cannot be described by the regression theorem.
First we derive the adjoint master equation, which is dual to the ordinary master equation.
Following this simple derivation, that of the two-time correlations is more straightforward.

\subsection{The Adjoint Master Equation}\label{sec:Ldagger}
First we would like to derive the (super) adjoint master equation for the dynamics of single-time operators
\begin{equation}
\mathbf{X}(t) = \left\langle \{\mathbf{X}\}(t) \right\rangle_{\!\mathrm{E}} = \left\langle \boldsymbol{\mathcal{G}}_\mathrm{C}^\dagger(t)\!\left\{ \mathbf{X} \right\} \right\rangle_{\!\mathrm{E}} \, .
\end{equation}
As we already know the ordinary master equation, no new information will be gained in doing this, but the basic procedure will remain largely the same for multi-time correlations.
The super-adjoint propagator satisfies the differential equation
\begin{align}
\frac{d}{dt} \boldsymbol{\mathcal{G}}_\mathrm{C}^\dagger(t) &= \boldsymbol{\mathcal{G}}_\mathrm{C}^\dagger(t) \, \boldsymbol{\mathcal{L}}_\mathrm{C}^\dagger(t) \, ,
\end{align}
where here in the unitary theory of the system + environment we have
\begin{align}
\boldsymbol{\mathcal{L}}_\mathrm{C}^\dagger(t) \, \mathbf{X} &= \left[ + \imath \, \mathbf{H}_\mathrm{C}(t) , \mathbf{X} \right] \, ,
\end{align}
and therefore our adjoint master equation can be expressed
\begin{equation}
\dot{\mathbf{X}}(t) = \boldsymbol{\mathcal{G}}^\dagger(t) \! \left\{ \boldsymbol{\mathcal{L}}_0^\dagger(t) \, \mathbf{X} \right\} + \left\langle \boldsymbol{\mathcal{G}}_\mathrm{C}^\dagger(t) \, \boldsymbol{\mathcal{L}}_\mathrm{I}^\dagger(t) \{ \mathbf{X} \} \right\rangle_{\!\mathrm{E}} \, ,
\end{equation}
where we have used the fact that $\boldsymbol{\mathcal{L}}_\mathrm{E}^\dagger(t) \{ \mathbf{X} \} = 0$ for system operators.
We want to express the right-hand side in terms of single-time operators, therefore the second term is cast into the form of an adjoint super-operator
\begin{align}
\dot{\mathbf{X}}(t) & = \boldsymbol{\mathcal{G}}^\dagger(t) \! \left\{ \boldsymbol{\mathcal{L}}_0^\dagger(t) \, \mathbf{X} \right\} + \boldsymbol{\mathcal{G}}^\dagger(t) \! \left\{ \boldsymbol{\delta\!\mathcal{L}}^\dagger(t) \, \mathbf{X} \right\} \, , \\
\boldsymbol{\delta\!\mathcal{L}}^\dagger(t) & \equiv \left\langle \boldsymbol{\mathcal{G}}_\mathrm{C}^\dagger(t) \right\rangle_{\!\mathrm{E}}^{\!\!-1} \left\langle \boldsymbol{\mathcal{G}}_\mathrm{C}^\dagger(t) \, \boldsymbol{\mathcal{L}}_\mathrm{I}^\dagger(t) \right\rangle_{\!\mathrm{E}}
\end{align}
which will be calculated perturbatively.
To do this, first we express the closed system propagator via the interaction picture where
\begin{align}
\boldsymbol{\mathcal{G}}_\mathrm{C}^\dagger(t) &= \underline{\boldsymbol{\mathcal{G}}}_\mathrm{C}^\dagger(t) \, \boldsymbol{\mathcal{G}}_\mathrm{F}^\dagger(t) \, , \\
\boldsymbol{\mathcal{L}}_\mathrm{I}^\dagger(t) &= \boldsymbol{\mathcal{G}}_\mathrm{F}^{-\dagger}(t) \, \underline{\boldsymbol{\mathcal{L}}}_\mathrm{I}^\dagger(t) \, \boldsymbol{\mathcal{G}}_\mathrm{F}^\dagger(t) \, .
\end{align}
The interaction picture propagator then has the integral equation of motion
\begin{align}
\underline{\boldsymbol{\mathcal{G}}}_\mathrm{C}^\dagger(t) &= \int_0^t \!\! d\tau \, \underline{\boldsymbol{\mathcal{G}}}_\mathrm{C}^\dagger(\tau) \, \underline{\boldsymbol{\mathcal{L}}}_\mathrm{I}^\dagger(t) \, ,
\end{align}
which is directly amenable to perturbation theory via Neumann series expansion.
Ordinary perturbation theory then reveals the lowest order correction
\begin{align}
\underline{\boldsymbol{\mathcal{L}}}_2^\dagger(t) &= \int_0^t \!\! d\tau \left\langle \underline{\boldsymbol{\mathcal{L}}}_\mathrm{I}^\dagger(\tau) \, \underline{\boldsymbol{\mathcal{L}}}_\mathrm{I}^\dagger(t) \right\rangle_{\!\mathrm{E}} \, ,
\end{align}
and one can see that this is indeed the super-adjoint of Eq.~\eqref{eq:WCL}.
More specifically, for the sum of separable couplings the super-adjoint master equation resolves to
\begin{align}
& \boldsymbol{\mathcal{L}}_2^\dagger(t) \, \mathbf{X} = \label{eq:adjointME} \\
& \sum_{nm} \left\{  (\mathbf{A}_{nm} \diamond \mathbf{L}_m)^\dagger \left[ \mathbf{X} , \mathbf{L}_n \right] + \left[ \mathbf{L}_n , \mathbf{X} \right] (\mathbf{A}_{nm} \diamond \mathbf{L}_m) \right\} \, . \nonumber
\end{align}

\subsection{Two-Time Correlations}
For two-time and other multi-time correlations, one should note that there will be no assurance of a Quantum Regression Theorem
\begin{equation}
\left\langle \dot{\mathbf{X}}_1(t_1) \, \mathbf{X}_2(t_2) \right\rangle_{\!\!\mathrm{E}} \neq \Bigl\langle \left\{ \boldsymbol{\mathcal{L}}^\dagger(t_1) \, \mathbf{X}_1 \right\}\!(t_1) \, \{ \mathbf{X}_2 \}(t_2) \Bigr\rangle_{\!\mathrm{E}}  \, ,
\end{equation}
given a non-Markovian process.
Here we look for perturbative corrections to the standard QRT formula.
After application of second-order perturbation theory, of the same form as was used for the adjoint master equation, the standard QRT terms on the right-hand side are seen to match the zeroth-order terms and most of the second-order terms of the left-hand side.
The second-order remainder is such that
\begin{align}
&& \left\langle \dot{\mathbf{X}}_1(t_1) \, \mathbf{X}_2(t_2) \right\rangle_{\!\!\mathrm{E}} = \Bigl\langle \left\{ \boldsymbol{\mathcal{L}}^\dagger(t_1) \, \mathbf{X}_1 \right\}\!(t_1) \, \{ \mathbf{X}_2 \}(t_2) \Bigr\rangle_{\!\mathrm{E}} \nonumber \\
&& +\int_0^{t_2} \!\!\! d\tau \left\langle \underline{\boldsymbol{\mathcal{L}}}_\mathrm{I}^\dagger(t_1) \!\left\{ \underline{\mathbf{X}}_1(t_1) \right\} \underline{\boldsymbol{\mathcal{L}}}_\mathrm{I}^\dagger(\tau) \!\left\{ \underline{\mathbf{X}}_2(t_2) \right\} \right\rangle_{\!\mathrm{E}} \, , \label{eq:NMQRT}
\end{align}
and it is also revealed, but not directly shown here, that it is imperative to cast the standard QRT terms inside a single environmental trace, as to cancel certain highly secular terms.

A general non-secular perturbative fit to this expression is not so obvious, so we will resort to a more specific model.
For a constant Hamiltonian with separable coupling to a stationary environment, the second-order non-Markovian corrections to the QRT evaluate to
\begin{equation}
- \sum_{nm} \left[ \underline{\mathbf{L}}_n(t_1) , \underline{\mathbf{X}}_1(t_1) \right] \left[ \int_0^{t_2} \!\!\! d\tau \, \alpha_{nm}(t_1,\tau) \, \underline{\mathbf{L}}_m(\tau) , \underline{\mathbf{X}}_2(t_2) \right] \, , \label{eq:NMC}
\end{equation}
in agreement with Ref.~\cite{Vega06}.
The two-time integral
\begin{equation}
\int_0^{t_2} \!\!\! d\tau \, \alpha_{nm}(t_1,\tau) \, \underline{\mathbf{L}}_m(\tau) = \boldsymbol{\mathcal{G}}_0^\dagger(t_1) \left\{ (\mathbf{A}_{nm} \diamond \mathbf{L}_m)(t_1,t_2) \right\} \, ,
\end{equation}
can be expressed in terms of ordinary second-order operators as
\begin{equation}
(\mathbf{A}_{nm} \diamond \mathbf{L}_m)(t_1,t_2) \equiv (\mathbf{A}_{nm} \diamond \mathbf{L}_m)(t_1) - (\mathbf{A}_{nm} \diamond \mathbf{L}_m)(t_1\!-\!t_2) \, , \label{eq:2tA}
\end{equation}
and it should be noted that this non-Markovian correction requires no renormalization as it involves the difference between two ordinary second-order operators.
The most natural non-secular expression to fit \eqref{eq:NMC} to is
\begin{equation}
 - \sum_{nm} \left\{ \left[ \mathbf{L}_n , \mathbf{X}_1 \right] \right\}\!(t_1) \Bigl[ \left\{ (\mathbf{A}_{nm} \diamond \mathbf{L}_m)(t_1,t_2) \right\}\!(t_1) , \left\{ \mathbf{X}_2 \right\}\!(t_2) \Bigr] \, ,
\end{equation}
which is obviously non-Markovian and cannot be expressed as a sum of two operators evaluated at two times.
Note that, although it was necessary to place the standard QRT terms inside a single environmental trace,
at least here at second order, the corrections may be considered as a product of single-time operators or a collective mutli-time correlation.
Higher-order perturbation theory might reveal which representation is superior (in the stronger-coupling regime) as second-order perturbation did for Eq.~\eqref{eq:NMQRT}.

These non-Markovian corrections do not vanish in the weak coupling limit (second-order) but, from Eq.~\eqref{eq:2tA}, if $t_1 > t_2$ they will vanish in the Markovian limit.
More generally, the two-time operator vanishes in the softer limit of $t_1 \gg t_2$, as compared to the system and environment timescales,
much in the same manner that the master equation takes its stationary limit.
The stationary limit of this expression does not merely require late times, but also a similarly lengthy span of time between the two times of the correlation.
So, even at late times when one has a time-homogeneous master equation, the non-Markovian property does not go away.
This should not be surprising as the noise process does not stop being non-Markovian.

\section{The Quantum Langevin Equation}\label{sec:QLE}
The quantum Langevin equation \cite{FordOconnell88} is well understood in the context of bilinear position-position couplings between the system and environment.
Here we would like to extend the simpler theory by considering quantum Langevin equations that correspond to the same class of Gaussian influences which appear in the second-order master-equation and influence-functional formalisms.
Our motivation is to compare behavior with the master equation, to better decompose and identify the roles of the environment's influence kernels,
to distinguish the stochastic and dissipative forces engendered upon the system from ordinary forces engendered upon the system,
to promote a natural notion of the open system's energy,
and finally to promote a proper criteria for renormalization.
In this sense the quantum Langevin equation is a powerful pedagogical instrument,
despite the fact that it is not as useful for practical quantum calculations.

We begin with the same Hamiltonian for our system and environment
\begin{align}
\mathbf{H}_\mathrm{C} &= \mathbf{H}_\mathrm{S} + \mathbf{H}_\mathrm{I} + \mathbf{H}_\mathrm{E} \, , \\
\mathbf{H}_\mathrm{I}(t) &= \sum_n \mathbf{L}_n(t) \otimes \mathbf{l}_n(t) \, ,
\end{align}
and the Heisenberg equations of motion for any system operator $\mathbf{X}(t)$, as driven by the environment, is therefore given by
\begin{align}
\dot{\mathbf{X}}(t) &= + \imath \left[ \mathbf{H}(t) , \mathbf{X}(t) \right] + \imath \sum_n \left[ \mathbf{l}_n(t) \otimes \mathbf{L}_n(t) , \mathbf{X}(t) \right] \, . \label{eq:LNL}
\end{align}
To generate a Gaussian influence, we specify the environment to be comprised of simple (non-parametric) oscillators linear in its driven dynamics
\begin{align}
\mathbf{H}_\mathrm{E} &= \sum_k \frac{1}{2} \left( \boldsymbol{\scpi}_k^2 + \omega_k^2 \, \mathbf{q}_k^2 \right) \, , \\
\mathbf{l}_n &= \sum_k \left[ g_{kn}^q(t) \, \mathbf{q}_k + g_{kn}^\pi(t) \, \boldsymbol{\scpi}_k \right] \, ,
\end{align}
and its coupling to be a general linear and time-dependent coupling.
Let us define the ``phase-space'' vectors
\begin{align}
\mathbf{z}_k &\equiv \left( \mathbf{q}_k , \boldsymbol{\scpi}_k \right) \, , \\
\mathbf{g}_{kn}(t) &\equiv \left( g_{kn}^q(t) , g_{kn}^\pi(t) \right) \, .
\end{align}
The Heisenberg equations of motion for the environment operators, as driven by the system, may then be expressed conveniently
\begin{align}
\dot{\mathbf{z}}_k(t) &= \left[ \begin{array}{cc} 0 & 1 \\ -\omega_k^2 & 0 \end{array} \right] \mathbf{z}_k(0) + \sum_m \left[ \begin{array}{cc} 0 & 1 \\ -1 & 0 \end{array} \right] \mathbf{g}_{km}(t) \, \mathbf{L}_m(t) \, ,
\end{align}
so that the driven solutions of the environment operators are given by
\begin{align}
\mathbf{z}_k(t) &= \boldsymbol{\Phi}_k(t) \, \mathbf{z}_k(0) \\
& + \int_0^t \!\! d\tau \sum_m \boldsymbol{\Phi}_k(t\!-\!\tau) \left[ \begin{array}{cc} 0 & 1 \\ -1 & 0 \end{array} \right] \mathbf{g}_{km}(t) \, \mathbf{L}_m(\tau) \, , \nonumber
\end{align}
in terms of the free environment transition matrix $\boldsymbol{\Phi}$ and Green function $f$, given by
\begin{align}
\boldsymbol{\Phi}_k(t) &= \left[ \begin{array}{cc} \dot{f}_k(t) & f_k(t) \\ \ddot{f}_k(t) & \dot{f}_k(t) \end{array}\right] \, , \\
f_k(t) &= \frac{\sin(\omega_k t)}{\omega_k} \, .
\end{align}
A straightforward calculation then reveals the collective coupling operator of the environment to take the driven form
\begin{align}
\mathbf{l}_n(t) &= \boldsymbol{\xi}_n(t) + 2 \sum_m \int_0^t \!\! d\tau \, \mu_{nm}(t,\tau) \, \mathbf{L}_m(\tau) \, , \label{eq:l(t)driven}
\end{align}
where, assuming the environment to be initially in a Gaussian state,
$\boldsymbol{\xi}_n(t)$ is an operator-valued Gaussian stochastic process, with two-time correlation function
\begin{align}
\left\langle \boldsymbol{\xi}_n(t) \, \boldsymbol{\xi}_m(\tau) \right\rangle_{\boldsymbol{\xi}} &= \mathbf{\alpha}_{nm}(t,\tau) = \nu_{nm}(t,\tau) + \imath \, \mu_{nm}(t,\tau) \, ,
\end{align}
in terms of the noise kernel $\nu$ and dissipation kernel $\mu$, given by
\begin{align}
\nu_{nm}(t,\tau) &= \left\langle \frac{1}{2} \left\{ \boldsymbol{\xi}_n(t) , \boldsymbol{\xi}_m(\tau) \right\} \right\rangle_{\!\!\boldsymbol{\xi}} \, , \\
\mu_{nm}(t,\tau) &= \left\langle \frac{1}{2\imath} \left[ \boldsymbol{\xi}_n(t) , \boldsymbol{\xi}_m(\tau) \right] \right\rangle_{\!\!\boldsymbol{\xi}} \, , \label{eq:lmuC}
\end{align}
which will be discussed more thoroughly in Sec.~\ref{sec:NoiseDecomp}.
Curiously, if we couple the system to a environment of parametric oscillators, then the memory kernel in Eq.~\eqref{eq:l(t)driven} will not be determined by the commutator in Eq.~\eqref{eq:lmuC},
whereas parametric coupling to a environment of simple harmonic oscillators has no effect upon this relation.
Therefore the ``canonical'' quantum Langevin equation we present corresponds to a particular microscopic interpretation of non-stationary correlations.
This also appears to be the case for the influence-functional formalism.

Substituting \eqref{eq:l(t)driven} back into \eqref{eq:LNL} results in a quantum Langevin equation for the open system.
However, due to the fact that $\mathbf{l}(t)$ commutes with system operators at all times, whereas the operator noise $\boldsymbol{\xi}(t)$ and dissipation (apart) do not,
there is no unique method of expressing the quantum Langevin equation in a manifestly Hermitian manner.
All representations generate equivalent solutions, but some are more robust with respect to approximation than others.
In particular, non-Hermitian representations are rather pathological as they only preserve Hermiticity upon noise averaging.
The second-order adjoint master equation \eqref{eq:adjointME} would appear to suggest the following representation
\begin{align}
\dot{\mathbf{X}}(t) &= + \imath \left[ \mathbf{H}(t) , \mathbf{X}(t) \right] + \frac{\imath}{2} \sum_n \left\{ \mathbf{l}_n(t) , \left[ \mathbf{L}_n(t) , \mathbf{X}(t) \right] \right\} \, , \label{eq:QLEf} \\
\mathbf{l}_n(t) &= \boldsymbol{\xi}_n(t) + 2 \sum_m \int_0^t \!\! d\tau \, \mu_{nm}(t,\tau) \, \mathbf{L}_m(\tau) \, . \label{eq:l(t)2}
\end{align}
Although we have derived this Langevin equation under the assumption of a linear environment, we may apply it self-consistently for any Gaussian influence.\footnote{Note that one needs to include inverted oscillators in the environment to mock general correlation functions.}
The quantum Langevin equation here is nonperturbative, however it only approximates non-Gaussian environments in a perturbative manner.
Therefore it parallels the second-order master equation quite well.
Finally note that in the corresponding classical Langevin equation, $\boldsymbol{\xi}_n(t)$ would be real Gaussian noise with two-time correlation $\nu$,
and the dissipation kernel would only appear as given in Eq.~\eqref{eq:l(t)2}.
That the dissipation kernel appears in the operator-noise average is a quantum feature which is made less than obvious with the choice of $\hbar = 1$.


\subsection{The Damping Kernel}
Here we will use the Langevin equation \eqref{eq:QLEf} to determine the role of the dissipation kernel and to motivate an appropriate method of renormalization.
The correlation function $\boldsymbol{\alpha}(t,\tau)$ is positive definite and therefore the noise kernel $\boldsymbol{\nu}(t,\tau)$ must also be positive definite.
However the dissipation kernel $\boldsymbol{\mu}(t,\tau)$ is not positive definite, but it is related to the damping kernel $\boldsymbol{\gamma}(t,\tau)$, which is given by
\begin{align}
\boldsymbol{\mu}(t,\tau) &= -\frac{\partial}{\partial \tau} \boldsymbol{\gamma}(t,\tau) \, , \label{eq:damping_kernel1}
\end{align}
and can be positive definite, negative definite or indefinite, as will be discussed in Sec.~\ref{sec:NoiseDecomp}.
For non-stationary noise, Eq.~\eqref{eq:damping_kernel1} is an incomplete definition, and constructing a symmetric damping kernel will require additional considerations.
Assuming relation \eqref{eq:damping_kernel1} is sufficient, which is the case for stationary correlations, the nonlocal term present in the Langevin equation can be represented as
\begin{align}
&\underbrace{ \int_0^t \!\! d\tau \, \mu_{nm}(t,\tau) \, \mathbf{L}_m(\tau) }_\mathrm{dissipation} = \underbrace{ \int_0^t \!\! d\tau \, \gamma_{nm}(t,\tau) \, \dot{\mathbf{L}}_m(\tau) }_\mathrm{damping} \nonumber \\
& -\underbrace{ \gamma_{nm}(t,t) \, \mathbf{L}_m(t) }_\mathrm{renormalizable} + \underbrace{ \gamma_{nm}(t,0) \, \mathbf{L}_m(0) }_\mathrm{slip} \, .
\end{align}
In the damping-kernel representation we now have three terms: nonlocal damping, renormalizable forces, and transient ``slip'' forces.
The nonlocal damping and renormalizable forces will be more thoroughly considered in the following sections.
The transient slip is a peculiarity associated with factorized initial conditions, and can be avoided with the consideration of a properly correlated initial state \cite{Correlations}.
In the most extreme case, for Ohmic coupling with a delta-correlated damping kernel $\gamma(t,\tau) = 2 \gamma_0 \delta(t-\tau)$,
the slip generates the instantaneous unitary transformation
\begin{align}
\mathbf{X} & \to e^{+\imath \gamma_0 \mathbf{L}^2} \, \mathbf{X} \, e^{-\imath \gamma_0 \mathbf{L}^2} \, , \\
\boldsymbol{\rho} & \to e^{-\imath \gamma_0 \mathbf{L}^2} \, \boldsymbol{\rho} \, e^{+\imath \gamma_0 \mathbf{L}^2} \, ,
\end{align}
which one can invert to identify a more properly correlated initial state, as discussed in \cite{QBM}.

Given a symmetric damping kernel, which is the case for stationary correlations, our quantum Langevin equation can then be expressed as
\begin{align}
& \dot{\mathbf{X}}(t) = + \imath \left[ \mathbf{H}_\mathrm{eff}(t) , \mathbf{X}(t) \right]
+ \frac{\imath}{2} \sum_n \left\{ \boldsymbol{\xi}_n(t) , \left[ \mathbf{L}_n(t) , \mathbf{X}(t) \right] \right\} \nonumber \\
& + \imath \sum_{nm} \int_0^t \!\! d\tau \, \gamma_{nm}(t,\tau)  \left\{ \dot{\mathbf{L}}_m(\tau) , \left[ \mathbf{L}_n(t) , \mathbf{X}(t) \right] \right\} \, , \label{eq:QLEg}
\end{align}
when discarding the transient slip and where the effective Hamiltonian is given by
\begin{align}
\mathbf{H}_\mathrm{eff}(t) &= \mathbf{H}(t) - \sum_{nm} \mathbf{L}_n(t) \, \gamma_{nm}(t,t) \, \mathbf{L}_m(t) \, .
\end{align}
We call this term the effective Hamiltonian and not the renormalized Hamiltonian, as the correction may contain both divergences which require renormalization and terms which describe completely new environmentally-induced forces.
A canonical example of this is non-equilibrium electrodynamics, wherein these corrections contain both the mass renormalization of the electron as well as the magnetostatic forces between electrons \cite{ADL}.

\subsubsection{Renormalization}\label{sec:renormalization1}
Given a symmetric damping kernel, one can see from Eq.~\eqref{eq:QLEg} that the open system experiences effective forces as generated by the Hamiltonian
\begin{align}
\mathbf{H}_\mathrm{eff}(t) &= \mathbf{H}(t) - \sum_{nm} \mathbf{L}_n(t) \, \gamma_{nm}(t,t) \, \mathbf{L}_m(t) \, . \label{eq:Eren}
\end{align}
For local damping this necessitates renormalization.
Moreover, it is physically imperative that renormalization is only performed among these terms in the Langevin equation.
This is not so apparent from the second-order master equation, as one has the unitary generator $\mathbf{H}  + \mathbf{V}$, with
\begin{equation}
\mathbf{V} = \frac{1}{2\imath} \sum_{nm} \left[ \mathbf{L}_n \, (\mathbf{A}_{nm}\! \diamond \mathbf{L}_m) - (\mathbf{A}_{nm}\! \diamond \mathbf{L}_m)^\dagger \, \mathbf{L}_n \right] \, , \label{eq:V}
\end{equation}
which contains a vast number of terms in addition to effective forces.
The effective-force terms will be more clearly delineated in Sec.~\ref{sec:renormalization}.
The following section on energy damping will also motivate the perspective that $\mathbf{H}_\mathrm{eff}(t)$ becomes the more relevant energy of the open system.

\subsubsection{The Energy-Damping Property}\label{sec:Edamping}
To demonstrate that a symmetric damping kernel can be given a physical interpretation we will consider the time evolution of the effective system Hamiltonian.
First we will require the time evolution of the system-coupling operators
\begin{align}
\dot{\mathbf{L}}_m(t) &= + \imath \left[ \mathbf{H}_\mathrm{eff}(t) , \mathbf{L}_m \right] \, .
\end{align}
This relation is exact if all of the system coupling operators $\mathbf{L}_n$ commute,
otherwise it is only perturbative.
Now we feed Eq.~\eqref{eq:Eren} into Eq.~\eqref{eq:QLEg}, apply the above relation, and integrate to obtain the open-system energy as a function of time
\begin{align}
\mathbf{H}_\mathrm{eff}(t) &= \mathbf{H}_\mathrm{eff}(0) - \mathbf{H}_\gamma(t) + \mathbf{H}_\xi(t) \, , \\
\mathbf{H}_\gamma(t) &= +\sum_{nm} \int_0^t \!\! d\tau  \int_0^t \!\!\, d\tau' \, \gamma_{nm}(\tau,\tau')  \left\{ \dot{\mathbf{L}}_n(\tau') , \dot{\mathbf{L}}_m(\tau) \right\} \, , \\
\mathbf{H}_\xi(t) &= -\sum_n \frac{1}{2} \left\{ \boldsymbol{\xi}_n(t) , \dot{\mathbf{L}}_n(t) \right\} \, ,
\end{align}
when neglecting the transient slip.
This relation reveals that a positive-definite damping kernel will only decrease the system energy,
whereas a negative-definite damping kernel will only increase the system energy.
The criteria for each conditions will be more thoroughly covered in Sec.~\ref{sec:Damping}.
This expression for $\mathbf{H}_\gamma(t)$ also contrasts \emph{nonlocal damping} to \emph{local damping}.
Evaluation with a delta correlated damping kernel yields damping which is strictly dissipative at every instant of time whereas nonlocal damping is only assured to be accumulatively dissipative in the full-time integral.

\section{Environmental Correlations}\label{sec:QNC}
\subsection{Classification of Correlations}
\subsubsection{Positivity and Decoherence}

The (multivariate) environmental correlation function, first defined in Eq.~\eqref{eq:alpha}, is Hermitian in the sense of
\begin{align}
\boldsymbol{\alpha}(t,\tau) &= \boldsymbol{\alpha}^{\!\dagger}\!(\tau,t) \, , \label{eq:alpha_her}
\end{align}
and also positive definite in the sense of
\begin{align}
\int_0^t \!\! d\tau_1 \! \int_0^t \!\! d\tau_2 \, \mathbf{f}^\dagger(\tau_1) \, \boldsymbol{\alpha}(\tau_1,\tau_2) \, \mathbf{f}(\tau_2) & \geq 0 \, , \label{eq:posdef1}
\end{align}
for all vector functions $\mathbf{f}(t)$ indexed by the noise.
All quantum correlations are at least \emph{nonlocally decoherent}:
any resultant algebraic dissipator will be positive definite for all time, $\boldsymbol{\Delta}(t) > \mathbf{0}$.
This property is required for completely-positive time evolution as proven in Sec.~\ref{sec:PosProof}.
Nonlocal decoherence only implies that there is more net decoherent evolution than recoherent evolution, as per Eq.~\eqref{eq:off-freq}.
The stricter property of decoherence at every instant in time, $\dot{\boldsymbol{\Delta}}(t) > \mathbf{0}$,
will only  be satisfied generally by delta correlations which exhibit \emph{local decoherence}.
Such correlations will always produce a Lindblad master equation, as can be inferred from Eq.~\eqref{eq:p-Lindblad}.
However some very restricted classes of system-environment interactions, such as the RWA-interaction Hamiltonian~\cite{RWA},
can be constrained by their coupling to be instantaneously decoherent.
This characterizes the class of systems with non-Markovian dynamics whose master equation is, nevertheless, naturally of Lindblad form, though not necessarily at all times.

Finally we note these stochastic processes are partially ordered in their \emph{decoherence strength}.
Given two correlation functions, one can sometimes order them $\boldsymbol{\alpha}_+(t,\tau) > \boldsymbol{\alpha}_-(t,\tau)$ according to the positivity relation \eqref{eq:posdef1}.
For instance, the set of univariate Markov processes is totally ordered by the scalar magnitude of the respective delta correlations,
e.g. $2 \, \delta(t\!-\!\tau) > 1 \, \delta(t\!-\!\tau)$.
This idea is given more consideration, including nontrivial examples, in Ref.~\cite{Decoherence}.

\subsubsection{Time-Dependence and Stability}
\label{sec:NoiseStability}
\emph{Stationary} correlations are defined by their invariance to time translation
\begin{align}
\boldsymbol{\alpha}(t,\tau) &= \boldsymbol{\alpha}(t\!-\!\tau) \, ,
\end{align}
and can produce asymptotically stationary (time-constant) master equations.
Such correlations are produced when the environment is in an initially stationary state
\begin{align}
\boldsymbol{\rho}_\mathrm{E}(0) &= \sum_i p_\mathrm{E}\!\left(\varepsilon_i\right) \ket{\varepsilon_i} \!\! \bra{ \varepsilon_i } \, ,
\end{align}
where $\ket{ \varepsilon }$ denotes the energy basis of the environment and $p_\mathrm{E}\!\left(\varepsilon\right)$ are its stationary probabilities at the initial time.
Furthermore the coupling operators must be constant in time, resulting in
\begin{equation}
\alpha_{nm}(t,\tau) = \sum_{ij} p_\mathrm{E}\!\left(\varepsilon_i\right) \bra{ \varepsilon_i }  \mathbf{l}_n \ket{ \varepsilon_{ij} } \overline{ \bra{ \varepsilon_i }  \mathbf{l}_m \ket{ \varepsilon_{ij} } } \, e^{+ \imath \varepsilon_j (t-\tau)} \, . \label{eq:alpha_stat}
\end{equation}
Comparing to the mode sum
\begin{align}
\boldsymbol{\alpha}(t) &= \frac{1}{2\pi} \int_{-\infty}^{+\infty} \!\!\! d\omega \, e^{+\imath \omega t} \, \tilde{\boldsymbol{\alpha}}(\omega) \, ,
\end{align}
the accompanying characteristic function can be readily identified as
\begin{equation}
\tilde{\alpha}_{nm}(\omega) \;\underline{\propto}\; 2 \pi \sum_i p_\mathrm{E}\!\left(\varepsilon_i\right) \bra{ \varepsilon_i }  \mathbf{l}_n \ket{ \varepsilon_{i}\!-\!\omega } \overline{ \bra{ \varepsilon_i }  \mathbf{l}_m \ket{ \varepsilon_{i}\!-\!\omega } } \, , \label{eq:ideal_bath}
\end{equation}
where the underscored proportionality here is strictly in reference to the continuum limit of the environment which relates environmental mode sums to integrals given the infinitesimal strength of individual environmental mode couplings.
This can be more rigorously defined through the use of a finite spectral-density function in place of the infinitesimal environment couplings.

Also of note are \emph{quasi-stationary} correlations such as of the form
\begin{align}
\boldsymbol{\alpha}(t,\tau) &= \boldsymbol{\alpha}_\mathrm{S}(t\!-\!\tau) + \boldsymbol{\delta \alpha}_\mathrm{NS}(t\!+\!\tau) \, ,
\end{align}
where $\boldsymbol{\alpha}_\mathrm{S}(t\!-\!\tau)$ denotes a stationary correlation
while $\boldsymbol{\delta\alpha}_\mathrm{NS}(t\!+\!\tau)$ is an additional non-stationary contribution.
Such correlations will result from linear coupling to an environment with non-stationary initial state.
Other kinds of quasi-stationary correlations can result from quadratic coupling, etc.
Due to their highly oscillatory behavior in the late-time limit, given the Fourier representation
\begin{equation}
\boldsymbol{\delta\alpha}_\mathrm{NS}(t\!+\!\tau) = \frac{1}{2\pi} \int_{-\infty}^{+\infty} \!\! d\varepsilon \, e^{+\imath \varepsilon (t+\tau)} \, \boldsymbol{\delta}\tilde{\boldsymbol{\alpha}}_\mathrm{NS}(\varepsilon) \, ,
\end{equation}
the non-stationary dynamical contributions typically lose effect therein.
Therefore quasi-stationary correlations can produce an asymptotically stationary master equation with equivalent asymptotics as generated by their corresponding stationary projection $\boldsymbol{\alpha}_\mathrm{S}$.

\emph{Cyclo-stationary} correlations are defined by their invariance to periodic translations
\begin{align}
\boldsymbol{\alpha}(t_1,t_2) &= \boldsymbol{\alpha}(t_1\!+\!\mathcal{T},t_2\!+\!\mathcal{T}) \, ,
\end{align}
and can produce asymptotically cyclo-stationary (periodic) master equations.
Such correlations are produced when the environment is in an initially stationary state and its coupling operators are periodic in time
\begin{align}
\mathbf{l}_n(t) &= \sum_{u} \mathbf{l}_n^{[u]} \, e^{+\imath u \Omega_\mathrm{H} t} \, ,
\end{align}
where $\Omega_\mathrm{H} = \frac{2\pi}{\mathcal{T}}$ is the interaction period.
The correlation function $\boldsymbol{\alpha}(t_1,t_2)$ can then be expressed as
\begin{equation}
\sum_{uv} \boldsymbol{\alpha}_{[uv]}(t_1\!-\!t_2) \, e^{+\imath \frac{u+v}{2} \Omega_\mathrm{H} (t_1-t_2)} \, e^{+\imath \frac{u-v}{2} \Omega_\mathrm{H} (t_1+t_2)} \, , \label{eq:CycloAlpha}
\end{equation}
in terms of the stationary kernels
\begin{equation}
\boldsymbol{\alpha}_{nm}^{[uv]}(t) = \sum_{ij} p_\mathrm{E}(\varepsilon_i) \bra{\varepsilon_i} \mathbf{l}_n^{[u]} \ket{\varepsilon_j} \overline{\bra{\varepsilon_i} \mathbf{l}_m^{[v]} \ket{\varepsilon_j}} \, e^{+\imath \varepsilon_{ij} t} \, . \label{eq:CycloStationary}
\end{equation}
and such that the non-stationary factors of the full correlation function are more obviously periodic.
Equivalent correlations are produced when the environment is in an initially cyclo-stationary state and its coupling operators are constant in time.
I.e. the environmental modes have Floquet-normal-form solutions such that the coupling operators admit the spectral decomposition
\begin{align}
\underline{\mathbf{l}}_n\!(t) &= \sum_{ij,u} \bra{\varepsilon_i} \mathbf{l}_n^{[u]} \ket{\varepsilon_{ij}} e^{+ \imath \varepsilon_j t} \, e^{+\imath u \Omega_\mathrm{H} t} \, , \label{eq:lspec}
\end{align}
where $\ket{ \varepsilon }$ denotes the pseudo-energy basis of the environment associated with the Floquet-normal-form solutions and $\Omega_\mathrm{H}$ is the period of the environment Hamiltonian.

\subsection{Correlation Function Decomposition}
\label{sec:NoiseDecomp}
Second-order correlation functions can always be decomposed into a real \emph{noise kernel} and \emph{dissipation kernel}.
\begin{align}
\underbrace{\boldsymbol{\alpha}(t,\tau)}_\mathrm{complex \, noise} &= \underbrace{\boldsymbol{\nu}(t,\tau)}_\mathrm{noise} + \imath\! \underbrace{\boldsymbol{\mu}(t,\tau)}_\mathrm{dissipation} \, , \label{eq:3kernels}
\end{align}
The dissipation kernel nomenclature is slightly misleading as this kernel may engender a host of effects, with dissipation (used in the physical sense to denote lost in energy , not states) key among them.
The Hermiticity stated in Eq.~\eqref{eq:alpha_her} leads to the relations
\begin{align}
\boldsymbol{\nu}(t,\tau) & \equiv \frac{1}{2} \left[ \boldsymbol{\alpha}(t,\tau) + \boldsymbol{\alpha}^{\!\mathrm{T}\!}(\tau,t) \right] \, , \\
\boldsymbol{\mu}(t,\tau) & \equiv \frac{1}{2\imath} \left[ \boldsymbol{\alpha}(t,\tau) - \boldsymbol{\alpha}^{\!\mathrm{T}\!}(\tau,t) \right] \label{eq:muKernel} \, .
\end{align}
The role of each kernel can be  inferred from the \emph{influence functional}~\cite{Feynman63,CaldeiraLeggett81} and \emph{quantum Langevin equation} (see Sec.~\ref{sec:QLE}).
The noise kernel $\boldsymbol{\nu}$ appears in the influence kernel as the correlation of an ordinary real stochastic source,
whereas the dissipation kernel $\boldsymbol{\mu}$ alone would produce a purely homogeneous (though not generally positivity-preserving) evolution.

For stationary correlations $\boldsymbol{\alpha}(t\!-\!\tau)$ with characteristic function (Fourier transform) $\tilde{\boldsymbol{\alpha}}(\omega)$, the real noise and damping kernels are then Hermitian in both noise index and frequency argument
\begin{align}
\tilde{\boldsymbol{\nu}}(\omega) &= \tilde{\boldsymbol{\nu}}^\dagger(\omega) = \tilde{\boldsymbol{\nu}}^*(-\omega) \, , \label{eq:noise_dher} \\
\tilde{\boldsymbol{\gamma}}(\omega) &= \tilde{\boldsymbol{\gamma}}^\dagger(\omega) = \tilde{\boldsymbol{\gamma}}^*(-\omega) \, , \label{eq:damp_dher}
\end{align}
where for stationary correlations the damping kernel $\boldsymbol{\gamma}$ is uniquely defined by the relation
\begin{align}
\boldsymbol{\mu}(t) &= \frac{d}{dt} \boldsymbol{\gamma}(t) \, ,
\end{align}
This kernel was first introduced for the Langevin equation in Sec.~\ref{sec:QLE} and will be further discussed momentarily.
One implication of this double Hermicity is that, in the noise index, their real symmetric parts are even functions of the frequency while their imaginary anti-symmetric parts are odd functions of the frequency.
More importantly, from Bochner's theorem both $\tilde{\boldsymbol{\alpha}}(\omega)$ and $\tilde{\boldsymbol{\nu}}(\omega)$ are positive-definite for all frequencies.
The damping kernel $\tilde{\boldsymbol{\gamma}}(\omega)$ may be positive definite, negative definite, or indefinite.

The two kernels naturally decompose the second-order operators, Eq.~\eqref{eq:WCOG}, into their Hermitian and anti-Hermitian parts, in the ordinary sense of Hilbert space operators.
\begin{align}
\mathbf{A}_{nm} &= \underbrace{\mathbf{N}_{nm}}_\mathrm{diffusion} + \imath\!\!\! \underbrace{\mathbf{M}_{nm}}_\mathrm{dissipation} \, , \\
(\mathbf{N}_{nm} \!\diamond \mathbf{L}_m)(t) & \equiv  \int_0^t \!\! d\tau \, \nu_{nm}(t,\tau) \, \left\{ \boldsymbol{\mathcal{G}}_0(t,\tau) \, \mathbf{L}_m(\tau) \right\} \, , \\
(\mathbf{M}_{nm} \!\diamond \mathbf{L}_m)(t) & \equiv  \int_0^t \!\! d\tau \, \mu_{nm}(t,\tau) \, \left\{ \boldsymbol{\mathcal{G}}_0(t,\tau) \, \mathbf{L}_m(\tau) \right\} \, ,
\end{align}
where again, there may be more effects in these ``dissipation'' coefficients than dissipation.
The second-order master equation can then be expressed entirely in terms of Hermitian operators
\begin{equation}
\boldsymbol{\mathcal{L}}_2 \, \boldsymbol{\rho} = - \sum_{nm} \left[ \mathbf{L}_n , \imath \left\{ (\mathbf{M}_{nm}\! \diamond \mathbf{L}_m) , \boldsymbol{\rho} \right\} + \left[ (\mathbf{N}_{nm} \!\diamond \mathbf{L}_m) , \boldsymbol{\rho} \right] \right] \, . \label{eq:GHPZ}
\end{equation}
Here the noise coefficients describe diffusion while the dissipation coefficients describe dissipation (or the opposite), renormalization, and other homogeneous dynamics.

The correlation function $\boldsymbol{\alpha}(t,\tau)$ is positive definite and therefore the noise kernel $\boldsymbol{\nu}(t,\tau)$ must also be positive definite.
The dissipation kernel $\boldsymbol{\mu}(t,\tau)$ is not positive definite, but it is related to the previously mentioned damping kernel $\boldsymbol{\gamma}(t,\tau)$, which is given by
\begin{align}
\boldsymbol{\mu}(t,\tau) &= -\frac{\partial}{\partial \tau} \boldsymbol{\gamma}(t,\tau) \, , \label{eq:damping_kernel}
\end{align}
and can be positive definite, negative definite, or indefinite.
For non-stationary noise, Eq.~\eqref{eq:damping_kernel} is an incomplete definition, and constructing a positive-definite damping kernel may require additional considerations.
Assuming relation \eqref{eq:damping_kernel} is sufficient, the dissipation kernel coefficients can then be expressed
\begin{align}
\underbrace{(\mathbf{M}_{nm} \diamond \mathbf{L}_m)(t)}_{\mathrm{dissipation}} =& \underbrace{(\boldsymbol{\Gamma}_{nm} \diamond \dot{\mathbf{L}}_m)(t)}_{\mathrm{damping}} - \underbrace{\gamma_{nm}(t,t) \, \mathbf{L}_m(t)}_{\mathrm{renormalizable}} \nonumber \\
& + \underbrace{\gamma_{nm}(t,0) \left\{ \boldsymbol{\mathcal{G}}_0(t) \, \mathbf{L}_m(0) \right\}}_{\mathrm{slip}} \, , \label{eq:mu-gamma}
\end{align}
in terms of damping kernel coefficients
\begin{align}
(\boldsymbol{\Gamma}_{nm} \diamond \dot{\mathbf{L}}_m)(t) & \equiv \int_0^t \!\! d\tau \, \gamma_{nm}(t,\tau) \, \left\{ \boldsymbol{\mathcal{G}}_0(t,\tau) \, \dot{\mathbf{L}}_m(\tau) \right\} \, , \\
\dot{\mathbf{L}}_m(t) & \equiv +\imath\left[ \mathbf{H}(t) , \mathbf{L}_m(t) \right] + \frac{\partial}{\partial t} \mathbf{L}_m(t) \, , \label{eq:Ldot}
\end{align}
which can be used to place Eq.~\eqref{eq:GHPZ} into a form much like the QBM master equation \cite{HPZ92,QBM}.
Note that if the $\mathbf{L}_n(t)$ do not commute, then Eq.~\eqref{eq:Ldot} is missing higher-order corrections.
The slip is a transient effect, a result of the factorized initial conditions, and will be avoided by the preparation of a properly correlated initial state \cite{Correlations}.
The renormalizable terms cause a long-lasting shift of the system Hamiltonian which diverge in the limit of local or simple damping.

\subsubsection{Renormalization}\label{sec:renormalization}
Note that from the master-equation perspective we have a host of unitary generators $\mathbf{V}$, given by Eq.~\eqref{eq:V}.
\begin{equation}
\mathbf{V} \equiv \frac{1}{2\imath} \sum_{nm} \left[ \mathbf{L}_n \, (\mathbf{A}_{nm}\! \diamond \mathbf{L}_m) - (\mathbf{A}_{nm}\! \diamond \mathbf{L}_m)^\dagger \, \mathbf{L}_n \right] \, ,
\end{equation}
to second order.
It is not advisable to renormalize all of these terms as (1) many of them are physical and (2) doing so will not produce a homogeneous dynamics of interaction system which are most similar to that of the free system.

The renormalizable terms identified above in Eq.~\eqref{eq:mu-gamma}, by guidance from the damping kernel, are equivalent to those more rigorously identified in the quantum Langevin equation (Sec.~\ref{sec:renormalization1}),
where the effective Hamiltonian for the open system is now identified as
\begin{align}
\mathbf{H}_\mathrm{eff}(t) &= \mathbf{H}(t) - \sum_{nm} \mathbf{L}_n(t) \, \gamma_{nm}(t,t) \, \mathbf{L}_m(t) \, .
\end{align}

%

\subsubsection{Classification of Damping Kernels}\label{sec:Damping}
From Sec.~\ref{sec:Edamping}, environments with positive-definite damping kernels are \emph{damping} or \emph{resistive} environments,
while those with negative-definite damping kernels are \emph{amplifying}.
If the coupling variables $\mathbf{L}_n$ are position variables, the damping terms correspond to forces linear in momentum (or velocity).
Stationary correlations are the easiest to dissect.
Their dissipation and damping kernels are related by
\begin{align}
\tilde{\boldsymbol{\mu}}(\varepsilon) &= \imath \varepsilon \, \tilde{\boldsymbol{\gamma}}(\varepsilon) \, ,
\end{align}
and from the definition of the dissipation kernel in Eq.~\eqref{eq:muKernel} and the double Hermiticity in Eq.~\eqref{eq:damp_dher}-\eqref{eq:noise_dher},
the damping kernel will be most-generally positive or negative definite if we have a strict inequality between positive and negative-energy argumented correlation functions.
\begin{align}
\tilde{\boldsymbol{\alpha}}(-|\omega|) &> \tilde{\boldsymbol{\alpha}}^*(+|\omega|)  \;\;\;\;\mathrm{(Damping)}\, , \\
\tilde{\boldsymbol{\alpha}}(-|\omega|) &< \tilde{\boldsymbol{\alpha}}^*(+|\omega|)  \;\;\;\;\mathrm{(Amplifying)}\, .
\end{align}
Using Eq.~\eqref{eq:ideal_bath}, one can show that damping environments result when the initial environmental state probability $p_\mathrm{E}(\varepsilon)$ is a monotonically decreasing function of the environment energy.
Amplifying environments result from monotonically increasing functions or \emph{population inversion}.
The most common example of each being positive and negative temperature reservoirs.

The justification for this notion of resistive and amplifying environments was given rigorously in Sec.~\ref{sec:Edamping} by means of the quantum Langevin equation,
however the second-order master equation gives a perturbatively consistent account.
Given our damping representation of the master-equation coefficients and the adjoint master equation \eqref{eq:adjointME},
one can determine the dynamics of the open-system energy (power) to be
\begin{equation}
\boldsymbol{\mathcal{L}}^\dagger \, \mathbf{H}_\mathrm{eff} = -\left\{ \dot{\mathbf{L}}_n , \left( \boldsymbol{\Gamma}_{nm} \diamond \dot{\mathbf{L}}_m \right) \right\} + \imath \left[  \dot{\mathbf{L}}_n , \left( \mathbf{N}_{nm} \diamond \mathbf{L}_m \right) \right] \, ,
\end{equation}
where we have neglected any power generated by the slip and time-dependence intrinsic to the coupling operators.
Using the zeroth-order solution $\boldsymbol{\rho}(t) = \boldsymbol{\mathcal{G}}_0(t) \, \boldsymbol{\rho}(0)$ and symmetries of the damping kernel,
the second-order expectation value for energy lost through damping can be represented as
\begin{align}
& \int_0^t \!\! d\tau \left\langle \boldsymbol{\mathcal{L}}_\gamma^\dagger(\tau) \, \mathbf{H}_\mathrm{eff}(\tau) \right\rangle =  \\
& -\int_0^t \!\! d\tau_1 \int_0^t \!\! d\tau_2 \sum_{nm} \gamma_{nm}(\tau_1,\tau_2) \, \mathrm{Tr}\! \left[ \underline{\dot{\mathbf{L}}}_n(\tau_1) \, \boldsymbol{\rho}(0) \, \underline{\dot{\mathbf{L}}}_m(\tau_2) \right] \, , \nonumber
\end{align}
which will be strictly dissipative for a positive-definite damping kernel.
This expression also contrasts \emph{nonlocal damping} to \emph{local damping}.
Evaluation with a delta correlated damping kernel yields damping which is strictly dissipative at every instant of time whereas nonlocal damping is only assured to be accumulatively dissipative in the full-time integral.

\subsubsection{Fluctuation- Dissipation Relations and Inequality}
\label{sec:FDR}
From the definitions of the noise and damping kernels, Eq.~\eqref{eq:3kernels}-\eqref{eq:muKernel}, one can easily prove the (stationary) \emph{fluctuation-dissipation inequality}:
\begin{align}
\tilde{\boldsymbol{\nu}}(\omega) &\geq \pm \, \omega \, \tilde{\boldsymbol{\gamma}}(\omega) \, . \label{eq:qnoise_bound}
\end{align}
To prove this one simply notes that the noise kernel is the sum of two positive-definite kernels whereas the dissipation kernel is given by their difference.
The essential point is that if there is any damping, or amplification, there will be quantum noise and Eq.~\eqref{eq:qnoise_bound} determines its lower bound.
This is quite a departure from classical physics where noise can be made to vanish in the zero temperature limit.
The FDI and its relation to the Heisenberg uncertainty principle is discussed more thoroughly in Ref.~\cite{FDR}.

For the case of one collective environment coupling, it is sufficient to define a fluctuation- dissipation relation
\begin{align}
\tilde{\nu}(\omega) &= \tilde{\kappa}(\omega) \, \tilde{\gamma}(\omega) \, , \\
\tilde{\kappa}(\omega) & \equiv \frac{\tilde{\nu}(\omega)}{\tilde{\gamma}(\omega)} \, ,
\end{align}
with $\tilde{\kappa}(\omega)$ the FDR integration kernel which relates fluctuations to dissipation.
For multivariate noise one could use the symmetrized product
\begin{align}
\tilde{\boldsymbol{\nu}}(\omega) &= \frac{1}{2} \left[ \tilde{\boldsymbol{\kappa}}(\omega) \, \tilde{\boldsymbol{\gamma}}(\omega) + \tilde{\boldsymbol{\gamma}}(\omega) \, \tilde{\boldsymbol{\kappa}}(\omega) \right] \, , \label{eq:multiFDR}
\end{align}
which would ensure $\tilde{\boldsymbol{\kappa}}(\omega)$ to be positive definite if $\tilde{\boldsymbol{\gamma}}(\omega)$ is,
in accord with this being a (continuous) Lyapunov equation \cite{Bhatia07}.
Inequality \eqref{eq:qnoise_bound} can then be restated as
\begin{align}
\tilde{\boldsymbol{\kappa}}(\omega) & \geq | \omega | \, ,
\end{align}
for damping environments.
Typically $\tilde{\boldsymbol{\kappa}}(\omega)$ will contain dependence upon the precise nature of environment couplings $\mathbf{l}_n(t)$,
in that if one changes the couplings then the FDR also changes.

\section{Thermal Reservoirs}\label{sec:Equilibrium}
In this section we describe many of the important properties of thermal reservoirs and their corresponding correlations.
In the first subsection we demonstrate the equivalences between \emph{the} fluctuation-dissipation relation, Boltzmann distribution, KMS relation~\cite{Kubo57,Martin59}, and detailed balance in the master equation.

\subsection{Derivation of Thermal Correlations}
\subsubsection{\emph{The} Fluctuation- Dissipation Relation}
As explained in Sec.~\ref{sec:FDR}, a fluctuation-dissipation relation can be almost any (possibly tautological) relation between the noise and damping kernels, though such relations are somewhat constrained by quantum mechanics.
However, if the FDR is to be independent of precisely how the system and environment are coupled, or $\mathbf{l}_n$,
then one can work out from Eq.~\eqref{eq:ideal_bath} that the FDR kernel $\boldsymbol{\kappa}$ must be a scalar quantity, directly related to the initial state of the environment by way of
\begin{align}
\frac{\tilde{\kappa}(\omega)}{\omega} &= \frac{p_\mathrm{E}(\varepsilon \!-\! \omega) + p_\mathrm{E}(\varepsilon)}{p_\mathrm{E}(\varepsilon \!-\! \omega) - p_\mathrm{E}(\varepsilon)} \, ,
\end{align}
for all $\varepsilon$.
But this implies the functional relation
\begin{align}
p_\mathrm{E}(\varepsilon \!-\! \omega) &= \left[\frac{\frac{\tilde{\kappa}(\omega)}{\omega}+1}{\frac{\tilde{\kappa}(\omega)}{\omega}-1}\right] p_\mathrm{E}(\varepsilon) \, ,
\end{align}
where the $\omega$ translations can factor out.
This factorization property is unique to exponential functions,
therefore only the thermal distribution $p_\mathrm{E}(\varepsilon) \propto e^{-\beta \varepsilon}$ can produce a fluctuation- dissipation relation which is generally coupling independent.
We then have the thermal FDR kernel
\begin{align}
\tilde{\kappa}(\omega) &= \omega \coth\!\left( \frac{\omega}{2T} \right) \, ,
\end{align}
which must be maintained no matter how the system is coupled to the environment.
Returning all dimensionful constants reveals the high-temperature and semi-classical FDR
\begin{equation}
\lim_{\hbar \to 0} \tilde{\kappa}(\omega) = \lim_{T \gg \omega} \tilde{\kappa}(\omega) = 2T \, .
\end{equation}

\subsubsection{KMS Relations}
Here we will detail the form of correlations that all thermal reservoirs generate.
We simply assume that our ideal reservoir is in a thermal state
\begin{align}
\boldsymbol{\rho}_B &= \frac{1}{Z} \sum_k e^{-\frac{\varepsilon_k}{T}} \ket{\varepsilon_k}\!\!\bra{ \varepsilon_k } \, , \\
Z &\equiv \sum_k e^{-\frac{\varepsilon_k}{T}} \, .
\end{align}
Next we perform the change of variables $\varepsilon_k \to \varepsilon_k + \frac{\omega}{2}$ on Eq.~\eqref{eq:ideal_bath} to expose the inherent symmetry in all thermal coefficients.
\begin{align}
\tilde{\alpha}_{nm}(\omega) &\;\underline{\propto}\; \frac{2\pi}{Z} e^{-\frac{\omega}{2T}} \sum_k e^{-\frac{\varepsilon_k}{T}} \times \label{eq:Aa} \\
& \bra{ \varepsilon_k \!+\! \frac{\omega}{2} } \! \mathbf{l}_n \! \ket{ \varepsilon_k \!-\! \frac{\omega}{2} } \overline{ \bra{ \varepsilon_k \!+\! \frac{\omega}{2} } \! \mathbf{l}_m \! \ket{ \varepsilon_k \!-\! \frac{\omega}{2} } } \, , \nonumber
\end{align}
where the last factor defines a positive-definite matrix (in the noise index) which is Hermitian in both the noise index and frequency argument.
The damping and noise kernels exhibit double Hermiticity [see Sec.~\eqref{sec:NoiseDecomp}],
whereas the thermal correlation function exhibits ordinary Hermiticity in the noise index and the thermal (a)symmetry
\begin{equation}
\tilde{\boldsymbol{\alpha}}(+\omega) = \tilde{\boldsymbol{\alpha}}^{\!\mathrm{T}\!}(-\omega) \, e^{-\frac{\omega}{T}} = \tilde{\boldsymbol{\alpha}}^{*}(-\omega) \, e^{-\frac{\omega}{T}} \, , \label{eq:thermalsym}
\end{equation}
in its frequency argument.

\subsubsection{Detailed Balance} \label{sec:DetailedBalance}
Equilibrium states are stationary states, though not all stationary states are equilibrium states.
To lowest order in the coupling, stationary states $\mathbf{p}$ are diagonal in the energy basis.
The stationary constraint is given by the Pauli master equation $\mathbf{W}$ \eqref{eq:PauliME}, and takes the form $\mathbf{W} \, \vec{\mathbf{p}} = 0$,
where $\vec{\mathbf{p}}$ denotes the diagonal entries of the stationary state.
In terms of the environment correlations, this works out to be the vanishing of all sums
\begin{equation}
\sum_{jnm} \bra{\omega_i} \mathbf{L}_m \ket{\omega_j} \left[ \tilde{\alpha}_{mn}(\omega_{ji}) \, p_{i} - \tilde{\alpha}_{nm}(\omega_{ij}) \, p_{j} \right] \overline{ \bra{\omega_i} \mathbf{L}_n \ket{\omega_j}} \, .
\end{equation}
This constraint must be satisfied by any stationary state of the system.
Moreover, if the thermal state is to be insensitive to the precise nature of the system coupling, then the more strict constraint
\begin{align}
\frac{p_{i}}{p_{j}} &= \frac{\tilde{\alpha}_{nm}(\omega_{ij})}{\tilde{\alpha}_{mn}(\omega_{ji})} \, , \label{eq:DetBal1}
\end{align}
must be satisfied; this stronger constraint of term-by-term cancellation is known as \emph{detailed balance of the solutions}.
If this condition is met then the system has reached equilibrium with its environment.
If there is not much detail in the system to be balanced, then Eq.~\eqref{eq:DetBal1} can be satisfied trivially.
Such is the case with a system consisting of a single oscillator which can only asymptote into a Gaussian state with linear coupling \cite{FDR}.
But for detailed balance to be attainable in any system, certain transitivity properties must be present in the environmental correlations, e.g.
\begin{align}
\frac{\tilde{\alpha}_{nm}(\omega_{ij})}{\tilde{\alpha}_{mn}(\omega_{ji})} &= \frac{\tilde{\alpha}_{nm}(\omega_{ik})}{\tilde{\alpha}_{mn}(\omega_{ki})} \frac{\tilde{\alpha}_{nm}(\omega_{kj})}{\tilde{\alpha}_{mn}(\omega_{jk})} \, ,
\end{align}
which is known as \emph{detailed balance of the master equation} or equivalently \emph{detailed balance of the correlations.}
Transitivity is a unique property of the exponential function, and therefore only Maxwell-Boltzmann states
\begin{align}
\boldsymbol{\rho}_T &\propto e^{-\beta\mathbf{H}} \, ,
\end{align}
and thermal correlations, Eq.~\eqref{eq:thermalsym},
can non-trivially satisfy this constraint for asymptotic states which are insensitive to the precise nature of the system-environment coupling.
Given that thermal reservoirs generate thermal correlations,
it can be said that the thermal state is the only state which is \emph{self-replicating}.
Any other stationary reservoir will induce a stationary state upon the system which does not generally resemble that of the environment.

Furthermore we can also say something about dynamics of thermalization with an old proof from the theory of Markov chains and Pauli master equations.
Notice that given detailed balance, $\mathbf{W}$ is similar to a real, symmetric matrix
\begin{align}
\mathbf{W} &= \boldsymbol{\rho}_T^{+\frac{1}{2}} \, \mathbf{S} \, \boldsymbol{\rho}_T^{-\frac{1}{2}} \, .
\end{align}
Being real and symmetric, $\mathbf{S}$ must have real eigen-values.
Given the similarity transform, $\mathbf{W}$ must share the same real eigen-values.
The physical implication is that in this weak coupling limit, relaxation generated by $\mathbf{W}$ is purely decay without oscillation.
This property will not hold when near resonance.

\subsection{Properties of Thermal Correlations}
From any of the above expressions, all thermal correlations and noise kernels can be expressed
\begin{align}
\tilde{\boldsymbol{\alpha}}(\omega) &= \tilde{\boldsymbol{\gamma}}(\omega) \, \omega \left[ \coth\!\left(\frac{\omega}{2T}\right) - 1 \right] \, , \\
\tilde{\boldsymbol{\nu}}(\omega) &= \tilde{\boldsymbol{\gamma}}(\omega) \, \omega \coth\!\left(\frac{\omega}{2T}\right) \, , \label{eq:nuThermal}
\end{align}
in terms of the damping kernel $\boldsymbol{\gamma}$.
Alternatively the thermal correlation and damping kernel can be expressed
\begin{align}
\tilde{\boldsymbol{\alpha}}(\omega) &= \tilde{\boldsymbol{\nu}}(\omega) \left[ 1 - \tanh\!\left(\frac{\omega}{2T}\right) \right] \, , \label{eq:alphaThermal} \\
\tilde{\boldsymbol{\gamma}}(\omega) &= \tilde{\boldsymbol{\nu}}(\omega) \frac{1}{\omega} \tanh\!\left(\frac{\omega}{2T}\right) \, ,
\end{align}
in terms of the noise kernel $\boldsymbol{\nu}$.

For positive frequencies larger than the temperature the correlations vanish, but not necessarily for large negative frequencies.
This is the asymmetry between thermal activation and freezing and its lopsidedness is most apparent in the zero-temperature limit.
\begin{align}
\lim_{T \to 0} \tilde{\boldsymbol{\alpha}}(\omega) &= 2 \left|\omega\right| \lim_{T \to 0} \tilde{\boldsymbol{\gamma}}(\omega) \mbox{~~} (\omega<0) \, . \label{eq:zeroA}
\end{align}

The \emph{Markovian regime} (complex white noise) corresponds both local damping and high temperature (local FDR kernel).
The nomenclature of ``thermal noise'' as being $f^0$ is not entirely correct.
$f^0$ correlations are Markovian and high temperature is necessary but not sufficient to be in the Markovian regime.
Non-Markovian thermal noise is perfectly capable of being $1/f^n$ noise.
The Markovian regime is reached when the system time scales are much slower than those of the environment, so that we can take the low-frequency approximation
\begin{align}
\lim_{\omega \to 0} \tilde{\boldsymbol{\gamma}}(\omega) &= \tilde{\boldsymbol{\gamma}}_0 \, , \\
\lim_{\omega \to 0} \tilde{\boldsymbol{\alpha}}(\omega) &= 2 \, \tilde{\boldsymbol{\gamma}}_0 \, T \, . \label{eq:beta_0}
\end{align}
These are the coefficients one would see in the Markov-Lindblad master equation as this is the limit in which the reservoir time scales are much more rapid than the system time scales.

\begin{center}
\begin{figure}[h]
\centering
\includegraphics[width=0.5\textwidth]{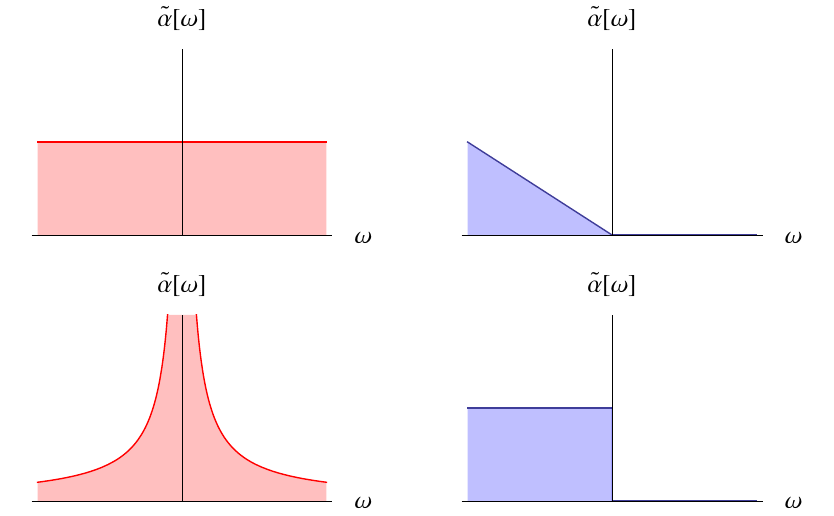}
\caption{Environmental correlations $\tilde{\alpha}(\omega)$ for high-temperature (left) and low-temperature (right) reservoirs, given simple damping $\tilde{\gamma}(\omega)$ (top) and simple dissipation $\tilde{\mu}(\omega)$ (bottom). \label{fig:white}}
\end{figure}
\end{center}

Occasionally in the literature, zero-temperature correlations with simple dissipation, $\tilde{\mu}(\omega)$ constant in frequency or $1/f$ damping, are also referred to as being ``white noise''.
In this case, the noise kernel $\boldsymbol{\nu}$ is local and thus the real noise is white,
however the environment correlations are nonlocal and thus the complex noise which unravels the system evolution is not white.
This ``white noise'' does not strictly correspond to the Markovian regime.
Corresponding to the lack of thermal activation and the lack of negative energy modes in the environment, these correlations only appear white in certain respects.
The upper left plot in Fig.~\ref{fig:white} shows high-temperature (complex) white noise, while the lower right plot shows low temperature ``white noise''.
For low-temperature noise, the small temperature T provides a very long time scale which determines the scale of drastic change between positive and negative frequency behavior.
Low-temperature ``white noise'' generally requires an infrared cutoff to be suitable for positive temperature.
High-temperature white noise generally requires an ultraviolet cutoff to be suitable for finite temperature.

Finally, given a simple damping kernel, it is convenient to specify the convolution in \eqref{eq:MNC2} for our stationary coefficients
using the rational expansion of the hyperbolic cotangent
\begin{align}
\coth\!{\left( \frac{\varepsilon}{2 T} \right)} &= \frac{2 T}{\varepsilon} + \frac{2}{\pi} \sum_{k=1}^\infty \frac{\frac{\varepsilon}{2 \pi T}}{k^2 + \left(\frac{\varepsilon}{2 \pi T}\right)^2} \label{eq:coth} \, ,
\end{align}
so that with proper choice of damping kernel, this contour integral may be evaluated via the residue theorem;
we remember that the $\pm \omega$ poles are regulated away and do not count towards the residue theorem.

Alternatively, given a simple noise kernel, it is convenient to specify the coefficients
using the rational expansion of the hyperbolic tangent
\begin{align}
\tanh\!{\left( \frac{\varepsilon}{2 T} \right)} &= \frac{2}{\pi} \sum_{k=1}^\infty \frac{\frac{\varepsilon}{2 \pi T}}{\left(k - \frac{1}{2}\right)^2 + \left(\frac{\varepsilon}{2 \pi T}\right)^2} \label{eq:tanh} \, .
\end{align}

\section{Discussion}
We have given a relatively simple perturbative formalism for the analysis of open-system dynamics and addressed concerns such as late-time convergence and complete positivity.
Master equations not of Lindblad form (thus describing non-Markovian dynamics) are often viewed with a level of suspicion as they would seem to lack some vital structure necessary to ensure positivity.
We have detailed specifically how this information is encoded in the time dependence of all microscopically derived coefficients.
Moreover, one has to temper expectations of complete positivity to the relevant perturbative order of accuracy generated by an inexact master equation.
It is not well known that the second-order master equation can only provide the diagonal components of the density matrix to zeroth-order at late times~\cite{Accuracy}.
Therefore when solving the second-order master equation, late-time positivity will only be preserved to zeroth-order in the late-time regime,
and in this regard the Lindlbad form is irrelevant.

It is often suspected that these perturbative master equations cannot be employed for significant lengths of time,
and it is certainly true that the late-time and weak-coupling limits do not always commute.
However we showed that the master equation can provide correct  late-time behavior by an expansion in orders of the  coupling, provided that the functional distribution of noise is Gaussian (as to provide cancelation of certain potentially secular terms) and that the environmental correlations are sufficiently localized.
Moreover, one must also be wary of what order of accuracy the master equation is actually capable of providing at late times.
The late-time accuracy is strictly less than the early-time accuracy.

There has also been some debate as to the superiority of the time-local versus time-nonlocal master equation representations.
As we have rigorously shown, both representations result in the same asymptotic behavior, assuming the perturbative formalism is valid.

In conclusion, in this work we have given a new derivation and analysis of the perturbative time-local and nonlocal master equation formalism, including solutions,
non-Markovian quantum regression theorem (QRT) corrections,
nonlinear quantum Langevin equation formalism,
and we have derived a three-way correspondence between the fluctuation-dissipation relation (FDR), Kubo-Martin-Schwinger (KMS) relation and the detailed balance condition.


\section*{Acknowledgement}
We would like to thank Albert Roura for discussions pertaining to multi-time correlations and for works in derivative papers \cite{Decoherence,FDR,Correlations}.
We would also like to thank Nick I. Cummings for discussions pertaining to open-system equilibrium states and for works in derivative papers \cite{Accuracy}.

\appendix

\bibliography{bib}{}

\begin{thebibliography}{55}%
\makeatletter
\providecommand \@ifxundefined [1]{%
 \@ifx{#1\undefined}
}%
\providecommand \@ifnum [1]{%
 \ifnum #1\expandafter \@firstoftwo
 \else \expandafter \@secondoftwo
 \fi
}%
\providecommand \@ifx [1]{%
 \ifx #1\expandafter \@firstoftwo
 \else \expandafter \@secondoftwo
 \fi
}%
\providecommand \natexlab [1]{#1}%
\providecommand \enquote  [1]{``#1''}%
\providecommand \bibnamefont  [1]{#1}%
\providecommand \bibfnamefont [1]{#1}%
\providecommand \citenamefont [1]{#1}%
\providecommand \href@noop [0]{\@secondoftwo}%
\providecommand \href [0]{\begingroup \@sanitize@url \@href}%
\providecommand \@href[1]{\@@startlink{#1}\@@href}%
\providecommand \@@href[1]{\endgroup#1\@@endlink}%
\providecommand \@sanitize@url [0]{\catcode `\\12\catcode `\$12\catcode
  `\&12\catcode `\#12\catcode `\^12\catcode `\_12\catcode `\%12\relax}%
\providecommand \@@startlink[1]{}%
\providecommand \@@endlink[0]{}%
\providecommand \url  [0]{\begingroup\@sanitize@url \@url }%
\providecommand \@url [1]{\endgroup\@href {#1}{\urlprefix }}%
\providecommand \urlprefix  [0]{URL }%
\providecommand \Eprint [0]{\href }%
\providecommand \doibase [0]{http://dx.doi.org/}%
\providecommand \selectlanguage [0]{\@gobble}%
\providecommand \bibinfo  [0]{\@secondoftwo}%
\providecommand \bibfield  [0]{\@secondoftwo}%
\providecommand \translation [1]{[#1]}%
\providecommand \BibitemOpen [0]{}%
\providecommand \bibitemStop [0]{}%
\providecommand \bibitemNoStop [0]{.\EOS\space}%
\providecommand \EOS [0]{\spacefactor3000\relax}%
\providecommand \BibitemShut  [1]{\csname bibitem#1\endcsname}%
\let\auto@bib@innerbib\@empty
\bibitem [{\citenamefont {Kampen}\ and\ \citenamefont
  {Oppenheim}(1997)}]{Kampen97}%
  \BibitemOpen
  \bibfield  {author} {\bibinfo {author} {\bibfnamefont {N.}~\bibnamefont
  {Kampen}}\ and\ \bibinfo {author} {\bibfnamefont {I.}~\bibnamefont
  {Oppenheim}},\ }\href {\doibase 10.1007/BF02181287} {\bibfield  {journal}
  {\bibinfo  {journal} {J. Stat. Phys.}\ }\textbf {\bibinfo {volume} {87}},\
  \bibinfo {pages} {1325} (\bibinfo {year} {1997})}\BibitemShut {NoStop}%
\bibitem [{\citenamefont {Breuer}\ \emph {et~al.}(2003)\citenamefont {Breuer},
  \citenamefont {Ma},\ and\ \citenamefont {Petruccione}}]{Breuer03}%
  \BibitemOpen
  \bibfield  {author} {\bibinfo {author} {\bibfnamefont {H.~P.}\ \bibnamefont
  {Breuer}}, \bibinfo {author} {\bibfnamefont {A.}~\bibnamefont {Ma}}, \ and\
  \bibinfo {author} {\bibfnamefont {F.}~\bibnamefont {Petruccione}},\ }in\
  \href@noop {} {\emph {\bibinfo {booktitle} {Quantum Computing and Quantum
  Bits in Mesoscopic Systems}}},\ \bibinfo {editor} {edited by\ \bibinfo
  {editor} {\bibfnamefont {A.~J.}\ \bibnamefont {Leggett}}, \bibinfo {editor}
  {\bibfnamefont {B.}~\bibnamefont {Ruggiero}}, \ and\ \bibinfo {editor}
  {\bibfnamefont {P.}~\bibnamefont {Silvestrini}}}\ (\bibinfo  {publisher}
  {Kluwer},\ \bibinfo {address} {Dordrecht},\ \bibinfo {year} {2003})\ \Eprint
  {http://arxiv.org/abs/quant-ph/0209153} {quant-ph/0209153} \BibitemShut
  {NoStop}%
\bibitem [{\citenamefont {Strunz}\ and\ \citenamefont {Yu}(2004)}]{Strunz04}%
  \BibitemOpen
  \bibfield  {author} {\bibinfo {author} {\bibfnamefont {W.~T.}\ \bibnamefont
  {Strunz}}\ and\ \bibinfo {author} {\bibfnamefont {T.}~\bibnamefont {Yu}},\
  }\href {\doibase 10.1103/PhysRevA.69.052115} {\bibfield  {journal} {\bibinfo
  {journal} {Phys. Rev. A}\ }\textbf {\bibinfo {volume} {69}},\ \bibinfo
  {pages} {052115} (\bibinfo {year} {2004})}\BibitemShut {NoStop}%
\bibitem [{\citenamefont {Pollard}\ \emph {et~al.}(1997)\citenamefont
  {Pollard}, \citenamefont {Felts},\ and\ \citenamefont
  {Friesner}}]{Pollard97}%
  \BibitemOpen
  \bibfield  {author} {\bibinfo {author} {\bibfnamefont {W.~T.}\ \bibnamefont
  {Pollard}}, \bibinfo {author} {\bibfnamefont {A.~K.}\ \bibnamefont {Felts}},
  \ and\ \bibinfo {author} {\bibfnamefont {R.~A.}\ \bibnamefont {Friesner}},\
  }\href {\doibase 10.1002/9780470141526.ch3} {\bibfield  {journal} {\bibinfo
  {journal} {Adv. Chem. Phys.}\ }\textbf {\bibinfo {volume} {93}},\ \bibinfo
  {pages} {77} (\bibinfo {year} {1997})}\BibitemShut {NoStop}%
\bibitem [{\citenamefont {Carmichael}(1999)}]{Carmichael99}%
  \BibitemOpen
  \bibfield  {author} {\bibinfo {author} {\bibfnamefont {H.~J.}\ \bibnamefont
  {Carmichael}},\ }\href@noop {} {\emph {\bibinfo {title} {Statistical Methods
  in Quantum Optics I}}}\ (\bibinfo  {publisher} {Springer},\ \bibinfo
  {address} {New York},\ \bibinfo {year} {1999})\BibitemShut {NoStop}%
\bibitem [{\citenamefont {Breuer}\ and\ \citenamefont
  {Petruccione}(2002)}]{Breuer02}%
  \BibitemOpen
  \bibfield  {author} {\bibinfo {author} {\bibfnamefont {H.~P.}\ \bibnamefont
  {Breuer}}\ and\ \bibinfo {author} {\bibfnamefont {F.}~\bibnamefont
  {Petruccione}},\ }\href {\doibase 10.1093/acprof:oso/9780199213900.001.0001}
  {\emph {\bibinfo {title} {The Theory of Open Quantum Systems}}}\ (\bibinfo
  {publisher} {Oxford University Press},\ \bibinfo {address} {New York},\
  \bibinfo {year} {2002})\BibitemShut {NoStop}%
\bibitem [{\citenamefont {Kampen}(2007)}]{Kampen07}%
  \BibitemOpen
  \bibfield  {author} {\bibinfo {author} {\bibfnamefont {N.~G.~V.}\
  \bibnamefont {Kampen}},\ }\href {\doibase 10.1002/9780470142530.ch5}
  {\bibfield  {journal} {\bibinfo  {journal} {Adv. Chem. Phys.}\ }\textbf
  {\bibinfo {volume} {34}},\ \bibinfo {pages} {245} (\bibinfo {year}
  {2007})}\BibitemShut {NoStop}%
\bibitem [{\citenamefont {Lindblad}(1976)}]{Lindblad76}%
  \BibitemOpen
  \bibfield  {author} {\bibinfo {author} {\bibfnamefont {G.}~\bibnamefont
  {Lindblad}},\ }\href {\doibase 10.1007/BF01608499} {\bibfield  {journal}
  {\bibinfo  {journal} {Comm. Math. Phys.}\ }\textbf {\bibinfo {volume} {48}},\
  \bibinfo {pages} {119} (\bibinfo {year} {1976})}\BibitemShut {NoStop}%
\bibitem [{\citenamefont {Gorini}\ \emph {et~al.}(1976)\citenamefont {Gorini},
  \citenamefont {Kossakowski},\ and\ \citenamefont {Sudarshan}}]{Gorini76}%
  \BibitemOpen
  \bibfield  {author} {\bibinfo {author} {\bibfnamefont {V.}~\bibnamefont
  {Gorini}}, \bibinfo {author} {\bibfnamefont {A.}~\bibnamefont {Kossakowski}},
  \ and\ \bibinfo {author} {\bibfnamefont {E.~C.~G.}\ \bibnamefont
  {Sudarshan}},\ }\href {\doibase 10.1063/1.522979} {\bibfield  {journal}
  {\bibinfo  {journal} {J. Math. Phys.}\ }\textbf {\bibinfo {volume} {17}},\
  \bibinfo {pages} {821} (\bibinfo {year} {1976})}\BibitemShut {NoStop}%
\bibitem [{\citenamefont {Fleming}\ \emph
  {et~al.}(2010{\natexlab{a}})\citenamefont {Fleming}, \citenamefont {Hu},\
  and\ \citenamefont {Roura}}]{Decoherence}%
  \BibitemOpen
  \bibfield  {author} {\bibinfo {author} {\bibfnamefont {C.~H.}\ \bibnamefont
  {Fleming}}, \bibinfo {author} {\bibfnamefont {B.~L.}\ \bibnamefont {Hu}}, \
  and\ \bibinfo {author} {\bibfnamefont {A.}~\bibnamefont {Roura}},\
  }\href@noop {} {\enquote {\bibinfo {title} {Decoherence strength of multiple
  non-{M}arkovian environments},}\ } (\bibinfo {year} {2010}{\natexlab{a}}),\
  \Eprint {http://arxiv.org/abs/1011.3286} {arXiv:1011.3286 [quant-ph]}
  \BibitemShut {NoStop}%
\bibitem [{\citenamefont {Fleming}\ \emph
  {et~al.}(2010{\natexlab{b}})\citenamefont {Fleming}, \citenamefont {Hu},\
  and\ \citenamefont {Roura}}]{FDR}%
  \BibitemOpen
  \bibfield  {author} {\bibinfo {author} {\bibfnamefont {C.~H.}\ \bibnamefont
  {Fleming}}, \bibinfo {author} {\bibfnamefont {B.~L.}\ \bibnamefont {Hu}}, \
  and\ \bibinfo {author} {\bibfnamefont {A.}~\bibnamefont {Roura}},\
  }\href@noop {} {\enquote {\bibinfo {title} {Non-equilibrium
  fluctuation-dissipation inequality, and non-equilibrium uncertainty
  principle},}\ } (\bibinfo {year} {2010}{\natexlab{b}}),\ \Eprint
  {http://arxiv.org/abs/1012.0681} {arXiv:1012.0681 [quant-ph]} \BibitemShut
  {NoStop}%
\bibitem [{\citenamefont {Fleming}\ \emph
  {et~al.}(2010{\natexlab{c}})\citenamefont {Fleming}, \citenamefont
  {Cummings}, \citenamefont {Anastopoulos},\ and\ \citenamefont {Hu}}]{RWA}%
  \BibitemOpen
  \bibfield  {author} {\bibinfo {author} {\bibfnamefont {C.~H.}\ \bibnamefont
  {Fleming}}, \bibinfo {author} {\bibfnamefont {N.~I.}\ \bibnamefont
  {Cummings}}, \bibinfo {author} {\bibfnamefont {C.}~\bibnamefont
  {Anastopoulos}}, \ and\ \bibinfo {author} {\bibfnamefont {B.~L.}\
  \bibnamefont {Hu}},\ }\href
  {http://stacks.iop.org/1751-8121/43/i=40/a=405304} {\bibfield  {journal}
  {\bibinfo  {journal} {J. Phys. A}\ }\textbf {\bibinfo {volume} {43}},\
  \bibinfo {pages} {405304} (\bibinfo {year} {2010}{\natexlab{c}})}\BibitemShut
  {NoStop}%
\bibitem [{\citenamefont {Fleming}\ and\ \citenamefont
  {Cummings}(2011)}]{Accuracy}%
  \BibitemOpen
  \bibfield  {author} {\bibinfo {author} {\bibfnamefont {C.~H.}\ \bibnamefont
  {Fleming}}\ and\ \bibinfo {author} {\bibfnamefont {N.~I.}\ \bibnamefont
  {Cummings}},\ }\href {\doibase 10.1103/PhysRevE.83.031117} {\bibfield
  {journal} {\bibinfo  {journal} {Phys. Rev. E}\ }\textbf {\bibinfo {volume}
  {83}},\ \bibinfo {pages} {031117} (\bibinfo {year} {2011})}\BibitemShut
  {NoStop}%
\bibitem [{\citenamefont {Fleming}\ \emph
  {et~al.}(2011{\natexlab{a}})\citenamefont {Fleming}, \citenamefont {Roura},\
  and\ \citenamefont {Hu}}]{Correlations}%
  \BibitemOpen
  \bibfield  {author} {\bibinfo {author} {\bibfnamefont {C.~H.}\ \bibnamefont
  {Fleming}}, \bibinfo {author} {\bibfnamefont {A.}~\bibnamefont {Roura}}, \
  and\ \bibinfo {author} {\bibfnamefont {B.~L.}\ \bibnamefont {Hu}},\ }\href
  {\doibase 10.1103/PhysRevE.84.021106} {\bibfield  {journal} {\bibinfo
  {journal} {Phys. Rev. E}\ }\textbf {\bibinfo {volume} {84}},\ \bibinfo
  {pages} {021106} (\bibinfo {year} {2011}{\natexlab{a}})}\BibitemShut
  {NoStop}%
\bibitem [{\citenamefont {Fleming}\ \emph
  {et~al.}(2010{\natexlab{d}})\citenamefont {Fleming}, \citenamefont
  {Cummings}, \citenamefont {Anastopoulos},\ and\ \citenamefont {Hu}}]{Dipole}%
  \BibitemOpen
  \bibfield  {author} {\bibinfo {author} {\bibfnamefont {C.~H.}\ \bibnamefont
  {Fleming}}, \bibinfo {author} {\bibfnamefont {N.~I.}\ \bibnamefont
  {Cummings}}, \bibinfo {author} {\bibfnamefont {C.}~\bibnamefont
  {Anastopoulos}}, \ and\ \bibinfo {author} {\bibfnamefont {B.~L.}\
  \bibnamefont {Hu}},\ }\href@noop {} {\enquote {\bibinfo {title}
  {Non-{M}arkovian entanglement of two-level atoms in an electromagnetic
  field},}\ } (\bibinfo {year} {2010}{\natexlab{d}}),\ \Eprint
  {http://arxiv.org/abs/1012.5067} {arXiv:1012.5067 [quant-ph]} \BibitemShut
  {NoStop}%
\bibitem [{\citenamefont {Subasi}\ \emph {et~al.}(2011)\citenamefont {Subasi},
  \citenamefont {Fleming}, \citenamefont {Taylor},\ and\ \citenamefont
  {Hu}}]{Equilibrium}%
  \BibitemOpen
  \bibfield  {author} {\bibinfo {author} {\bibfnamefont {Y.}~\bibnamefont
  {Subasi}}, \bibinfo {author} {\bibfnamefont {C.~H.}\ \bibnamefont {Fleming}},
  \bibinfo {author} {\bibfnamefont {J.~M.}\ \bibnamefont {Taylor}}, \ and\
  \bibinfo {author} {\bibfnamefont {B.~L.}\ \bibnamefont {Hu}},\ }\href@noop {}
  {\enquote {\bibinfo {title} {The equilibrium states of open systems},}\ }
  (\bibinfo {year} {2011}),\ \bibinfo {note} {\emph{in
  preparation}}\BibitemShut {NoStop}%
\bibitem [{\citenamefont {Swain}(1981)}]{Swain81}%
  \BibitemOpen
  \bibfield  {author} {\bibinfo {author} {\bibfnamefont {S.}~\bibnamefont
  {Swain}},\ }\href {http://stacks.iop.org/0305-4470/14/i=10/a=013} {\bibfield
  {journal} {\bibinfo  {journal} {J. Phys. A}\ }\textbf {\bibinfo {volume}
  {14}},\ \bibinfo {pages} {2577} (\bibinfo {year} {1981})}\BibitemShut
  {NoStop}%
\bibitem [{\citenamefont {Fleming}\ \emph
  {et~al.}(2011{\natexlab{b}})\citenamefont {Fleming}, \citenamefont {Roura},\
  and\ \citenamefont {Hu}}]{QBM}%
  \BibitemOpen
  \bibfield  {author} {\bibinfo {author} {\bibfnamefont {C.~H.}\ \bibnamefont
  {Fleming}}, \bibinfo {author} {\bibfnamefont {A.}~\bibnamefont {Roura}}, \
  and\ \bibinfo {author} {\bibfnamefont {B.~L.}\ \bibnamefont {Hu}},\ }\href
  {\doibase DOI:10.1016/j.aop.2010.12.003} {\bibfield  {journal} {\bibinfo
  {journal} {Ann. Phys.}\ }\textbf {\bibinfo {volume} {326}},\ \bibinfo {pages}
  {1207 } (\bibinfo {year} {2011}{\natexlab{b}})}\BibitemShut {NoStop}%
\bibitem [{\citenamefont {Scala}\ \emph
  {et~al.}(2007{\natexlab{a}})\citenamefont {Scala}, \citenamefont {Militello},
  \citenamefont {Messina}, \citenamefont {Maniscalco}, \citenamefont {Piilo},\
  and\ \citenamefont {Suominen}}]{Scala07}%
  \BibitemOpen
  \bibfield  {author} {\bibinfo {author} {\bibfnamefont {M.}~\bibnamefont
  {Scala}}, \bibinfo {author} {\bibfnamefont {B.}~\bibnamefont {Militello}},
  \bibinfo {author} {\bibfnamefont {A.}~\bibnamefont {Messina}}, \bibinfo
  {author} {\bibfnamefont {S.}~\bibnamefont {Maniscalco}}, \bibinfo {author}
  {\bibfnamefont {J.}~\bibnamefont {Piilo}}, \ and\ \bibinfo {author}
  {\bibfnamefont {K.}~\bibnamefont {Suominen}},\ }\href@noop {} {\bibfield
  {journal} {\bibinfo  {journal} {J. of Phys. A}\ }\textbf {\bibinfo {volume}
  {40}},\ \bibinfo {pages} {14527} (\bibinfo {year}
  {2007}{\natexlab{a}})}\BibitemShut {NoStop}%
\bibitem [{\citenamefont {Hu}\ \emph {et~al.}(1992)\citenamefont {Hu},
  \citenamefont {Paz},\ and\ \citenamefont {Zhang}}]{HPZ92}%
  \BibitemOpen
  \bibfield  {author} {\bibinfo {author} {\bibfnamefont {B.~L.}\ \bibnamefont
  {Hu}}, \bibinfo {author} {\bibfnamefont {J.~P.}\ \bibnamefont {Paz}}, \ and\
  \bibinfo {author} {\bibfnamefont {Y.}~\bibnamefont {Zhang}},\ }\href
  {\doibase 10.1103/PhysRevD.45.2843} {\bibfield  {journal} {\bibinfo
  {journal} {Phys. Rev. D}\ }\textbf {\bibinfo {volume} {45}},\ \bibinfo
  {pages} {2843} (\bibinfo {year} {1992})}\BibitemShut {NoStop}%
\bibitem [{\citenamefont {Xu}\ \emph {et~al.}(2003)\citenamefont {Xu},
  \citenamefont {Mo}, \citenamefont {Cui}, \citenamefont {Lin},\ and\
  \citenamefont {Yan}}]{Xu03}%
  \BibitemOpen
  \bibfield  {author} {\bibinfo {author} {\bibfnamefont {R.~X.}\ \bibnamefont
  {Xu}}, \bibinfo {author} {\bibfnamefont {Y.}~\bibnamefont {Mo}}, \bibinfo
  {author} {\bibfnamefont {P.}~\bibnamefont {Cui}}, \bibinfo {author}
  {\bibfnamefont {S.~H.}\ \bibnamefont {Lin}}, \ and\ \bibinfo {author}
  {\bibfnamefont {Y.~J.}\ \bibnamefont {Yan}},\ }\href@noop {} {\bibfield
  {journal} {\bibinfo  {journal} {Prog. Theo. Chem. Phys.}\ }\textbf {\bibinfo
  {volume} {12}},\ \bibinfo {pages} {7} (\bibinfo {year} {2003})}\BibitemShut
  {NoStop}%
\bibitem [{\citenamefont {de~Vega}\ and\ \citenamefont
  {Alonso}(2006)}]{Vega06}%
  \BibitemOpen
  \bibfield  {author} {\bibinfo {author} {\bibfnamefont {I.}~\bibnamefont
  {de~Vega}}\ and\ \bibinfo {author} {\bibfnamefont {D.}~\bibnamefont
  {Alonso}},\ }\href {\doibase 10.1103/PhysRevA.73.022102} {\bibfield
  {journal} {\bibinfo  {journal} {Phys. Rev. A}\ }\textbf {\bibinfo {volume}
  {73}},\ \bibinfo {pages} {022102} (\bibinfo {year} {2006})}\BibitemShut
  {NoStop}%
\bibitem [{\citenamefont {Scala}\ \emph
  {et~al.}(2007{\natexlab{b}})\citenamefont {Scala}, \citenamefont {Militello},
  \citenamefont {Messina}, \citenamefont {Piilo},\ and\ \citenamefont
  {Maniscalco}}]{Turku07}%
  \BibitemOpen
  \bibfield  {author} {\bibinfo {author} {\bibfnamefont {M.}~\bibnamefont
  {Scala}}, \bibinfo {author} {\bibfnamefont {B.}~\bibnamefont {Militello}},
  \bibinfo {author} {\bibfnamefont {A.}~\bibnamefont {Messina}}, \bibinfo
  {author} {\bibfnamefont {J.}~\bibnamefont {Piilo}}, \ and\ \bibinfo {author}
  {\bibfnamefont {S.}~\bibnamefont {Maniscalco}},\ }\href {\doibase
  10.1103/PhysRevA.75.013811} {\bibfield  {journal} {\bibinfo  {journal} {Phys.
  Rev. A}\ }\textbf {\bibinfo {volume} {75}},\ \bibinfo {pages} {013811}
  (\bibinfo {year} {2007}{\natexlab{b}})}\BibitemShut {NoStop}%
\bibitem [{\citenamefont {Scala}\ \emph
  {et~al.}(2007{\natexlab{c}})\citenamefont {Scala}, \citenamefont {Militello},
  \citenamefont {Messina}, \citenamefont {Maniscalco}, \citenamefont {Piilo},\
  and\ \citenamefont {Suominen}}]{Turku07b}%
  \BibitemOpen
  \bibfield  {author} {\bibinfo {author} {\bibfnamefont {M.}~\bibnamefont
  {Scala}}, \bibinfo {author} {\bibfnamefont {B.}~\bibnamefont {Militello}},
  \bibinfo {author} {\bibfnamefont {A.}~\bibnamefont {Messina}}, \bibinfo
  {author} {\bibfnamefont {S.}~\bibnamefont {Maniscalco}}, \bibinfo {author}
  {\bibfnamefont {J.}~\bibnamefont {Piilo}}, \ and\ \bibinfo {author}
  {\bibfnamefont {K.~A.}\ \bibnamefont {Suominen}},\ }\href
  {http://stacks.iop.org/1751-8121/40/i=48/a=015} {\bibfield  {journal}
  {\bibinfo  {journal} {J. Phys. A}\ }\textbf {\bibinfo {volume} {40}},\
  \bibinfo {pages} {14527} (\bibinfo {year} {2007}{\natexlab{c}})}\BibitemShut
  {NoStop}%
\bibitem [{\citenamefont {Pechukas}(1994)}]{Pechukas94}%
  \BibitemOpen
  \bibfield  {author} {\bibinfo {author} {\bibfnamefont {P.}~\bibnamefont
  {Pechukas}},\ }\href {\doibase 10.1103/PhysRevLett.73.1060} {\bibfield
  {journal} {\bibinfo  {journal} {Phys. Rev. Lett.}\ }\textbf {\bibinfo
  {volume} {73}},\ \bibinfo {pages} {1060} (\bibinfo {year}
  {1994})}\BibitemShut {NoStop}%
\bibitem [{\citenamefont {Shaji}\ and\ \citenamefont
  {Sudarshan}(2005)}]{Sudarshan05}%
  \BibitemOpen
  \bibfield  {author} {\bibinfo {author} {\bibfnamefont {A.}~\bibnamefont
  {Shaji}}\ and\ \bibinfo {author} {\bibfnamefont {E.}~\bibnamefont
  {Sudarshan}},\ }\href {\doibase DOI:10.1016/j.physleta.2005.04.029}
  {\bibfield  {journal} {\bibinfo  {journal} {Phys. Lett. A}\ }\textbf
  {\bibinfo {volume} {341}},\ \bibinfo {pages} {48 } (\bibinfo {year}
  {2005})}\BibitemShut {NoStop}%
\bibitem [{\citenamefont {Ling}\ \emph {et~al.}(2008)\citenamefont {Ling},
  \citenamefont {Nie}, \citenamefont {Qi},\ and\ \citenamefont {Ye}}]{Ling08}%
  \BibitemOpen
  \bibfield  {author} {\bibinfo {author} {\bibfnamefont {C.}~\bibnamefont
  {Ling}}, \bibinfo {author} {\bibfnamefont {J.}~\bibnamefont {Nie}}, \bibinfo
  {author} {\bibfnamefont {L.}~\bibnamefont {Qi}}, \ and\ \bibinfo {author}
  {\bibfnamefont {Y.}~\bibnamefont {Ye}},\ }\href@noop {} {\enquote {\bibinfo
  {title} {Bi-quadratic optimization over unit spheres and semidefinite
  programming relaxations},}\ } (\bibinfo {year} {2008})\BibitemShut {NoStop}%
\bibitem [{\citenamefont {Choi}(1975)}]{Choi75}%
  \BibitemOpen
  \bibfield  {author} {\bibinfo {author} {\bibfnamefont {M.-D.}\ \bibnamefont
  {Choi}},\ }\href {\doibase DOI:10.1016/0024-3795(75)90075-0} {\bibfield
  {journal} {\bibinfo  {journal} {Lin. Alg. \& App.}\ }\textbf {\bibinfo
  {volume} {10}},\ \bibinfo {pages} {285 } (\bibinfo {year}
  {1975})}\BibitemShut {NoStop}%
\bibitem [{\citenamefont {Kraus}(1983)}]{Kraus83}%
  \BibitemOpen
  \bibfield  {author} {\bibinfo {author} {\bibfnamefont {K.}~\bibnamefont
  {Kraus}},\ }\href@noop {} {\emph {\bibinfo {title} {States, Effects and
  Operations: Fundamental Notions of Quantum Theory}}}\ (\bibinfo  {publisher}
  {Springer-Verlag},\ \bibinfo {address} {Berlin},\ \bibinfo {year}
  {1983})\BibitemShut {NoStop}%
\bibitem [{\citenamefont {Kossakowski}(1972)}]{Kossakowski72}%
  \BibitemOpen
  \bibfield  {author} {\bibinfo {author} {\bibfnamefont {A.}~\bibnamefont
  {Kossakowski}},\ }\href {\doibase DOI:10.1016/0034-4877(72)90010-9}
  {\bibfield  {journal} {\bibinfo  {journal} {Rep. Math. Phys.}\ }\textbf
  {\bibinfo {volume} {3}},\ \bibinfo {pages} {247 } (\bibinfo {year}
  {1972})}\BibitemShut {NoStop}%
\bibitem [{\citenamefont {Davies}(1974)}]{Davies74}%
  \BibitemOpen
  \bibfield  {author} {\bibinfo {author} {\bibfnamefont {E.~B.}\ \bibnamefont
  {Davies}},\ }\href {\doibase 10.1007/BF01608389} {\bibfield  {journal}
  {\bibinfo  {journal} {Comm. Math. Phys.}\ }\textbf {\bibinfo {volume} {39}},\
  \bibinfo {pages} {91} (\bibinfo {year} {1974})}\BibitemShut {NoStop}%
\bibitem [{\citenamefont {Davies}(1976)}]{Davies76a}%
  \BibitemOpen
  \bibfield  {author} {\bibinfo {author} {\bibfnamefont {E.~B.}\ \bibnamefont
  {Davies}},\ }\href {\doibase 10.1007/BF01351898} {\bibfield  {journal}
  {\bibinfo  {journal} {Math. Ann.}\ }\textbf {\bibinfo {volume} {219}},\
  \bibinfo {pages} {147} (\bibinfo {year} {1976})}\BibitemShut {NoStop}%
\bibitem [{\citenamefont {Davies}(1977)}]{Davies77}%
  \BibitemOpen
  \bibfield  {author} {\bibinfo {author} {\bibfnamefont {E.~B.}\ \bibnamefont
  {Davies}},\ }\href {\doibase DOI:10.1016/0034-4877(77)90059-3} {\bibfield
  {journal} {\bibinfo  {journal} {Rep. Math. Phys.}\ }\textbf {\bibinfo
  {volume} {11}},\ \bibinfo {pages} {169 } (\bibinfo {year}
  {1977})}\BibitemShut {NoStop}%
\bibitem [{\citenamefont {Alicki}\ and\ \citenamefont
  {Lendi}(2007)}]{Alicki07}%
  \BibitemOpen
  \bibfield  {author} {\bibinfo {author} {\bibfnamefont {R.}~\bibnamefont
  {Alicki}}\ and\ \bibinfo {author} {\bibfnamefont {K.}~\bibnamefont {Lendi}},\
  }\href@noop {} {\emph {\bibinfo {title} {Quantum Dynamical Semigroups and
  Applications}}}\ (\bibinfo  {publisher} {Springer},\ \bibinfo {year}
  {2007})\BibitemShut {NoStop}%
\bibitem [{\citenamefont {Accardi}\ \emph {et~al.}(2002)\citenamefont
  {Accardi}, \citenamefont {Lu},\ and\ \citenamefont {Volovich}}]{Accardi02}%
  \BibitemOpen
  \bibfield  {author} {\bibinfo {author} {\bibfnamefont {L.}~\bibnamefont
  {Accardi}}, \bibinfo {author} {\bibfnamefont {Y.~G.}\ \bibnamefont {Lu}}, \
  and\ \bibinfo {author} {\bibfnamefont {I.~V.}\ \bibnamefont {Volovich}},\
  }\href@noop {} {\emph {\bibinfo {title} {Quantum Theory and Its Stochastic
  Limit}}}\ (\bibinfo  {publisher} {Springer},\ \bibinfo {year}
  {2002})\BibitemShut {NoStop}%
\bibitem [{\citenamefont {Attal}\ \emph {et~al.}(2006)\citenamefont {Attal},
  \citenamefont {Joye},\ and\ \citenamefont {Pillet}}]{Attal06}%
  \BibitemOpen
  \bibfield  {author} {\bibinfo {author} {\bibfnamefont {S.}~\bibnamefont
  {Attal}}, \bibinfo {author} {\bibfnamefont {A.}~\bibnamefont {Joye}}, \ and\
  \bibinfo {author} {\bibfnamefont {C.-A.}\ \bibnamefont {Pillet}},\
  }\href@noop {} {\emph {\bibinfo {title} {Open Quantum Systems II: The
  Markovian Approach}}}\ (\bibinfo  {publisher} {Springer},\ \bibinfo {year}
  {2006})\BibitemShut {NoStop}%
\bibitem [{\citenamefont {Ingarden}\ \emph {et~al.}(1997)\citenamefont
  {Ingarden}, \citenamefont {Kossakowski},\ and\ \citenamefont
  {Ohya}}]{Ingarden97}%
  \BibitemOpen
  \bibfield  {author} {\bibinfo {author} {\bibfnamefont {R.~S.}\ \bibnamefont
  {Ingarden}}, \bibinfo {author} {\bibfnamefont {A.}~\bibnamefont
  {Kossakowski}}, \ and\ \bibinfo {author} {\bibfnamefont {M.}~\bibnamefont
  {Ohya}},\ }\href@noop {} {\emph {\bibinfo {title} {Information Dynamics and
  Open Systems: Classical and Quantum Approach}}}\ (\bibinfo  {publisher}
  {Kluwer Academic},\ \bibinfo {address} {Dordrecht},\ \bibinfo {year}
  {1997})\BibitemShut {NoStop}%
\bibitem [{\citenamefont {Lindblad}(1983)}]{Lindblad83}%
  \BibitemOpen
  \bibfield  {author} {\bibinfo {author} {\bibfnamefont {G.}~\bibnamefont
  {Lindblad}},\ }\href@noop {} {\emph {\bibinfo {title} {Non-Equilibrium
  Entropy and Irreversibility}}}\ (\bibinfo  {publisher} {D. Reidel},\ \bibinfo
  {address} {Dordrecht},\ \bibinfo {year} {1983})\BibitemShut {NoStop}%
\bibitem [{\citenamefont {Weiss}(1993)}]{Weiss93}%
  \BibitemOpen
  \bibfield  {author} {\bibinfo {author} {\bibfnamefont {U.}~\bibnamefont
  {Weiss}},\ }\href@noop {} {\emph {\bibinfo {title} {Quantum Dissipative
  Systems}}}\ (\bibinfo  {publisher} {World Scientific},\ \bibinfo {address}
  {Singapore},\ \bibinfo {year} {1993})\BibitemShut {NoStop}%
\bibitem [{\citenamefont {Petz}\ and\ \citenamefont
  {Sud\'{a}r}(1996)}]{Petz96}%
  \BibitemOpen
  \bibfield  {author} {\bibinfo {author} {\bibfnamefont {D.}~\bibnamefont
  {Petz}}\ and\ \bibinfo {author} {\bibfnamefont {C.}~\bibnamefont
  {Sud\'{a}r}},\ }\href {\doibase 10.1063/1.531535} {\bibfield  {journal}
  {\bibinfo  {journal} {J. Math. Phys.}\ }\textbf {\bibinfo {volume} {37}},\
  \bibinfo {pages} {2662} (\bibinfo {year} {1996})}\BibitemShut {NoStop}%
\bibitem [{\citenamefont {Feynman}\ and\ \citenamefont
  {Vernon}(1963)}]{Feynman63}%
  \BibitemOpen
  \bibfield  {author} {\bibinfo {author} {\bibfnamefont {R.~P.}\ \bibnamefont
  {Feynman}}\ and\ \bibinfo {author} {\bibfnamefont {F.~L.}\ \bibnamefont
  {Vernon}},\ }\href {\doibase DOI:10.1016/0003-4916(63)90068-X} {\bibfield
  {journal} {\bibinfo  {journal} {Ann. Phys.}\ }\textbf {\bibinfo {volume}
  {24}},\ \bibinfo {pages} {118 } (\bibinfo {year} {1963})}\BibitemShut
  {NoStop}%
\bibitem [{\citenamefont {Magnus}(1954)}]{Magnus54}%
  \BibitemOpen
  \bibfield  {author} {\bibinfo {author} {\bibfnamefont {W.}~\bibnamefont
  {Magnus}},\ }\href {\doibase 10.1002/cpa.3160070404} {\bibfield  {journal}
  {\bibinfo  {journal} {Comm. Pure \& Appl. Maths}\ }\textbf {\bibinfo {volume}
  {7}},\ \bibinfo {pages} {649} (\bibinfo {year} {1954})}\BibitemShut {NoStop}%
\bibitem [{\citenamefont {Blanes}\ \emph {et~al.}(1998)\citenamefont {Blanes},
  \citenamefont {Casas}, \citenamefont {Oteo},\ and\ \citenamefont
  {Ros}}]{Blanes98}%
  \BibitemOpen
  \bibfield  {author} {\bibinfo {author} {\bibfnamefont {S.}~\bibnamefont
  {Blanes}}, \bibinfo {author} {\bibfnamefont {F.}~\bibnamefont {Casas}},
  \bibinfo {author} {\bibfnamefont {J.~A.}\ \bibnamefont {Oteo}}, \ and\
  \bibinfo {author} {\bibfnamefont {J.}~\bibnamefont {Ros}},\ }\href
  {http://stacks.iop.org/0305-4470/31/i=1/a=023} {\bibfield  {journal}
  {\bibinfo  {journal} {J. Phys. A}\ }\textbf {\bibinfo {volume} {31}},\
  \bibinfo {pages} {259} (\bibinfo {year} {1998})}\BibitemShut {NoStop}%
\bibitem [{\citenamefont {Fleming}\ \emph
  {et~al.}(2011{\natexlab{c}})\citenamefont {Fleming}, \citenamefont
  {Johnson},\ and\ \citenamefont {Hu}}]{ADL}%
  \BibitemOpen
  \bibfield  {author} {\bibinfo {author} {\bibfnamefont {C.~H.}\ \bibnamefont
  {Fleming}}, \bibinfo {author} {\bibfnamefont {P.~R.}\ \bibnamefont
  {Johnson}}, \ and\ \bibinfo {author} {\bibfnamefont {B.~L.}\ \bibnamefont
  {Hu}},\ }\href@noop {} {\enquote {\bibinfo {title} {Nonequilibrium dynamics
  of charged-particles in an electromagnetic field: causal and stable dynamics
  from $1/c$ expansion of {QED}},}\ } (\bibinfo {year} {2011}{\natexlab{c}}),\
  \Eprint {http://arxiv.org/abs/1106.1886} {arXiv:1106.1886 [quant-ph]}
  \BibitemShut {NoStop}%
\bibitem [{\citenamefont {Fer}(1958)}]{Fer58}%
  \BibitemOpen
  \bibfield  {author} {\bibinfo {author} {\bibfnamefont {F.}~\bibnamefont
  {Fer}},\ }\href@noop {} {\bibfield  {journal} {\bibinfo  {journal} {Bull.
  Classe Sci. Acad. Roy. Belg.}\ }\textbf {\bibinfo {volume} {44}},\ \bibinfo
  {pages} {818} (\bibinfo {year} {1958})}\BibitemShut {NoStop}%
\bibitem [{\citenamefont {Paz}\ and\ \citenamefont {Zurek}(1999)}]{Paz99}%
  \BibitemOpen
  \bibfield  {author} {\bibinfo {author} {\bibfnamefont {J.~P.}\ \bibnamefont
  {Paz}}\ and\ \bibinfo {author} {\bibfnamefont {W.~H.}\ \bibnamefont
  {Zurek}},\ }\href {\doibase 10.1103/PhysRevLett.82.5181} {\bibfield
  {journal} {\bibinfo  {journal} {Phys. Rev. Lett.}\ }\textbf {\bibinfo
  {volume} {82}},\ \bibinfo {pages} {5181} (\bibinfo {year}
  {1999})}\BibitemShut {NoStop}%
\bibitem [{\citenamefont {Briegel}\ and\ \citenamefont
  {Englert}(1993)}]{Briegel93}%
  \BibitemOpen
  \bibfield  {author} {\bibinfo {author} {\bibfnamefont {H.~J.}\ \bibnamefont
  {Briegel}}\ and\ \bibinfo {author} {\bibfnamefont {B.~G.}\ \bibnamefont
  {Englert}},\ }\href {\doibase 10.1103/PhysRevA.47.3311} {\bibfield  {journal}
  {\bibinfo  {journal} {Phys. Rev. A}\ }\textbf {\bibinfo {volume} {47}},\
  \bibinfo {pages} {3311} (\bibinfo {year} {1993})}\BibitemShut {NoStop}%
\bibitem [{\citenamefont {Fleming}(2011)}]{Floquet}%
  \BibitemOpen
  \bibfield  {author} {\bibinfo {author} {\bibfnamefont {C.~H.}\ \bibnamefont
  {Fleming}},\ }\href@noop {} {\enquote {\bibinfo {title} {Canonical-like
  perturbation theory with time dependence},}\ } (\bibinfo {year} {2011}),\
  \bibinfo {note} {\emph{in preparation}}\BibitemShut {NoStop}%
\bibitem [{\citenamefont {Nakajima}(1958)}]{Nakajima58}%
  \BibitemOpen
  \bibfield  {author} {\bibinfo {author} {\bibfnamefont {S.}~\bibnamefont
  {Nakajima}},\ }\href {\doibase 10.1143/PTP.20.948} {\bibfield  {journal}
  {\bibinfo  {journal} {Prog. Theo. Phys.}\ }\textbf {\bibinfo {volume} {20}},\
  \bibinfo {pages} {948} (\bibinfo {year} {1958})}\BibitemShut {NoStop}%
\bibitem [{\citenamefont {Zwanzig}(1960)}]{Zwanzig60}%
  \BibitemOpen
  \bibfield  {author} {\bibinfo {author} {\bibfnamefont {R.}~\bibnamefont
  {Zwanzig}},\ }\href {\doibase 10.1063/1.1731409} {\bibfield  {journal}
  {\bibinfo  {journal} {J. Chem. Phys.}\ }\textbf {\bibinfo {volume} {33}},\
  \bibinfo {pages} {1338} (\bibinfo {year} {1960})}\BibitemShut {NoStop}%
\bibitem [{\citenamefont {Ford}\ \emph {et~al.}(1988)\citenamefont {Ford},
  \citenamefont {Lewis},\ and\ \citenamefont {O'Connell}}]{FordOconnell88}%
  \BibitemOpen
  \bibfield  {author} {\bibinfo {author} {\bibfnamefont {G.~W.}\ \bibnamefont
  {Ford}}, \bibinfo {author} {\bibfnamefont {J.~T.}\ \bibnamefont {Lewis}}, \
  and\ \bibinfo {author} {\bibfnamefont {R.~F.}\ \bibnamefont {O'Connell}},\
  }\href {\doibase 10.1103/PhysRevA.37.4419} {\bibfield  {journal} {\bibinfo
  {journal} {Phys. Rev. A}\ }\textbf {\bibinfo {volume} {37}},\ \bibinfo
  {pages} {4419} (\bibinfo {year} {1988})}\BibitemShut {NoStop}%
\bibitem [{\citenamefont {Caldeira}\ and\ \citenamefont
  {Leggett}(1981)}]{CaldeiraLeggett81}%
  \BibitemOpen
  \bibfield  {author} {\bibinfo {author} {\bibfnamefont {A.~O.}\ \bibnamefont
  {Caldeira}}\ and\ \bibinfo {author} {\bibfnamefont {A.~J.}\ \bibnamefont
  {Leggett}},\ }\href {\doibase 10.1103/PhysRevLett.46.211} {\bibfield
  {journal} {\bibinfo  {journal} {Phys. Rev. Lett.}\ }\textbf {\bibinfo
  {volume} {46}},\ \bibinfo {pages} {211} (\bibinfo {year} {1981})}\BibitemShut
  {NoStop}%
\bibitem [{\citenamefont {Bhatia}(2007)}]{Bhatia07}%
  \BibitemOpen
  \bibfield  {author} {\bibinfo {author} {\bibfnamefont {R.}~\bibnamefont
  {Bhatia}},\ }\href@noop {} {\emph {\bibinfo {title} {Positive Definite
  Matrices}}}\ (\bibinfo  {publisher} {Princeton University Press},\ \bibinfo
  {address} {Princeton and Oxford},\ \bibinfo {year} {2007})\BibitemShut
  {NoStop}%
\bibitem [{\citenamefont {Kubo}(1957)}]{Kubo57}%
  \BibitemOpen
  \bibfield  {author} {\bibinfo {author} {\bibfnamefont {R.}~\bibnamefont
  {Kubo}},\ }\href {\doibase 10.1143/JPSJ.12.570} {\bibfield  {journal}
  {\bibinfo  {journal} {J. Phys. Soc. Jap.}\ }\textbf {\bibinfo {volume}
  {12}},\ \bibinfo {pages} {570} (\bibinfo {year} {1957})}\BibitemShut
  {NoStop}%
\bibitem [{\citenamefont {Martin}\ and\ \citenamefont
  {Schwinger}(1959)}]{Martin59}%
  \BibitemOpen
  \bibfield  {author} {\bibinfo {author} {\bibfnamefont {P.~C.}\ \bibnamefont
  {Martin}}\ and\ \bibinfo {author} {\bibfnamefont {J.}~\bibnamefont
  {Schwinger}},\ }\href {\doibase 10.1103/PhysRev.115.1342} {\bibfield
  {journal} {\bibinfo  {journal} {Phys. Rev.}\ }\textbf {\bibinfo {volume}
  {115}},\ \bibinfo {pages} {1342} (\bibinfo {year} {1959})}\BibitemShut
  {NoStop}%
\end{thebibliography}%
\bibliographystyle{apsrev4-1}

\end{document}